\documentclass[floatfix,twocolumn,showpacs,preprintnumbers,amsmath,amssymb,pra,superscriptaddress,longbibliography]{revtex4-1}
\usepackage{color}
\usepackage[usenames,dvipsnames,svgnames,table]{xcolor}
\usepackage[colorlinks=true,linkcolor=blue,urlcolor=blue,citecolor=blue]{hyperref}
\usepackage{mathtools}
\usepackage{graphicx}
\usepackage{dcolumn}
\usepackage{array}
\usepackage{lipsum}
\usepackage{bm}
\usepackage{subfigure}
\usepackage{amssymb}
\usepackage{multirow}
\usepackage{tabularx}
\usepackage{amsmath}
\usepackage{braket}
\usepackage{csquotes}
\graphicspath{{plots/}}
 \usepackage{lipsum}
\usepackage{mathrsfs}
\usepackage{placeins}

\newcommand{\beq}{\begin{equation}}
\newcommand{\eeq}{\end{equation}}
\newcommand{\bea}{\begin{eqnarray}}
\newcommand{\eea}{\end{eqnarray}}

\begin{document}
\title{ Extraction of the frequency moments of spectral densities from imaginary-time correlation function data
}

\author{Tobias Dornheim}
\email{t.dornheim@hzdr.de}

\affiliation{Center for Advanced Systems Understanding (CASUS), D-02826 G\"orlitz, Germany}
\affiliation{Helmholtz-Zentrum Dresden-Rossendorf (HZDR), D-01328 Dresden, Germany}

\author{Damar C.~Wicaksono}

\affiliation{Center for Advanced Systems Understanding (CASUS), D-02826 G\"orlitz, Germany}
\affiliation{Helmholtz-Zentrum Dresden-Rossendorf (HZDR), D-01328 Dresden, Germany}

\author{Juan~E.~Suarez-Cardona}

\affiliation{Center for Advanced Systems Understanding (CASUS), D-02826 G\"orlitz, Germany}
\affiliation{Helmholtz-Zentrum Dresden-Rossendorf (HZDR), D-01328 Dresden, Germany}
\affiliation{Technische  Universit\"at  Dresden,  D-01062  Dresden,  Germany}

\author{Panagiotis Tolias}
\affiliation{Space and Plasma Physics, Royal Institute of Technology (KTH), Stockholm, SE-100 44, Sweden}

\author{Maximilian P.~B\"ohme}

\affiliation{Center for Advanced Systems Understanding (CASUS), D-02826 G\"orlitz, Germany}
\affiliation{Helmholtz-Zentrum Dresden-Rossendorf (HZDR), D-01328 Dresden, Germany}
\affiliation{Technische  Universit\"at  Dresden,  D-01062  Dresden,  Germany}

\author{Zhandos~A.~Moldabekov}

\affiliation{Center for Advanced Systems Understanding (CASUS), D-02826 G\"orlitz, Germany}
\affiliation{Helmholtz-Zentrum Dresden-Rossendorf (HZDR), D-01328 Dresden, Germany}

\author{Michael Hecht}

\affiliation{Center for Advanced Systems Understanding (CASUS), D-02826 G\"orlitz, Germany}
\affiliation{Helmholtz-Zentrum Dresden-Rossendorf (HZDR), D-01328 Dresden, Germany}

\author{Jan Vorberger}
\affiliation{Helmholtz-Zentrum Dresden-Rossendorf (HZDR), D-01328 Dresden, Germany}

\begin{abstract}
We introduce an exact framework to compute the positive frequency moments $M^{(\alpha)}(\mathbf{q})=\braket{\omega^\alpha}$ of different dynamic properties from imaginary-time quantum Monte Carlo data. As a practical example, we obtain the first five moments of the dynamic structure factor $S(\mathbf{q},\omega)$ of the uniform electron gas at the electronic Fermi temperature based on \emph{ab initio} path integral Monte Carlo simulations. We find excellent agreement with known sum rules for $\alpha=1,3$, and, to our knowledge, present the first results for $\alpha=2,4,5$. Our idea can be straightforwardly generalized to other dynamic properties such as the single-particle spectral function $A(\mathbf{q},\omega)$, and will be useful for a number of applications, including the study of ultracold atoms, exotic warm dense matter, and condensed matter systems.
\end{abstract}
\maketitle

\section{Introduction\label{sec:introduction}}

The accurate understanding of interacting quantum many-body systems constitutes a highly active frontier in physics, quantum chemistry, and related fields. Current challenges include the understanding of the energy loss dynamics of a projectile in a medium~\cite{Nagy_PRA_1989,Balzer_PRB_2016}, photoionization processes in atoms and molecules~\cite{Blaga2009,HOCHSTUHL2010513}, and energy relaxation towards a state of equilibrium~\cite{transfer1,ma15051902}. The accurate description of such nonequilibrium dynamics constitutes a most formidable challenge~\cite{bonitz_book,stefanucci2013nonequilibrium}. Indeed, there as of yet exists no reliable method that is available for all systems and parameters of interest. 
Instead, one usually introduces approximations with respect to the coupling strength.

In thermodynamic equilibrium, different variants of the \emph{ab initio} quantum Monte Carlo (QMC) paradigm~\cite{anderson2007quantum} are, in principle, capable of exactly taking into account the full complex interplay between nonideality (i.e., coupling) and quantum effects. Moreover, the widely used path integral Monte Carlo (PIMC) method~\cite{cep,Berne_JCP_1982,Takahashi_Imada_PIMC_1984} allows to further include thermal excitations without any approximation. Unfortunately, by construction, most QMC methods are limited to the imaginary time domain and, thus, cannot be used in a direct way to compute dynamic properties of interest. On the other hand, many imaginary-time correlation functions (ITCF)~\cite{Berne_JCP_1983,Dornheim_JCP_ITCF_2021} are connected to a dynamic spectral function via an integral expression. For example, the dynamic structure factor (DSF) $S(\mathbf{q},\omega)$ is connected to the imaginary-time density--density correlation function via a two-sided Laplace transform
\begin{eqnarray}\label{eq:analytic_continuation}
F(\mathbf{q},\tau) = \int_{-\infty}^\infty \textnormal{d}\omega\ S(\mathbf{q},\omega)\ e^{-\tau\omega} =: \mathcal{L}\left[S(\mathbf{q},\omega)\right]\ ,
\end{eqnarray}
with $-i\hbar\tau\in -i\hbar[0,\beta]$ the imaginary time argument. In practice, the LHS of Eq.~(\ref{eq:analytic_continuation}) is known with high accuracy from \emph{ab initio} QMC simulations~\cite{Vitali_PRB_2010,Filinov_PRA_2012,Boninsegni_maximum_entropy,dornheim_dynamic,Ferre_PRB_2016,Motta_JCP_2015,Filinov_PRA_2016,Dornheim_SciRep_2022,Dornheim_insight_2022}; the task at hand is thus to numerically invert Eq.~(\ref{eq:analytic_continuation}) to obtain $S(\mathbf{q},\omega)$. This so-called \emph{analytic continuation} is ubiquitous within different fields of physics, including the study of ultracold atoms~\cite{Boninsegni1996,Vitali_PRB_2010,Filinov_PRA_2012,Filinov_PRA_2016,Ferre_PRB_2016,Dornheim_SciRep_2022} and exotic warm dense matter~\cite{dornheim_dynamic,dynamic_folgepaper,Dornheim_PRE_2020}.
In particular, it is of high importance within condensed matter physics~\cite{Mishchenko_PRB_2000,Silver_PRB_1990,Gull_PRL_2021} and constitutes an important ingredient to dynamical mean-field theory simulations~\cite{RevModPhys.78.865,Georges_RMP_1996}. Yet, the analytic continuation constitutes a notoriously difficult problem~\cite{JARRELL1996133,Goulko_PRB_2017}. Indeed, it is ill-posed with respect to the Monte Carlo error bars of $F(\mathbf{q},\tau)$ and subject to a number of practical instabilities. 

Due to the pressing need for an accurate dynamic description of interacting quantum many-body systems, a number of methods have been suggested to deal with the above problem. For example, maximum entropy methods~\cite{PhysRevB.41.2380,Boninsegni_maximum_entropy,Fuchs_PRE_2010} are based on Bayes' theorem and have been successfully applied in different contexts. Yet, the thus reconstructed spectral properties might be biased by the prior model function, although improvements over the original idea are continually being developed~\cite{Boninsegni_maximum_entropy}.
A second line of thought is based on averaging over a large number of noisy random trial solutions~\cite{Sandvik,Vitali_PRB_2010,Mishchenko_PRB_2000,Ferre_PRB_2016}, which includes the genetic inversion by falsification of theories (GIFT) method by Vitali and co-workers~\cite{Vitali_PRB_2010,Bertaina_GIFT_2017} and the stochastic optimization method~\cite{KRIVENKO2019166} introduced by Mishchenko \emph{et al.}~\cite{Mishchenko_PRB_2000}. While being computationally more expensive, such methods have the advantage that no prior information about the spectrum of interest is required. 
Finally, we mention the sparse-modeling technique by Otsuki \emph{et al.}~\cite{Otsuki_PRE_2017,Otsuki_JPSJ_2020,PhysRevB.105.035139}, which is capable of efficiently filtering out the relevant information from the noisy QMC input data.

Despite the aforementioned considerable methodological advances,
a direct analytic continuation only based on Eq.~(\ref{eq:analytic_continuation}) is often insufficient to capture all physical features~\cite{Filinov_PRA_2012,dornheim_dynamic}. Therefore, one must consider additional information such as the frequency moments 
\begin{eqnarray}\label{eq:moments}
M^{(\alpha)}_{\mathrm{S}}(\mathbf{q})=\braket{\omega^\alpha}_{S}=\int_{-\infty}^\infty \textnormal{d}\omega\ S(\mathbf{q},\omega)\ \omega^\alpha\ 
\end{eqnarray}
to further constrain the analytic continuation.
Hitherto, four moments have been known for interacting quantum systems: the normalization $M^{(0)}_{\mathrm{S}}(\mathbf{q})=S(\mathbf{q})$ that is given by the static structure factor, the inverse moment $M^{(-1)}_{\mathrm{S}}(\mathbf{q})$ that is determined by the imaginary-time version of the fluctuation--dissipation theorem~\cite{Dornheim_insight_2022}, and the cases $\alpha=1,3$ that can be evaluated from commutator expressions, known as sum rules~\cite{Mihara_Puff_PR_1968}. In fact, it is possible to reconstruct the DSF from its moments $M^{(\alpha)}_{\mathrm{S}}(\mathbf{q})$, which is known as the \emph{Hamburger problem} in the literature~\cite{tkachenko_book}. This formalism has been successfully utilized by Tkachenko and co-workers to estimate the dynamic structure factor of a number of classical and quantum systems~\cite{Vorberger_PRL_2012,Arkhipov_CPP_2018,Ara_proceeding_2022}. Yet, it is clear that the lack of accurate data for the even moments (except $\alpha=0$) constitutes a substantial bottleneck both with respect to the Hamburger problem, and to constrain the traditional analytic continuation based on Eq.~(\ref{eq:analytic_continuation}).

In this work, we overcome this fundamental limitation by introducing a new exact approach to estimate the positive integer (even and odd) frequency moments based on imaginary-time QMC data. As a practical example, we consider the uniform electron gas (UEG)~\cite{review,loos,quantum_theory}, also known as jellium or quantum one-component plasma in the literature, at the electronic Fermi temperature~\cite{Ott2018} $\Theta=k_\textnormal{B}T/E_\textnormal{F}=1$, where $T$ is the temperature and $E_\textnormal{F}$ is the usual Fermi energy. This system has attracted considerable interest over the last decade~\cite{Brown_PRB_2013,Brown_PRL_2013,Dutta_2013,stls2,dornheim_prl,groth_prl,Malone_PRL_2016,ksdt,Karasiev_status_2019,Dornheim_PRL_2020,dornheim_electron_liquid,dornheim_HEDP,Dornheim_HEDP_2022,castello2021classical,Tolias_JCP_2021,Dornheim_Nature_2022} due to its fundamental importance for so-called warm dense matter~\cite{wdm_book,new_POP,fortov_review,drake2018high}---an exotic state that naturally occurs in astrophysical objects like giant planet interiors~\cite{Militzer_2008,Benuzzi_Mounaix_2014} and is realized in experiments for example in the context of inertial confinement fusion~\cite{hu_ICF,Betti2016}.

In particular, we find good agreement with the sum rules for $\alpha=1$ and $\alpha=3$, and present the first results for $M^{(2)}_{\mathrm{S}}(\mathbf{q})$, $M^{(4)}_{\mathrm{S}}(\mathbf{q})$, and $M^{(5)}_{\mathrm{S}}(\mathbf{q})$ over a wide range of wave numbers. Our results can directly be used as input for an improved prediction of $S(\mathbf{q},\omega)$ via different methods, and constitute an valuable benchmark for the development of new dynamic simulation methods and approximations. Moreover, our idea can be straightforwardly generalized to other dynamic properties such as the single-particle spectral function $A(\mathbf{q},\omega)$, as we demonstrate in Sec.~\ref{sec:summary}. Therefore, we expect it to be of use in a number of fields including the study of ultracold atoms, warm dense matter, and condensed matter physics. Finally, we note that the extraction of physical properties from imaginary-time correlation functions is interesting in its own right, and has given important insights both from a theoretical perspective~\cite{Dornheim_insight_2022,Dornheim_PTR_2022}, and also for the interpretation of X-ray Thomson scattering experiments~\cite{Dornheim_T_2022,Dornheim_T2_2022}.

The paper is organized as follows: In Sec.~\ref{sec:theory}, we introduce the required theoretical background, including a brief introduction to the PIMC estimation of the ITCF (\ref{sec:PIMC}), its connection to the frequency moments of the dynamic structure factor (\ref{sec:DSF}), the frequency moment sum rules (\ref{sec:sum}), and some basic relations from linear response theory (\ref{sec:LRT}). Sec.~\ref{sec:results} is devoted to the presentation of our numerical results, starting with an overview of the UEG model in Sec.~\ref{sec:UEG} and a discussion of the polynomial fitting procedure in Sec.~\ref{sec:canonical}. 
In Sec.~\ref{sec:moments}, we present our new data for the first six (i.e., $\alpha=0,\dots,5$) frequency moments of the DSF and compare them to various theoretical estimates. The paper is concluded by a summary and outlook in Sec.~\ref{sec:summary}.

\section{Theory\label{sec:theory}}

We assume Hartree atomic units throughout this work.

\subsection{Path integral Monte Carlo\label{sec:PIMC}}

It is well established that different QMC methods~\cite{anderson2007quantum} allow for the highly accurate computation of different imaginary-time correlation functions. In this work, we focus on the path integral Monte Carlo approach, which operates at finite temperatures $T$ and gives one straightforward access~\cite{Berne_JCP_1983} to both the Matsubara Green function~\cite{boninsegni1,Filinov_PRA_2012} and also the imaginary-time density--density correlation function that is explored in this work, and that is defined as
\begin{eqnarray}\label{eq:define_F}
F(\mathbf{q},\tau) = \braket{\hat{n}(\mathbf{q},\tau)\hat{n}(-\mathbf{q},0)}_0\ .
\end{eqnarray}
Since a detailed introduction to the PIMC method has been presented elsewhere~\cite{cep,review}, we here restrict ourselves to a brief discussion of its main features that are of relevance for the evaluation of Eq.~(\ref{eq:define_F}). The basic idea behind PIMC is the celebrated classical isomorphism~\cite{Chandler_JCP_1981}, where the complex, nonideal quantum many-body system of interest is mapped onto a classical ensemble of interacting ring polymers. In particular, each quantum particle is represented by an entire path of particle coordinates on each of the $P$ imaginary-time slices, that are separated by an interval of $\epsilon=\beta/P$.

\begin{figure}\centering
\includegraphics[width=0.475\textwidth]{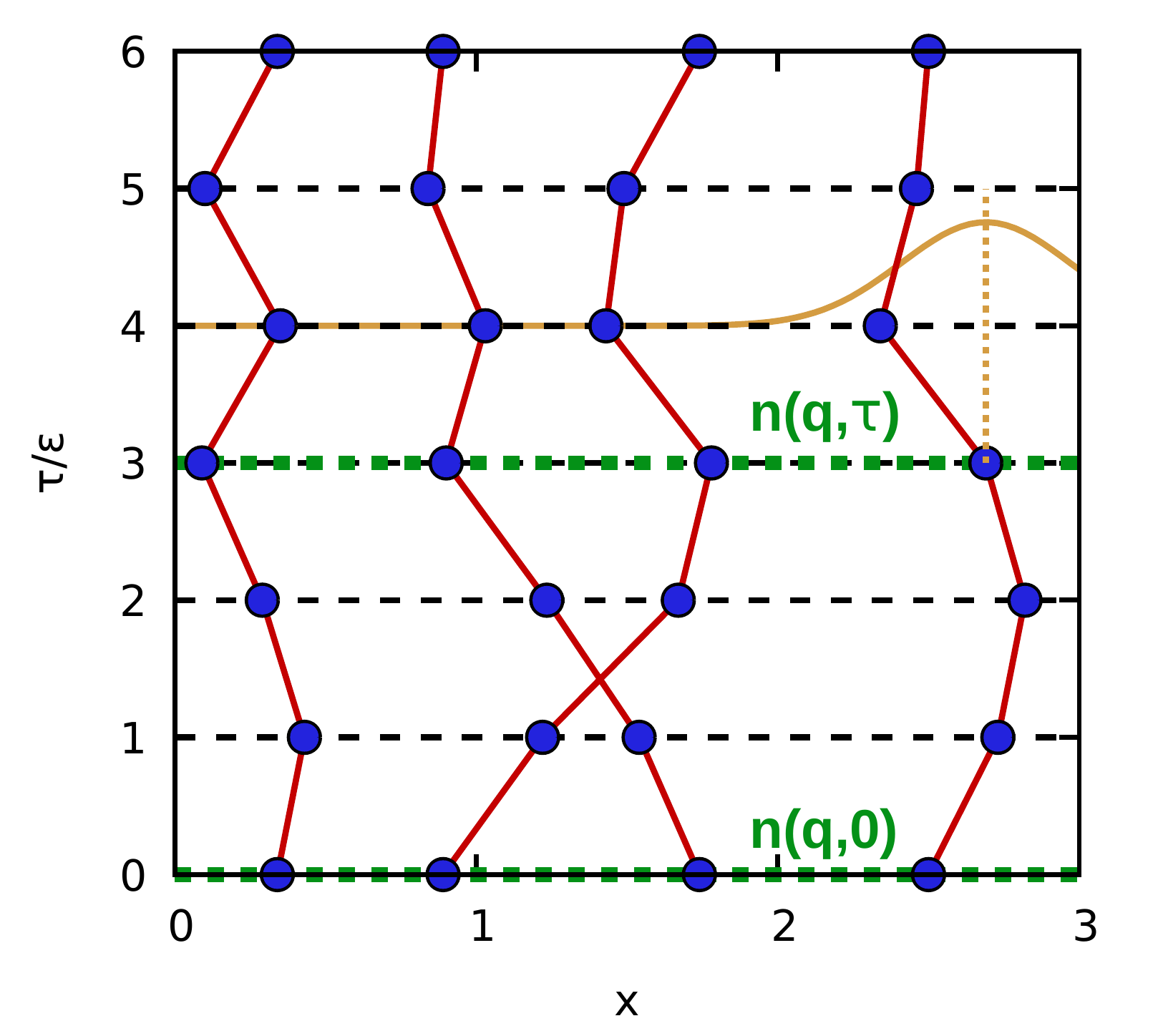}
\caption{\label{fig:PIMC}
Schematic illustration of the PIMC method and the estimation of the ITCF $F(\mathbf{q},\tau)$. Shown is a path-integral configuration with $N=4$ electrons in the $\tau$-$x$-plane. The single pair exchange of the two paths in the center leads to a negative contribution in the case of fermions, thereby contributing to the notorious fermion sign problem~\cite{dornheim_sign_problem}. The dotted green lines illustrate the correlated evaluation of the density in reciprocal space at different imaginary-time arguments for the estimation of $F(\mathbf{q},\tau)$ via Eq.~(\ref{eq:define_F}). The yellow Gaussian on the RHS illustrates the kinetic contribution to the thermal density matrix, which effectively connects beads on adjacent imaginary-time slices via a harmonic spring potential.
Taken from Ref.~\cite{Dornheim_insight_2022} with the permission of the authors.
}
\end{figure} 

This is schematically illustrated in Fig.~\ref{fig:PIMC}, where we show an example configuration of $N=4$ electrons in the $x$-$\tau$-plane. The basic idea of PIMC is to randomly generate a Markov chain of such path configurations that are taken into account according to their appropriate configuration weight; this can be done efficiently based on modern implementations of the Metropolis algorithm~\cite{metropolis}. An additional detail originates from the indistinguishability of fermions and bosons under the exchange of particle coordinates, which requires extending the usual partition function by a summation over all possible permutations. In the path-integral picture, such permutations manifest as so-called exchange-cycles~\cite{Dornheim_permutation_cycles}, which are trajectories with more than a single particle in it; see the two intermediate paths in Fig.~\ref{fig:PIMC} that form such a cycle. While the sampling of all possible permutation topologies within a PIMC simulation is not trivial, it can be efficiently accomplished via the worm algorithm introduced in Refs.~\cite{boninsegni1,boninsegni2}.

A particular challenge is given by the PIMC simulation of quantum degenerate Fermi systems, as the sign of the respective contribution to the total partition function alternates with each pair exchange. This is the root cause of the notorious fermion sign problem, which leads to an exponential increase in the required compute time with important system parameters such as the temperature $T$ or the system size $N$; a topical review on the sign problem in PIMC has been presented in Refs.~\cite{dornheim_sign_problem,Dornheim_2021}. On the one hand, the sign problem can be formally avoided by imposing a nodal restriction on the thermal density matrix. Yet, this simplification comes at the cost of an uncontrolled approximation in practice~\cite{Schoof_PRL_2015}, as the exact nodal structure of an interacting quantum many-body system is not known. On the other hand, such a nodal restriction would destroy the imaginary-time translation invariance in any case, which prevents the straightforward estimation of imaginary-time correlation functions. Therefore, we perform direct PIMC simulations in this work that are computationally demanding due to the sign problem, but exact within the given Monte Carlo error bars. 

In addition, the direct PIMC method allows for a straightforward estimation of $F(\mathbf{q},\tau)$ via Eq.~(\ref{eq:define_F}) in terms of the correlated evaluation of the density in reciprocal space on different imaginary-time slices; see the dashed green lines in Fig.~\ref{fig:PIMC}.

\subsection{Dynamic structure factor and frequency moments\label{sec:DSF}}

The dynamic structure factor $S(\mathbf{q},\omega)$ constitutes the central property in scattering experiments and is given by the Fourier transform of the intermediate scattering function $F(\mathbf{q},t)$~\cite{sheffield2010plasma},
\begin{eqnarray}\label{eq:DSF}
S(\mathbf{q},\omega) = \int_{-\infty}^\infty \textnormal{d}t\ e^{i\omega t} F(\mathbf{q},t)\ ,
\end{eqnarray}
with the latter being defined as
\begin{eqnarray}\label{eq:ISF2}
F(\mathbf{q},t) = \braket{\hat n(\mathbf{q},t)\hat n(-\mathbf{q},0)}_0\ .
\end{eqnarray}
Naturally, the direct evaluation of Eqs.~(\ref{eq:DSF}) and (\ref{eq:ISF2}) requires the availability of dynamic simulations; this is relatively straightforward for classical systems e.g.~via molecular dynamics (MD) simulations~\cite{Baus_Hansen_OCP,hansen2013theory}, but constitutes a most formidable challenge for interacting quantum many-body systems. Similarly, the numerical inversion of Eq.~(\ref{eq:analytic_continuation}) to compute $S(\mathbf{q},\omega)$ based on QMC results for the ITCF $F(\mathbf{q},\tau)$ constitutes an ill-posed problem as it has been explained in the introduction. In lieu of the full DSF $S(\mathbf{q},\omega)$, we will show here how one can still obtain dynamic properties of the given system of interest in the form of the frequency moments $M^{(\alpha)}_{\mathrm{S}}(\mathbf{q})$ defined in Eq.~(\ref{eq:moments}) above.

Let us start by considering the derivative of the ITCF with respect to the imaginary-time argument $\tau$
\beq
\frac{\partial^\alpha}{\partial\tau^\alpha}F(\mathbf{q},\tau)=(-1)^\alpha\int\limits_{-\infty}^{\infty} d\omega\, \omega^\alpha e^{-\tau\omega}S(\mathbf{q},\omega)\ .
\eeq
In particular, the $\tau$-derivative at the origin is given by
\beq\label{eq:derivative}
\frac{\partial^\alpha}{\partial\tau^\alpha}F(\mathbf{q},\tau)\Big|_{\tau=0}=(-1)^\alpha\int\limits_{-\infty}^{\infty} d\omega\, \omega^\alpha S(\mathbf{q},\omega)\ ,
\eeq
thereby giving one direct access to all positive frequency moments of the DSF without the need for an explicit analytic continuation. While being formally exact, we note that the numerical differentiation inherent to Eq.~(\ref{eq:derivative}) is cumbersome in practice and exclusively relies on high-quality information around $\tau=0$.
A better alternative is to perform a Taylor expansion of $F(\mathbf{q},\tau)$ around $\tau=0$,
\begin{eqnarray}
F(\mathbf{q},\tau) &=& \sum_{\alpha=0}^\infty \left\{
\frac{1}{\alpha!} \left. \frac{\partial^\alpha F(\mathbf{q},\tau)}{\partial \tau^\alpha}\right|_{\tau=0}\tau^\alpha
\right\} \\
&=& \sum_{\alpha=0}^\infty c_\alpha(\mathbf{q})\tau^\alpha \ . \label{eq:Taylor}
\end{eqnarray}
In practice, we truncate Eq.~(\ref{eq:Taylor}) at a finite degree $\alpha_\textnormal{max}$ and perform a corresponding polynomial fit to our PIMC data for $F(\mathbf{q},\tau)$. Combining Eqs.~(\ref{eq:derivative}) and (\ref{eq:Taylor}), we thus obtain the frequency moments of the DSF as
\begin{eqnarray}\label{eq:final}
M^{(\alpha)}_{\mathrm{S}}(\mathbf{q})=\left(-1\right)^\alpha \alpha!\ c_\alpha(\mathbf{q})\ .
\end{eqnarray}

\subsection{Frequency moment sum rules\label{sec:sum}}

In what follows, we briefly summarize the established literature background on the DSF frequency moment sum rules. From a practical perspective, their main utility lies in the computation of the various $\langle\omega^{\alpha}\rangle_{S}$ purely on the basis of equilibrium expectation values and, therefore, without the need for dynamic simulations or of an analytic continuation.

The \emph{inverse frequency moment} is easily calculated by combining the static limit of the Kramers--Kronig relation for the real part of the density response function (see also Sec.~\ref{sec:LRT} below) with the fluctuation--dissipation theorem and the detailed balance condition. It is given by~\cite{kugler_bounds,mahan1990many} 
\begin{equation}
\langle\omega^{-1}\rangle_{S}=-\frac{\chi(\boldsymbol{q})}{2n}\,,\label{eq:inverse2}
\end{equation}
where $\chi(\boldsymbol{q})\equiv\chi(\boldsymbol{q},0)$ is the static density response function that is a real quantity and $n=N/\mathcal{V}$ is the number density with $\mathcal{V}=L^3$ the volume of the PIMC simulation cell. Given its origin, it is no surprise that it also directly follows from the ITCF via the imaginary time version of the fluctuation--dissipation theorem~\cite{Dornheim_insight_2022,bowen2}
\begin{equation}
\chi(\boldsymbol{q})=-n\int_0^{\beta}d\tau{F}(\boldsymbol{q},\tau)\,,\label{eq:chi_static}
\end{equation}
leading to the equivalent relation
\begin{equation}
\langle\omega^{-1}\rangle_{S}=\frac{1}{2}\int_0^{\beta}d\tau{F}(\boldsymbol{q},\tau)\,.\label{eq:inverse}
\end{equation}
The \emph{zero frequency moment} is the normalization of the DSF and directly emerges from the static structure factor (SSF) definition $S(\boldsymbol{q})=\langle\hat{n}(\boldsymbol{q})\hat{n}(-\boldsymbol{q})\rangle_0=F(\boldsymbol{q},0)$. It simply reads~\cite{quantum_theory,hansen2013theory,ichimaru_bookII}
\begin{equation}
\langle\omega^{0}\rangle_{S}=S(\boldsymbol{q})\,.\label{eq:static}
\end{equation}
Recall that the combination of the zero frequency moment with the fluctuation--dissipation theorem is the major building block of all schemes of the self--consistent dielectric formalism and the reason that it has been dubbed as self--consistent~\cite{stls_original,SingwiTosi_Review,IIT,review}.

Odd frequency moments of the imaginary part of the density response function $\langle\omega^{2m+1}\rangle_{\mathrm{Im}\chi}$ can be expressed as the statistical averages of equal time commutators at equilibrium~\cite{quantum_theory,SingwiTosi_Review}. This general result stems from the high--frequency expansion of the Kramers--Kronig relation and the short time expansion of the standard definition $\chi(\boldsymbol{q},t)=-\imath\mathrm{H}(t)\langle[\hat{n}(\boldsymbol{q},t),\hat{n}(-\boldsymbol{q},0)]\rangle_0$, where $\mathrm{H}(.)$ is the Heaviside step function~\cite{kugler1,ichimaru_bookII}. Repeated application of the Heisenberg equation of motion converts the arbitrary order time derivatives into iterated commutators, reminiscent of those emerging in the Baker--Campbell--Hausdorff formula, with the number of Hamiltonian nests coinciding with the order of the frequency moment~\cite{quantum_theory}. The connection between $\langle\omega^{2m+1}\rangle_{\mathrm{Im}\chi}$ and $\langle\omega^{2m+1}\rangle_{S}$ is naturally established by the fluctuation--dissipation theorem and it reads as
\begin{equation}
\langle\omega^{2m+1}\rangle_{S}=-\frac{1}{2\pi{n}}\langle\omega^{2m+1}\rangle_{\mathrm{Im}\chi}\,.\label{eq:quantum_correspondence}
\end{equation}
The \emph{first frequency moment}, the universal f--sum rule, expresses particle number conservation~\cite{pines_nozieres_bookI} and is given by~\cite{mahan1990many,quantum_theory,ichimaru_bookII,pines_nozieres_bookI,PlaczekSumRule}
\begin{equation}
\langle\omega^{1}\rangle_{S}=\frac{{q}^2}{2}\,.\label{eq:first}
\end{equation}
The \emph{third frequency moment}, the cubic sum rule, involves the static structure factor (or the pair correlation function) and is given by~\cite{quantum_theory,ichimaru_bookII,PuffSumRule}
\begin{align}
\langle\omega^{3}\rangle_{S}&=\frac{{q}^2}{2}\left\{\frac{q^4}{4}+2q^2{K}+4\pi{n}+\nonumber\right.\\&\quad\left.\frac{4\pi}{\mathcal{V}}\sum_{\boldsymbol{k}\neq\boldsymbol{q},0}\left(\frac{\boldsymbol{q}\cdot\boldsymbol{k}}{qk}\right)^2\left[S(\boldsymbol{q}-\boldsymbol{k})-S(\boldsymbol{k})\right]\right\}\,,\label{eq:third}
\end{align}
where $K=\langle\sum_i\hat{p}_i^2/2\rangle_0$ is the total kinetic energy. The first two terms are kinetic and the last terms pair interacting in nature, with the third term being the Hartree contribution. In the literature, an equivalent form is also encountered that involves a symmetrized Coulomb pair interaction rather than the symmetrized SSF~\cite{Mihara_Puff_PR_1968,plasma2,Iwamoto_PRB_1984,KhodelReview}. It is worth noting that the third frequency moment is directly connected to the high--frequency limit of the dynamic local field correction (LFC) $G(\boldsymbol{q},\omega)$~\cite{NiklassonLimit,Iwamoto_PRB_1984}. This has been exploited for the construction of a static LFC functional of the SSF in the generalized random phase approximation of Pathak \& Vashishta~\cite{PathakVashishtaScheme} and for the construction of a dynamic LFC functional of the SSF in the dielectric scheme of Utsumi \& Ichimaru~\cite{UtsumiIchimaruI}. To our knowledge, the nested commutator formula has not been yet utilized for the computation of higher-order odd-frequency moments. Higher moments have only been reported within the classical limit~\cite{SixthMomentClassical,EighthMomentClassical,IchimaruClassicalSumrule,AilawadiReview}, where the commutators are replaced with Poisson brackets and the calculations simplify considerably after the application of Yvon's theorem~\cite{hansen2013theory}. Nevertheless, there should be a correspondence between the frequency order of the moments and the correlation order of the static distribution functions. For instance, the fifth frequency moment must involve the triplet structure factor $S^{(3)}(\boldsymbol{q},\boldsymbol{q}^{\prime})=\langle\hat{n}(\boldsymbol{q})\hat{n}(\boldsymbol{q}^{\prime})\hat{n}(-\boldsymbol{q}-\boldsymbol{q}^{\prime})\rangle_0$ in reciprocal space or the ternary correlation function $g^{(3)}(\boldsymbol{r},\boldsymbol{r}^{\prime})$ in real space~\cite{plasma2}.

Furthermore, it is important to point out that the even moments of the imaginary part of the density response function $\langle\omega^{2m}\rangle_{\mathrm{Im}\chi}$ are zero given the odd frequency parity of $\mathrm{Im}\chi(\boldsymbol{q},\omega)$~\cite{quantum_theory}. 
Consequently, the even moments of the DSF $\langle\omega^{2m}\rangle_{S}$ cannot be reduced to the equilibrium expectation value of an equal-time commutator. Finally, we point out that in the classical limit $\beta\hbar\omega\ll1$, the fluctuation--dissipation theorem establishes a different correspondence rule, this time between the even DSF moments and the odd $\mathrm{Im}\chi$ moments, that reads as~\cite{IchimaruClassicalSumrule,AilawadiReview,mcquarrie76a,kugler_classical}
\begin{equation}
\langle\omega^{2m}\rangle_{S}^{\mathrm{cl}}=-\frac{1}{\pi{n}\beta}\langle\omega^{2m-1}\rangle_{\mathrm{Im}\chi}^{\mathrm{cl}}\,,\label{eq:classical_correspondence}
\end{equation}
and the detailed balance condition collapses to the even frequency parity of the DSF, which implies that the odd DSF moments identically vanish, \emph{i.e.}~\cite{AilawadiReview,mcquarrie76a,kugler_classical}
\begin{equation}
\langle\omega^{2m+1}\rangle_{S}^{\mathrm{cl}}=0\,.\label{eq:classical_parity}
\end{equation}

\subsection{Linear response theory\label{sec:LRT}}

An alternative route towards the dynamic structure factor comes from linear response theory~\cite{quantum_theory,Dornheim_review}. More specifically, the well-known fluctuation--dissipation theorem relates the imaginary part of the dynamic linear density response function $\chi(\mathbf{q},\omega)$ to the DSF,
\begin{eqnarray}\label{eq:FDT}
S(\mathbf{q},\omega) = - \frac{\textnormal{Im}\chi(\mathbf{q},\omega)}{\pi n (1-e^{-\beta\omega})}\ .
\end{eqnarray}
Here $\chi(\mathbf{q},\omega)$ describes the response of a given system to an external perturbation of wave vector $\mathbf{q}$ and frequency $\omega$, see Ref.~\cite{Dornheim_review} for a recent comprehensive discussion. It is more convenient to utilize the following exact expression for $\chi(\mathbf{q},\omega)$~\cite{kugler1}
\begin{eqnarray}\label{eq:LFC}
\chi(\mathbf{q},\omega) = \frac{\chi_0(\mathbf{q},\omega)}{1-\frac{4\pi}{q^2}\left[1-G(\mathbf{q},\omega)\right]\chi_0(\mathbf{q},\omega)}\ ,
\end{eqnarray}
where $\chi_0(\mathbf{q},\omega)$ denotes the Lindhard function which describes the density response of a noninteracting Fermi gas and can be easily evaluated in practice~\cite{quantum_theory}. The complete wave-vector and frequency-resolved information about electronic exchange--correlation effects is encoded into the dynamic local field correction $G(\mathbf{q},\omega)$, which is formally equivalent to the exchange--correlation kernel $K_\textnormal{xc}(\mathbf{q},\omega)$ known from time-dependent density functional theory simulations~\cite{marques2012fundamentals}. While the exact $G(\mathbf{q},\omega)$
is generally unknown, Eqs.~(\ref{eq:FDT}) and (\ref{eq:LFC}) allow one to compute various approximations to $S(\mathbf{q},\omega)$. For example, when employing $G(\mathbf{q},\omega)\equiv0$, there is no polarization field and the random phase approximation (RPA) emerges that constitutes a mean-field description\,\cite{bonitz_book}. Or, when assuming that $G(\mathbf{q},\omega)\equiv1$, the polarization field exactly cancels out the mean field and the non-interacting density response $\chi(\mathbf{q},\omega)\equiv\chi_0(\mathbf{q},\omega)$ is retrieved. A sophisticated more accurate approach is given by the \emph{static approximation} $G(\mathbf{q},\omega)\equiv G(\mathbf{q},0)$, which has been shown~\cite{dornheim_dynamic,dynamic_folgepaper} to give highly accurate results for $S(\mathbf{q},\omega)$ in the regime of metallic densities $r_s\lesssim4$.
In practice, the \emph{static approximation} can be readily evaluated either using the neural-net representation of $G(\mathbf{q},0;r_s,\Theta)$ from Ref.~\cite{dornheim_ML}, or employing the analytic representation of $G(\mathbf{q},0;r_s,\Theta)$ from Ref.~\cite{Dornheim_PRB_ESA_2021}.

In the context of the present work, the main utility of Eqs.~(\ref{eq:FDT}) and (\ref{eq:LFC}) is a) to generate realistic synthetic data that can be used to verify our idea, and b) to generate approximate reference data for the frequency moments that we extract from our PIMC data for the UEG.

\section{Results\label{sec:results}}

All PIMC results that are presented in this work have been obtained using the extended ensemble approach introduced in Ref.~\cite{Dornheim_PRB_nk_2021}, which is a canonical adaption of the worm algorithm by Boninsegni \emph{et al.}~\cite{boninsegni1,boninsegni2}. More specifically, we employ a primitive factorization scheme with $P=200$ imaginary-time slices, and the convergence with $P$ has been carefully checked. Furthermore, we have carried out simulations with $N=34$ unpolarized electrons, and finite-size effects are known to be small in this regime~\cite{dornheim_prl,Dornheim_JCP_2021,Dornheim_PRE_2020}.

\subsection{Uniform electron gas model\label{sec:UEG}}

The UEG~\cite{loos,quantum_theory,review}, also known as \emph{jellium}, is the quantum version of the classical one-component plasma~\cite{Baus_Hansen_OCP,hansen2013theory,OCP_bridge_2022} and constitutes one of the most fundamental model systems in physics and related disciplines. From a theoretical perspective, it is convenient to characterize the UEG in terms of a few reduced parameters~\cite{Ott2018}. The density parameter serves as the quantum coupling parameter of the UEG and is defined as the ratio of the Wigner-Seitz radius to the Bohr radius, $r_s=d/a_\textnormal{B}$. In the limit of $r_s\to0$ (i.e., high density), the UEG becomes an ideal Fermi gas as the ratio of potential to kinetic energy vanishes proportionally to $r_s$ in this regime. Conversely, the UEG becomes a strongly coupled electron liquid~\cite{dornheim_dynamic,dornheim_electron_liquid,Tolias_JCP_2021}
for $r_s\gtrsim 10$, which gives rise to a number of interesting physical phenomena such as the roton minimum in the spectrum of density fluctuations~\cite{Dornheim_Nature_2022,Takada_PRB_2016}. 
In addition, the degeneracy temperature $\Theta=k_\textnormal{B}T/E_\textnormal{F}$ indicates the degree of quantum degeneracy, with $\Theta\ll1$ being fully degenerate and $\Theta\gg1$ being semi-classical~\cite{Dornheim_HEDP_2022}.
In principle, a third parameter is given by the spin-polarization $\xi=(N^\uparrow-N^\downarrow)/N$ with $N^\uparrow$ and $N^\downarrow$ being the number of electrons with majority and minority spin-orientation. In the present work, we restrict ourselves to the fully unpolarized, i.e., the paramagnetic case with $N^\uparrow=N^\downarrow$ and $\xi=0$. For completeness, we note that finite values of $0\leq\xi\leq1$ (see also Refs.~\cite{review,groth_prl,arora,Dornheim_PRR_2022,ksdt,Brown_PRL_2013,Tanaka_CPP_2017,Ceperley_Alder_PRL_1980,vwn}) are relevant for various applications such as spin-density functional theory calculations in quantum chemistry~\cite{fiolhais2008primer}.

In the ground-state limit with $\Theta=0$, the UEG constitutes a simple model for the conduction electrons in alkali metals~\cite{mahan1990many}. Moreover, the accurate parametrization of its properties~\cite{vwn,Perdew_Zunger_PRB_1981,Gori-Giorgi_PRB_2000,cdop} based on highly accurate ground-state QMC simulations~\cite{Ceperley_Alder_PRL_1980,Ortiz_PRB_1994,Ortiz_PRL_1999,Spink_PRB_2013} has facilitated many applications including the arguably unrivaled success of density functional theory with respect to the description of real materials~\cite{Jones_RMP_2015}.
In this work, we consider the case of $\Theta=1$ and $r_s\sim1$, which is commonly referred to as \emph{warm dense matter} (WDM) in the literature~\cite{wdm_book,new_POP,review}. These extreme conditions naturally occur in astrophysical objects such as giant planet interiors~\cite{Benuzzi_Mounaix_2014,fortov_review} and can be realized in the laboratory for example in inertial confinement fusion experiments~\cite{Betti2016,hu_ICF}. From a theoretical perspective, the accurate description of WDM is notoriously challenging due to the intriguingly intricate interplay of Coulomb coupling with strong thermal excitations and quantum degeneracy effects such as Pauli blocking and diffraction~\cite{new_POP,wdm_book}. Therefore, first accurate results for the warm dense UEG~\cite{review,status} have become available only recently based on different thermal QMC methods~\cite{Brown_PRL_2013,Schoof_PRL_2015,Malone_JCP_2015,Malone_PRL_2016,Dornheim_NJP_2015,Dornheim_JCP_2015,Joonho_JCP_2021,Yilmaz_JCP_2020}. In addition, we consider the case of $\Theta=1$ and $r_s=10$, which is located at the margin of the electron liquid regime. These conditions are particularly interesting as they exhibit a wealth of interesting phenomena, including the aforementioned \emph{roton minimum} in the dynamic structure factor~\cite{dornheim_dynamic,dynamic_folgepaper} which has been explained only recently~\cite{Dornheim_Nature_2022}.

\subsection{Canonical representation of the ITCF\label{sec:canonical}}

\begin{figure}\centering
\includegraphics[width=0.475\textwidth]{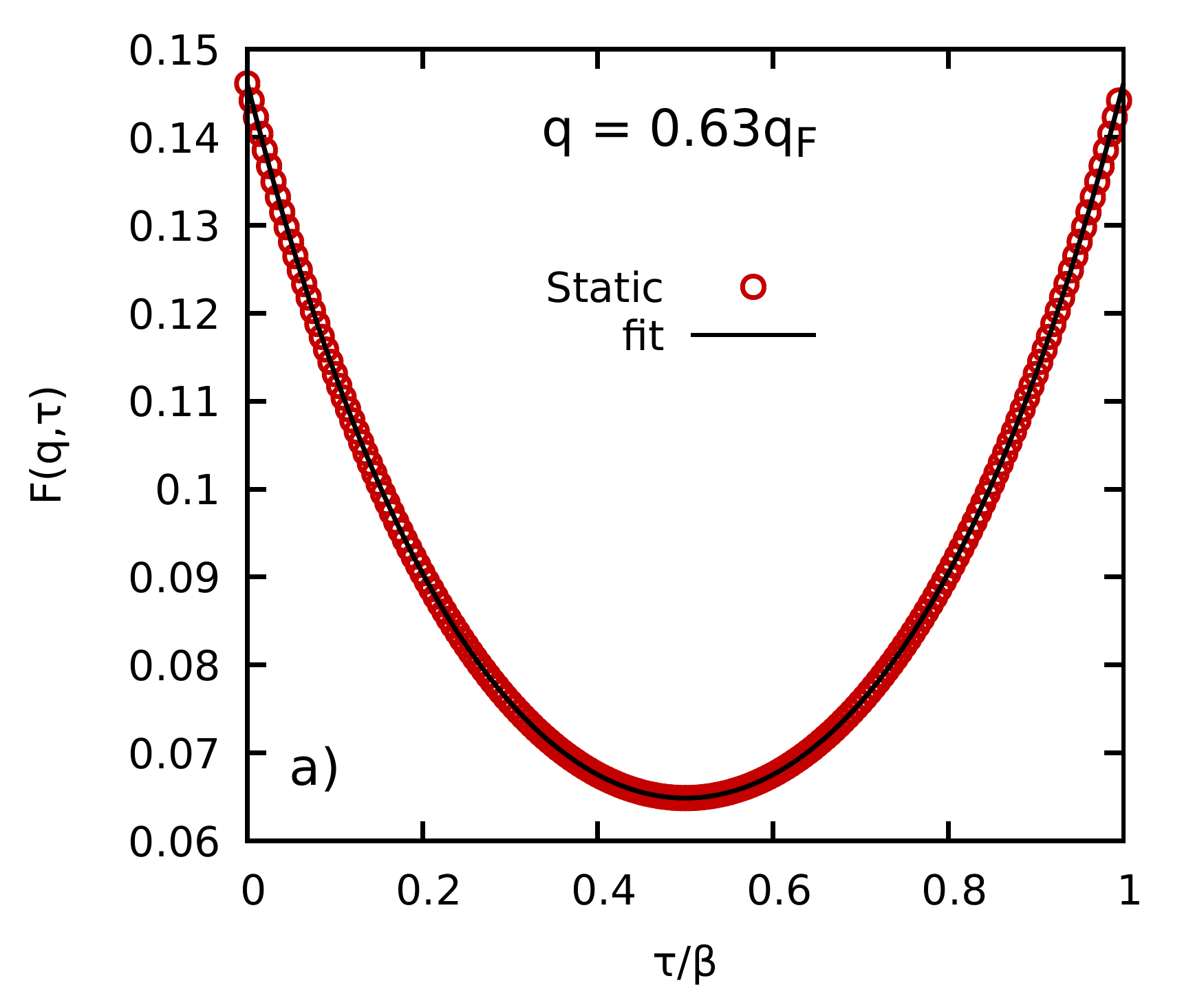}\\\vspace*{-0.43cm}\includegraphics[width=0.464\textwidth]{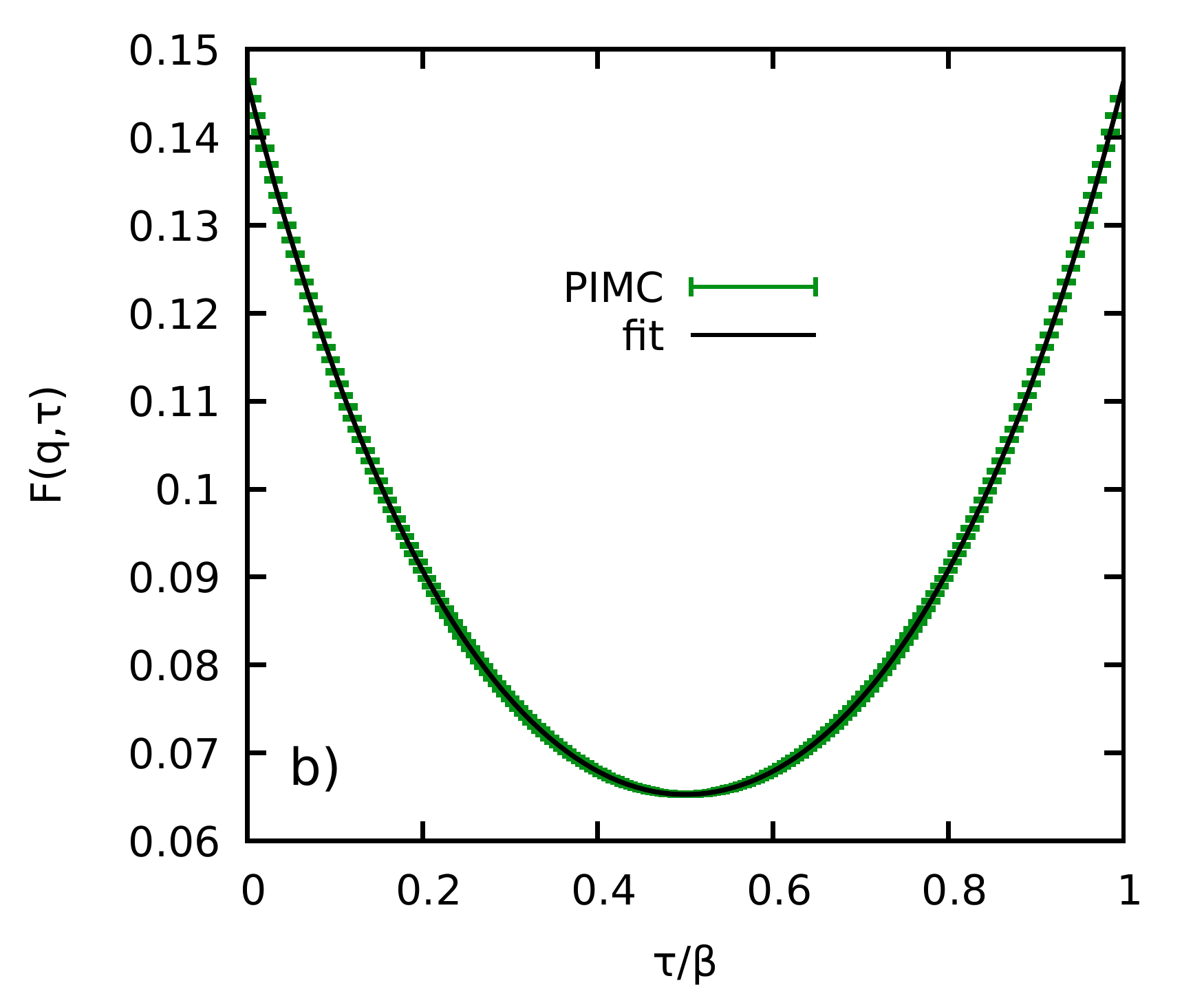}\\\vspace*{-0.43cm}\includegraphics[width=0.464\textwidth]{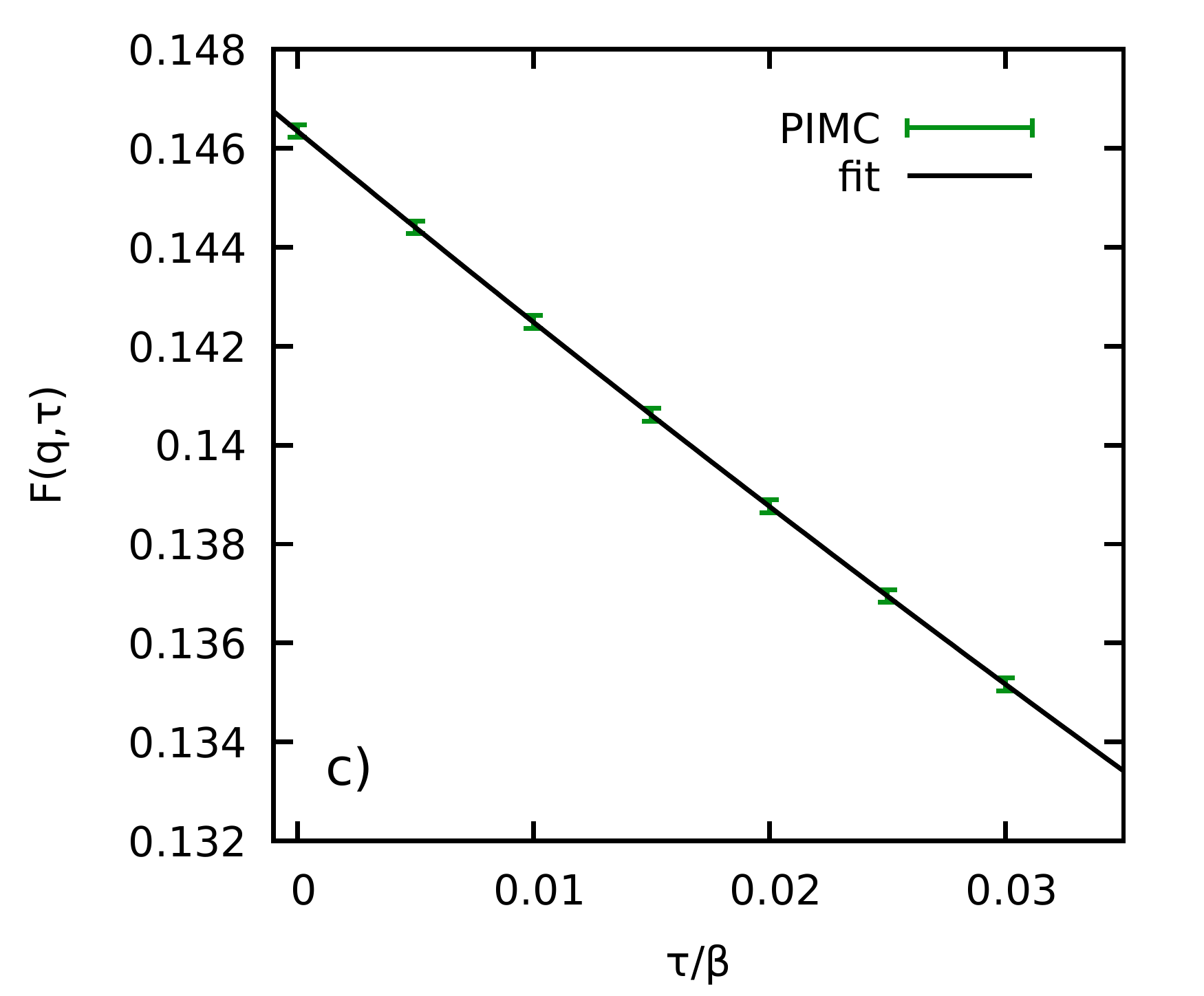}
\caption{\label{fig:F_rs10_theta1}
Illustration of the fitting procedure for the UEG at $r_s=10$ and $\Theta=1$ for $q=0.63q_\textnormal{F}$. a) Synthetic data for $F(\mathbf{q},\tau)$ within the \emph{static approximation} (red) and corresponding canonical fit of order $\alpha_\textnormal{max}=10$ (black). b) PIMC data for $F(\mathbf{q},\tau)$ (green) and corresponding fit with $\alpha_\textnormal{max}=6$; c) magnified segment showing the PIMC error bars.
}
\end{figure} 

In Fig.~\ref{fig:F_rs10_theta1}, we demonstrate the proposed fitting procedure for the UEG at $r_s=10$ and $\Theta=1$, and for the wave number $q=0.63q_\textnormal{F}$. The red circles in panel a) have been computed by obtaining $S(\mathbf{q},\omega)$ within the \emph{static approximation} (see Sec.~\ref{sec:LRT}), and subsequently evaluating the two-sided Laplace transform Eq.~(\ref{eq:analytic_continuation}) on a realistic $\tau$-grid ($P=200$). The solid black curve depicts a corresponding canonical fit according to Eq.~(\ref{eq:Taylor}). Empirically, we find a maximum significant order of $\alpha_\textnormal{max}$ for the polynomial expansion for this example, see Appendix~\ref{sec:details} for more details about the fitting procedure. For completeness, we note that $F(\mathbf{q},\tau)$ is symmetric around $\tau=\beta/2$, i.e., $F(\mathbf{q},\tau)=F(\mathbf{q},\beta-\tau)$. In practice, even though the information about the ITCF for $\beta/2<\tau\leq\beta$ is technically redundant, it is still strongly beneficial to fit Eq.~(\ref{eq:Taylor}) over the entire $\tau$-range as the symmetry condition is not automatically incorporated into the canonical representation of the polynomial. Therefore, the range $\beta/2 < \tau \leq\beta$ significantly helps to determine the coefficients $c_\alpha(\mathbf{q})$ and in this way to get more reliable information in particular about the higher frequency moments $M^{(\alpha)}_{\mathrm{S}}(\mathbf{q})$. We further note that the detailed discussion of the physical behavior of $F(\mathbf{q},\tau)$ is beyond the scope of the present work and has been presented in the recent Refs.~\cite{Dornheim_insight_2022,Dornheim_PTR_2022}.

\begin{figure*}\centering
\includegraphics[width=0.475\textwidth]{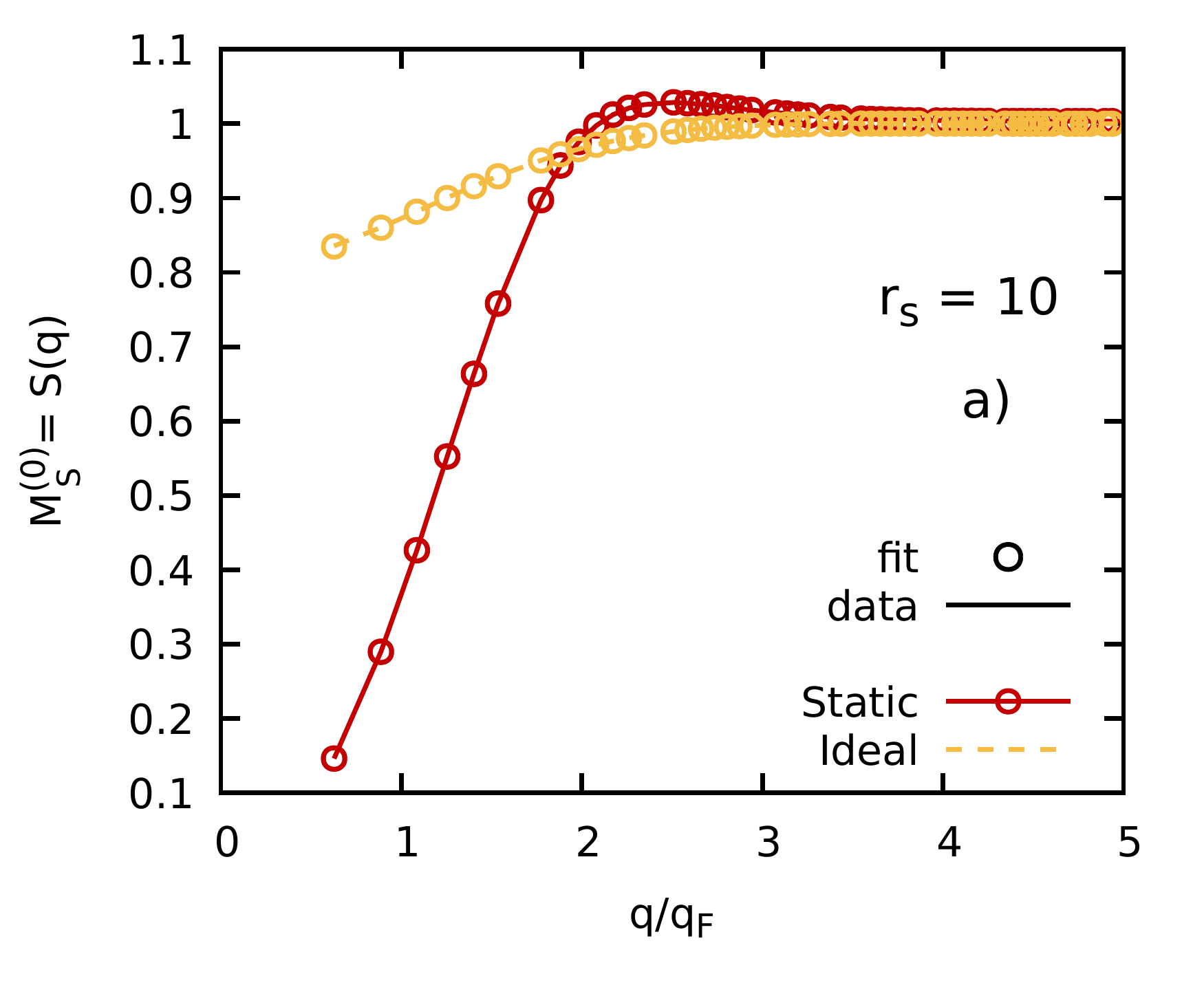}\includegraphics[width=0.475\textwidth]{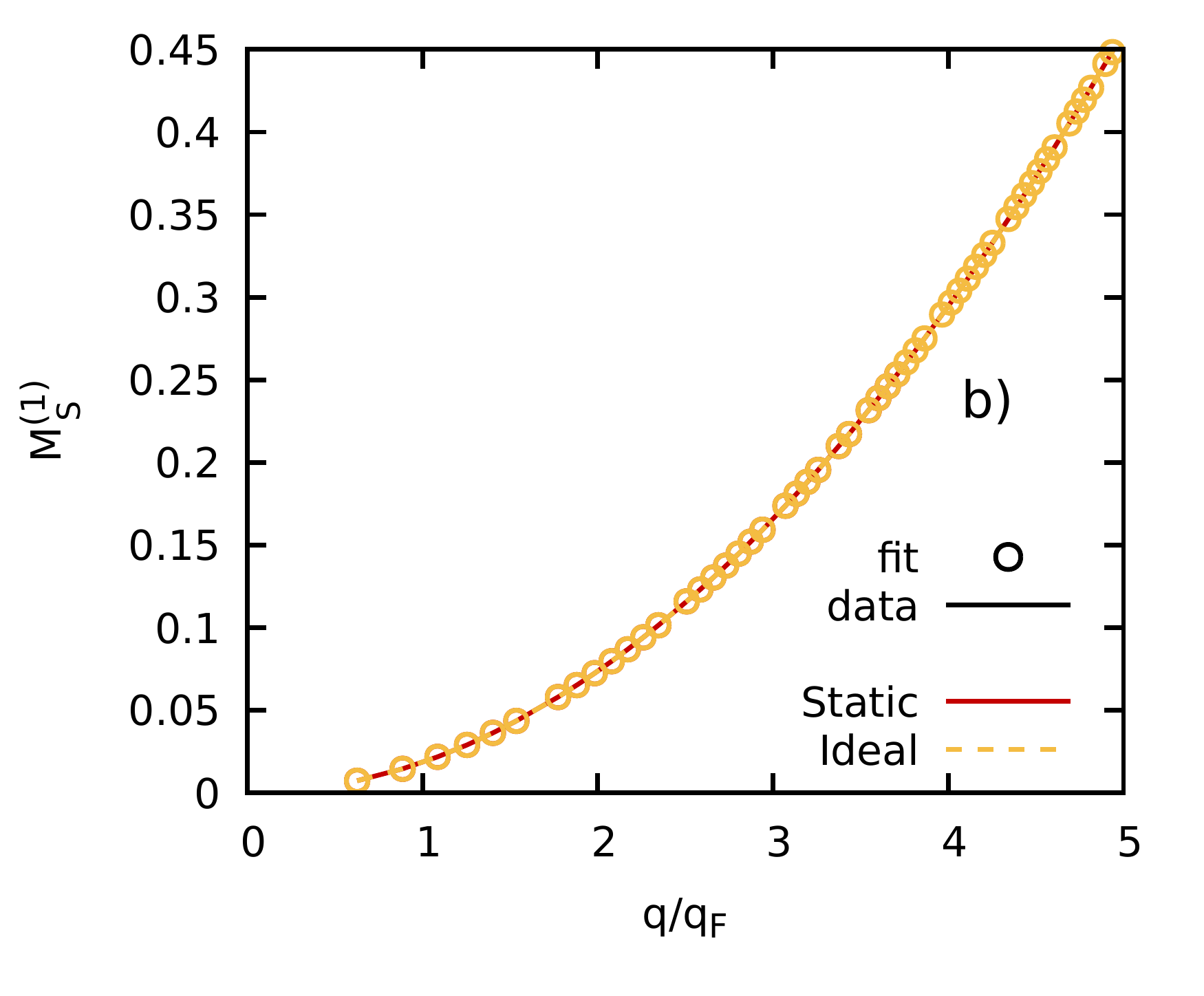}\\\vspace*{-0.3cm}\includegraphics[width=0.475\textwidth]{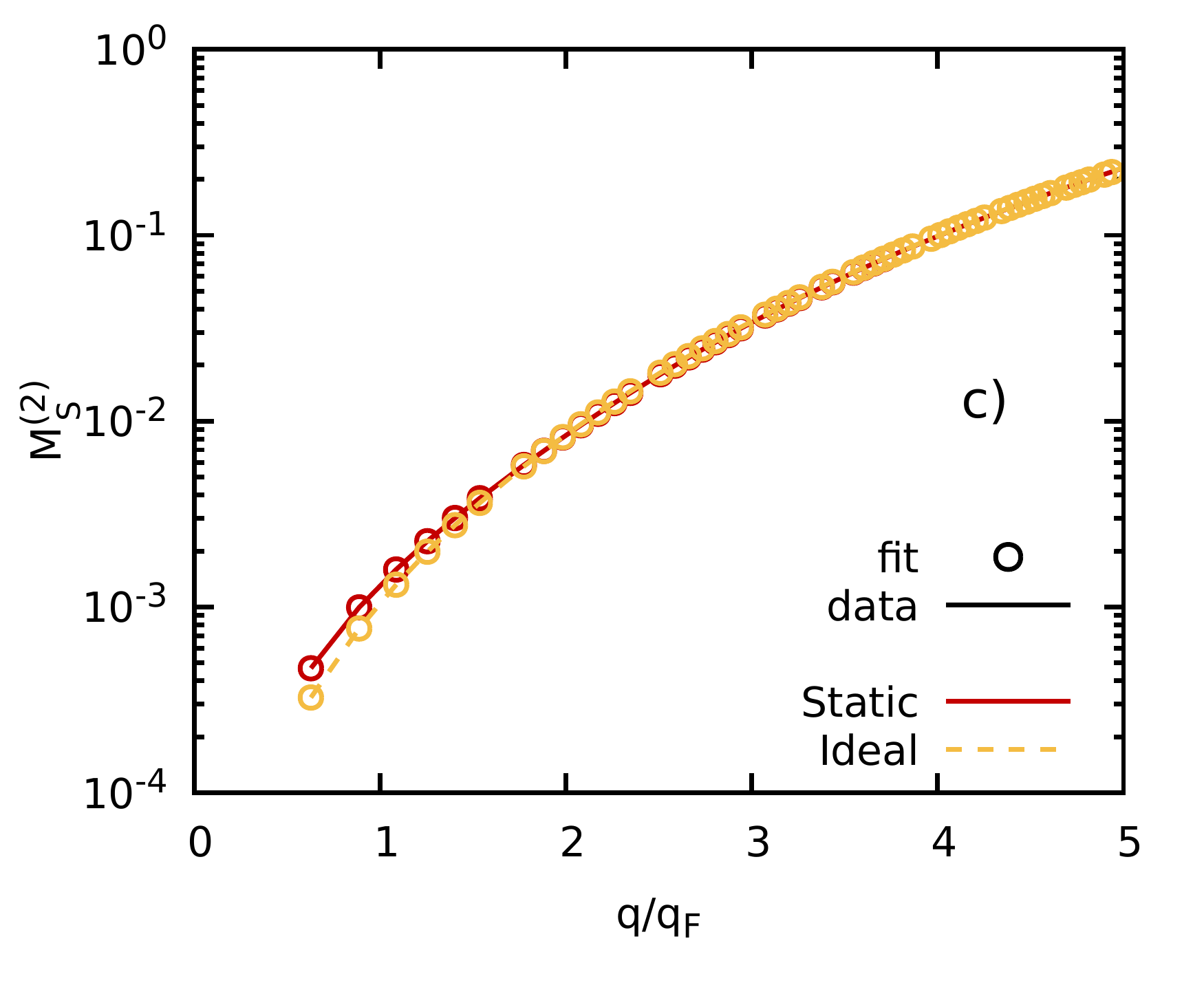}\includegraphics[width=0.475\textwidth]{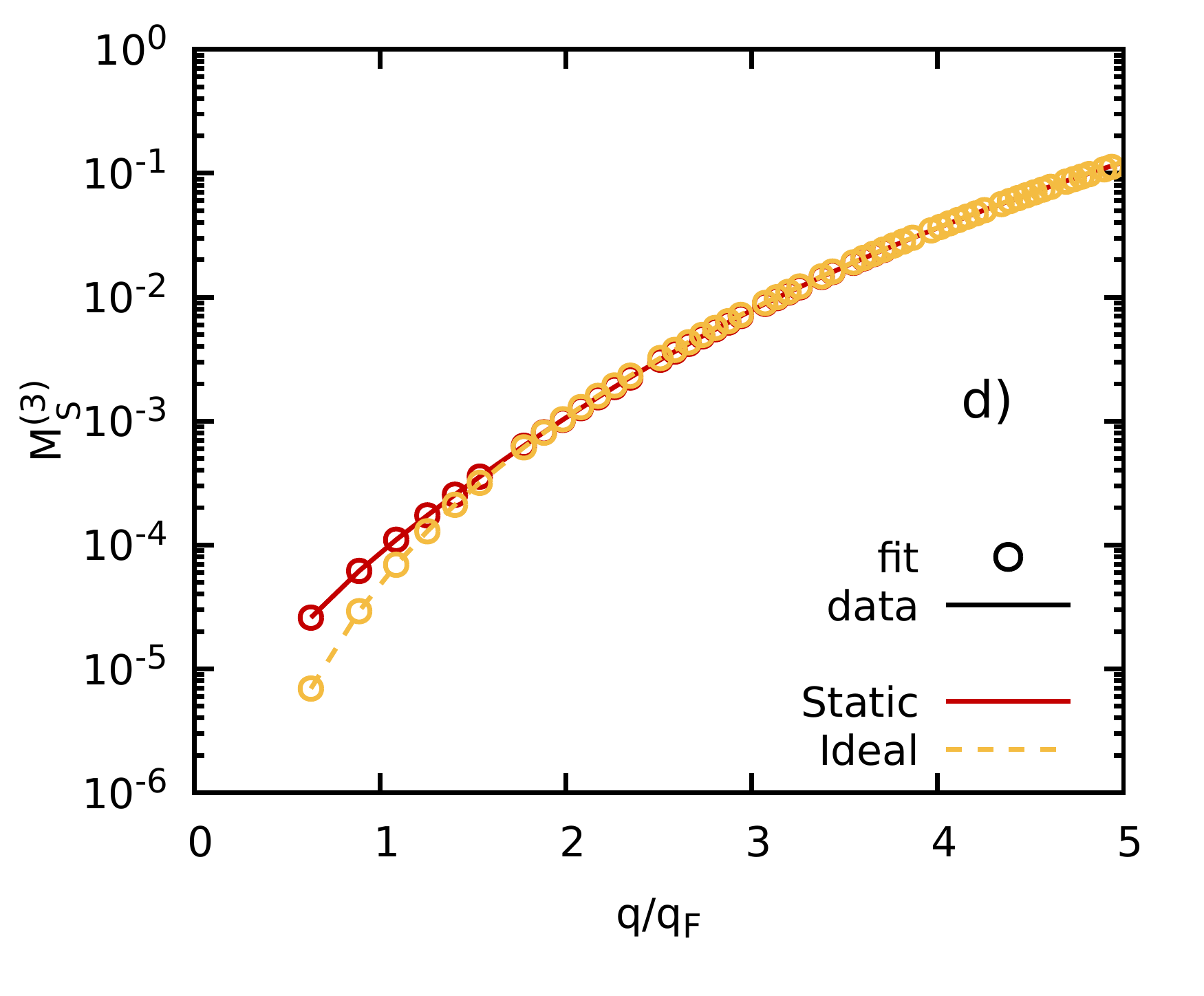}\\\vspace*{-0.3cm}\includegraphics[width=0.475\textwidth]{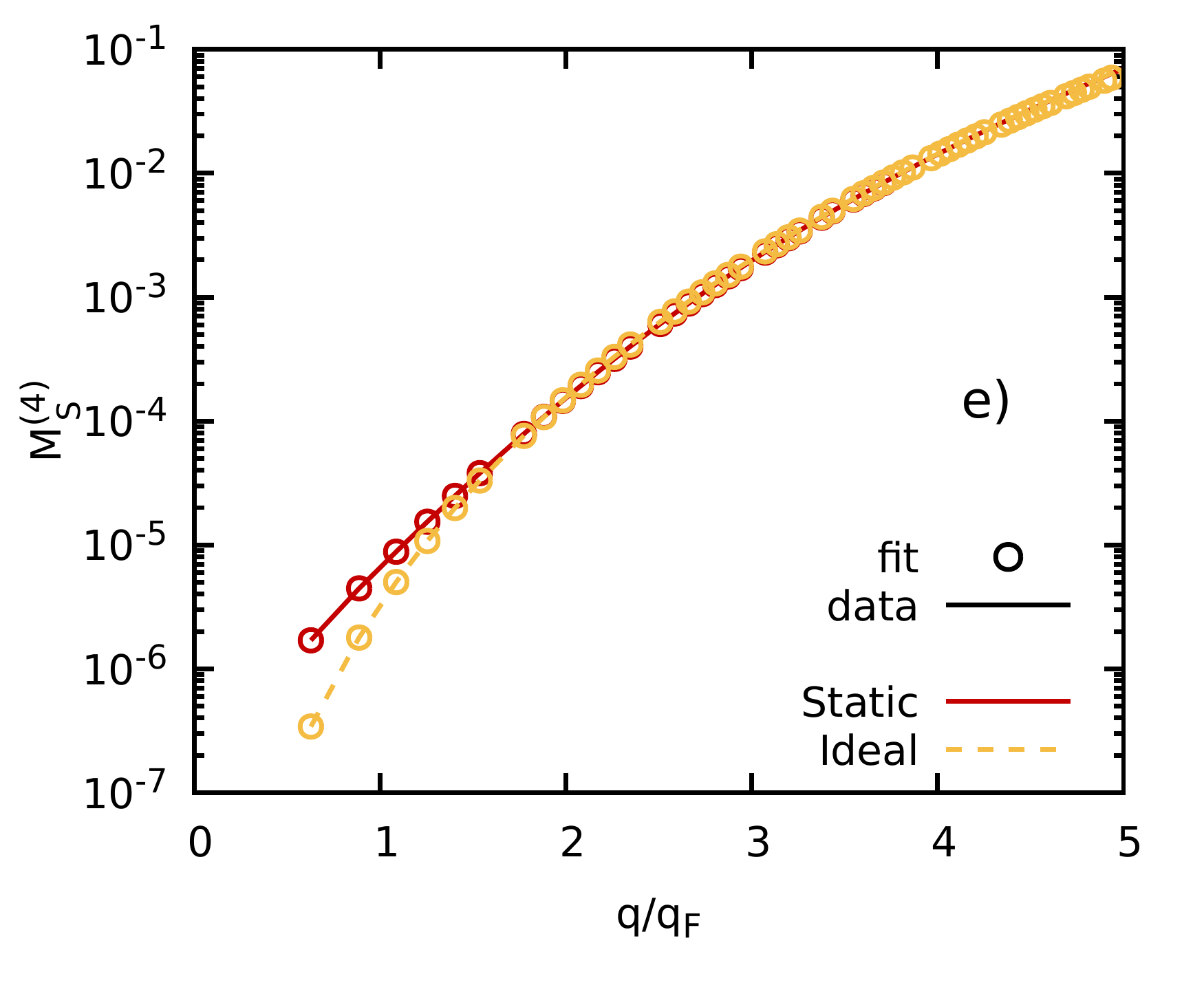}\includegraphics[width=0.475\textwidth]{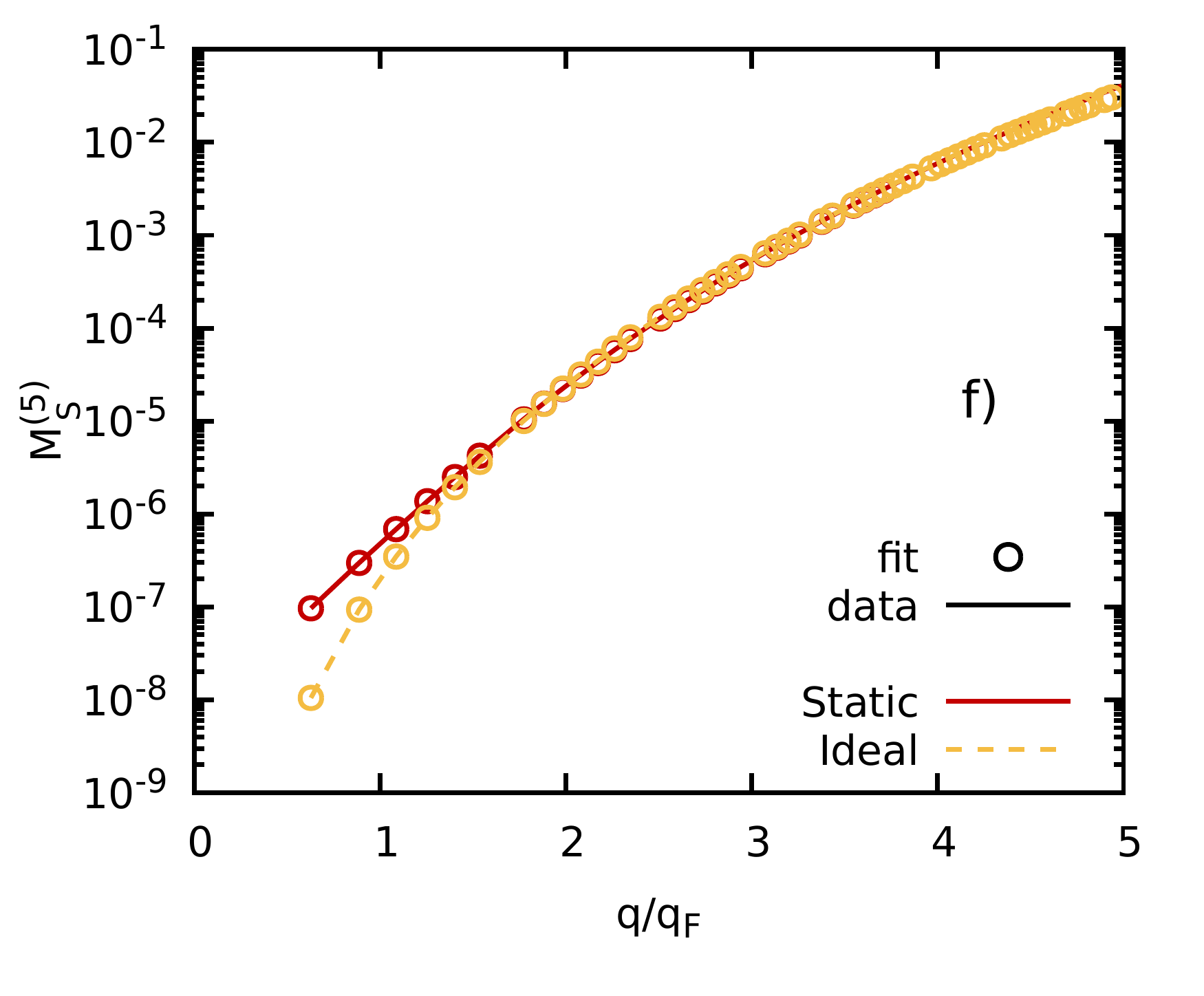}
\caption{\label{fig:synthetic_rs10}
Comparison of synthetic frequency moments (lines) computed directly from $S(\mathbf{q},\omega)$ via Eq.~(\ref{eq:moments}) for the ideal Fermi gas (yellow) and within the \emph{static approximation} (red) to results that have been extracted from the coefficients of canonical fits to the ITCF (circles) via Eq.~(\ref{eq:final}).
}
\end{figure*} 

\begin{figure}\centering
\includegraphics[width=0.475\textwidth]{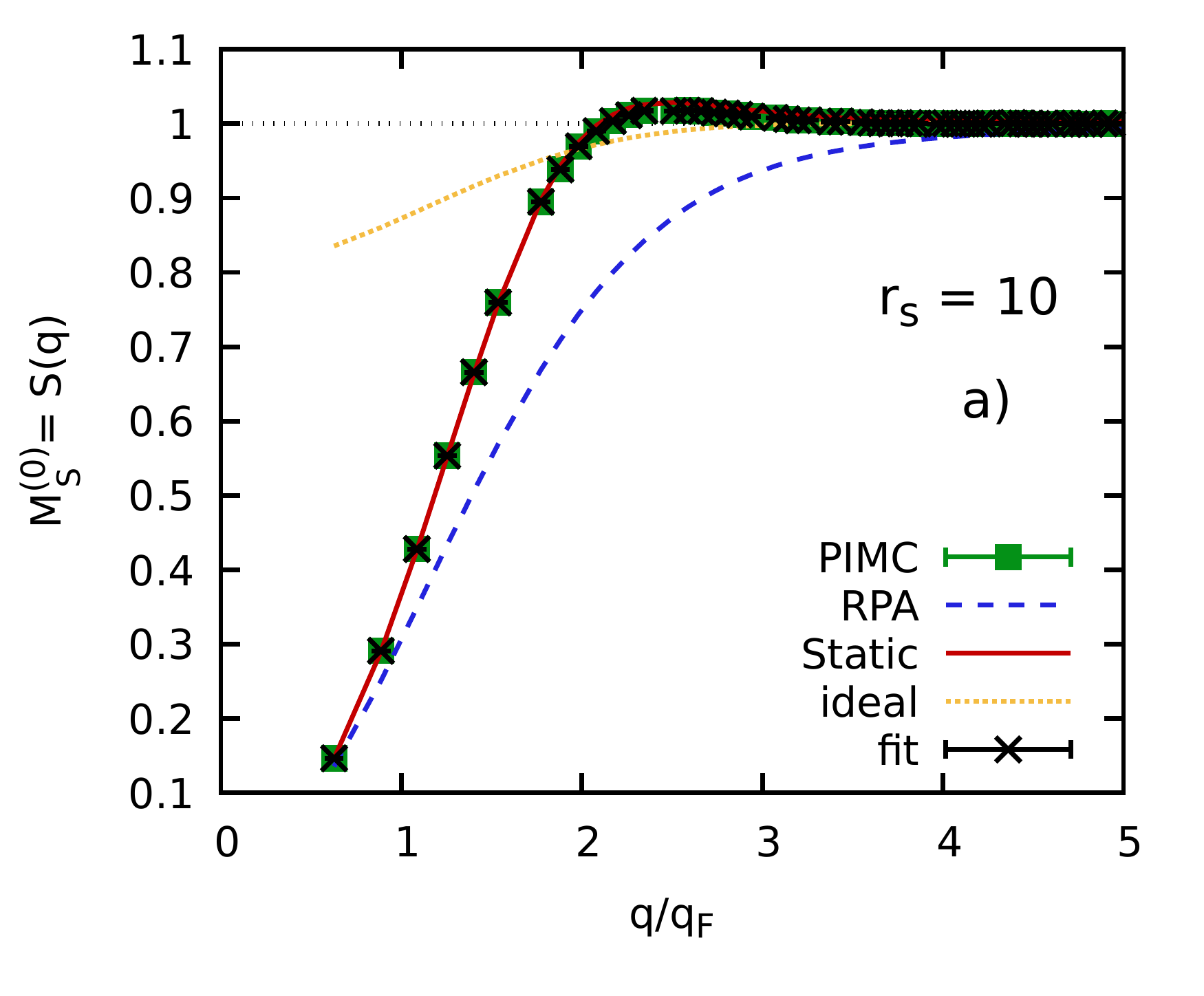}\\\vspace*{-1.3cm}\includegraphics[width=0.475\textwidth]{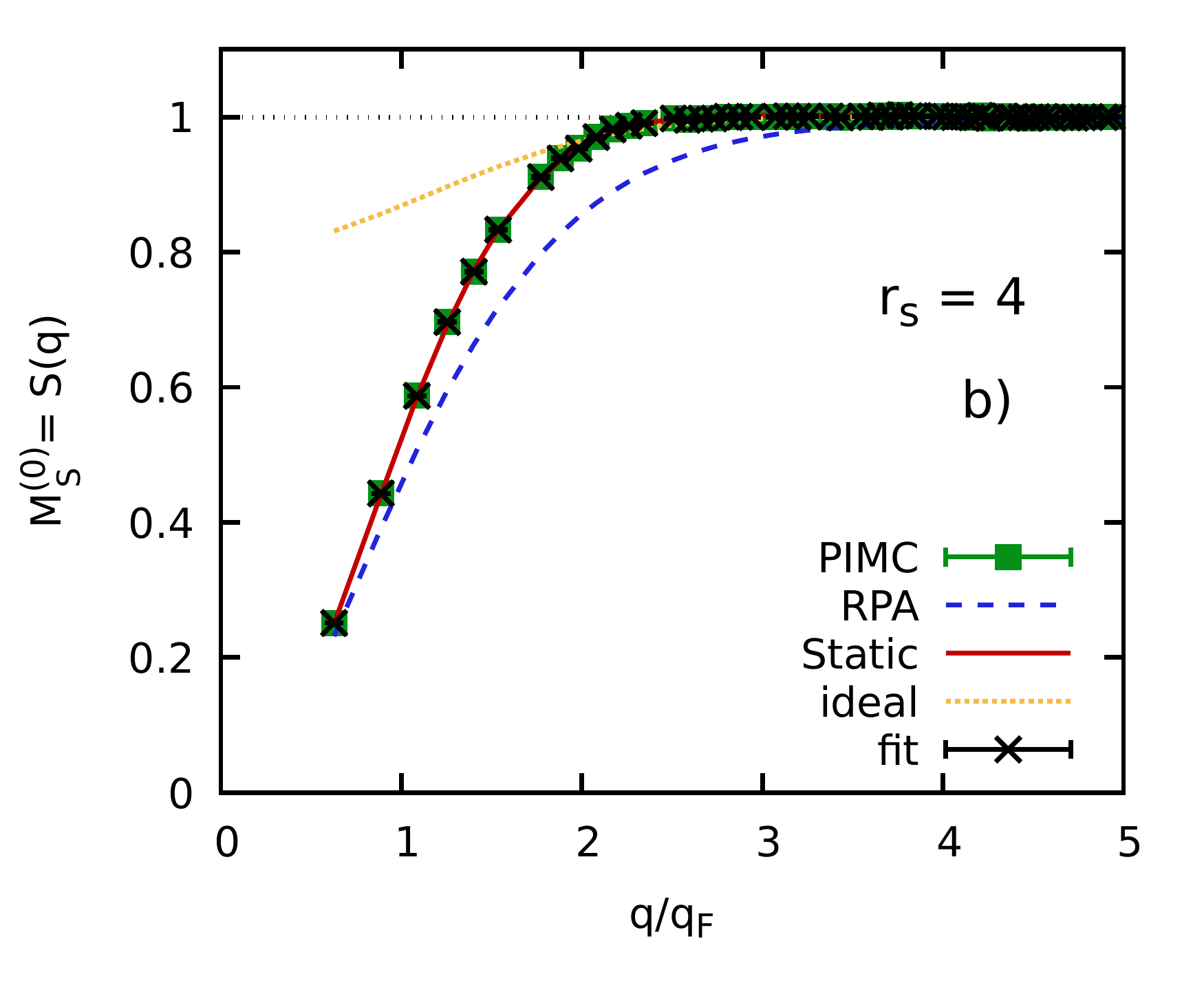}
\caption{\label{fig:Mom0}
The zero frequency moment $M^{(0)}_{S}(\mathbf{q})$ for the UEG at $\Theta=1$ for a) $r_s=10$ and b) $r_s=4$. Green squares: PIMC reference data for the static structure factor $S(\mathbf{q})=F(\mathbf{q},0)$; black crosses: moments extracted via Eq.~(\ref{eq:final}); dashed blue, solid red, and dotted yellow: reference data within RPA, \emph{static approximation}, and for the ideal Fermi gas computed from synthetic $S(\mathbf{q},\omega)$ directly via Eq.~(\ref{eq:moments}).
}
\end{figure} 

In Fig.~\ref{fig:synthetic_rs10}, we show the corresponding frequency moments that have been obtained from the canonical fitting coefficients via Eq.~(\ref{eq:final}) as a function of the wave number $q$. The panels a)-f) show the orders $\alpha=0,\dots,5$. The solid red and dashed yellow curves show exact reference data that have been directly computed via Eq.~(\ref{eq:moments}) from synthetic results for $S(\mathbf{q},\omega)$ within the \emph{static approximation} and for the ideal Fermi gas model, respectively. The corresponding circles show the frequency moments that have been extracted from the ITCF via Eq.~(\ref{eq:final}).
Clearly, the proposed extraction of the $M_\alpha$ from canonical fits to the ITCF works exceptionally well in all cases, and over the entire relevant range of wave numbers. This constitutes a strong empirical verification of our method and serves as an important benchmark for the following analysis of PIMC results, for which reliable benchmark data exist only for a subset of moments. In particular, this analysis of synthetic results demonstrates that even the extraction of the fifth moment $M^{(5)}_{S}(\mathbf{q})$ is, in principle, possible.

In Fig.~\ref{fig:F_rs10_theta1}b), we show the canonical fitting to our PIMC data for $F(\mathbf{q},\tau)$ for the same conditions as in panel a). In this case, the input data for the ITCF are afflicted with statistical error bars, see the magnified segment shown in panel c). Therefore, our fitting procedure gives access to a smaller number of polynomial coefficients compared to the synthetic data from the \emph{static approximation}, and we find $\alpha_\textnormal{max}=6$ in this case.

\subsection{Frequency moments\label{sec:moments}}

Let us begin our analysis of the frequency moments extracted from PIMC results for the ITCF with a discussion of $M^{(0)}_{S}(\mathbf{q})=S(\mathbf{q})=F(\mathbf{q},0)$ shown in Fig.~\ref{fig:Mom0}. In the following, all results have been obtained for $\Theta=1$, and panels a) and b) show results for $r_s=10$ and $r_s=4$. From a physical perspective, these cases correspond to an electron liquid~\cite{dornheim_dynamic} that exhibits interesting effects such as the roton feature in the dispersion~\cite{Dornheim_Nature_2022}, and to a metallic density that can be realized either in experiments with e.g.~sodium~\cite{Huotari_PRL_2010} or in hydrogen jets~\cite{Zastrau}. The green squares show our direct PIMC results for $S(\mathbf{q})$ and are in perfect agreement with the zero-order fitting coefficient $c_0(\mathbf{q})$ for both densities and over the entire range of wave numbers. As a reference, we also include synthetic data computed via Eqs.~(\ref{eq:FDT}) and (\ref{eq:LFC}) within the RPA (dashed blue), the \emph{static approximation} (solid red), and for the ideal Fermi gas (dotted yellow). Overall, the \emph{static approximation} that has been evaluated using the neural-net representation from Ref.~\cite{dornheim_ML} exhibits the highest degree of accuracy, as it is expected. The RPA and the ideal Fermi gas model are substantially less accurate and can easily be distinguished both from the exact PIMC results and from the extracted fitting coefficients.

\begin{figure}\centering
\includegraphics[width=0.475\textwidth]{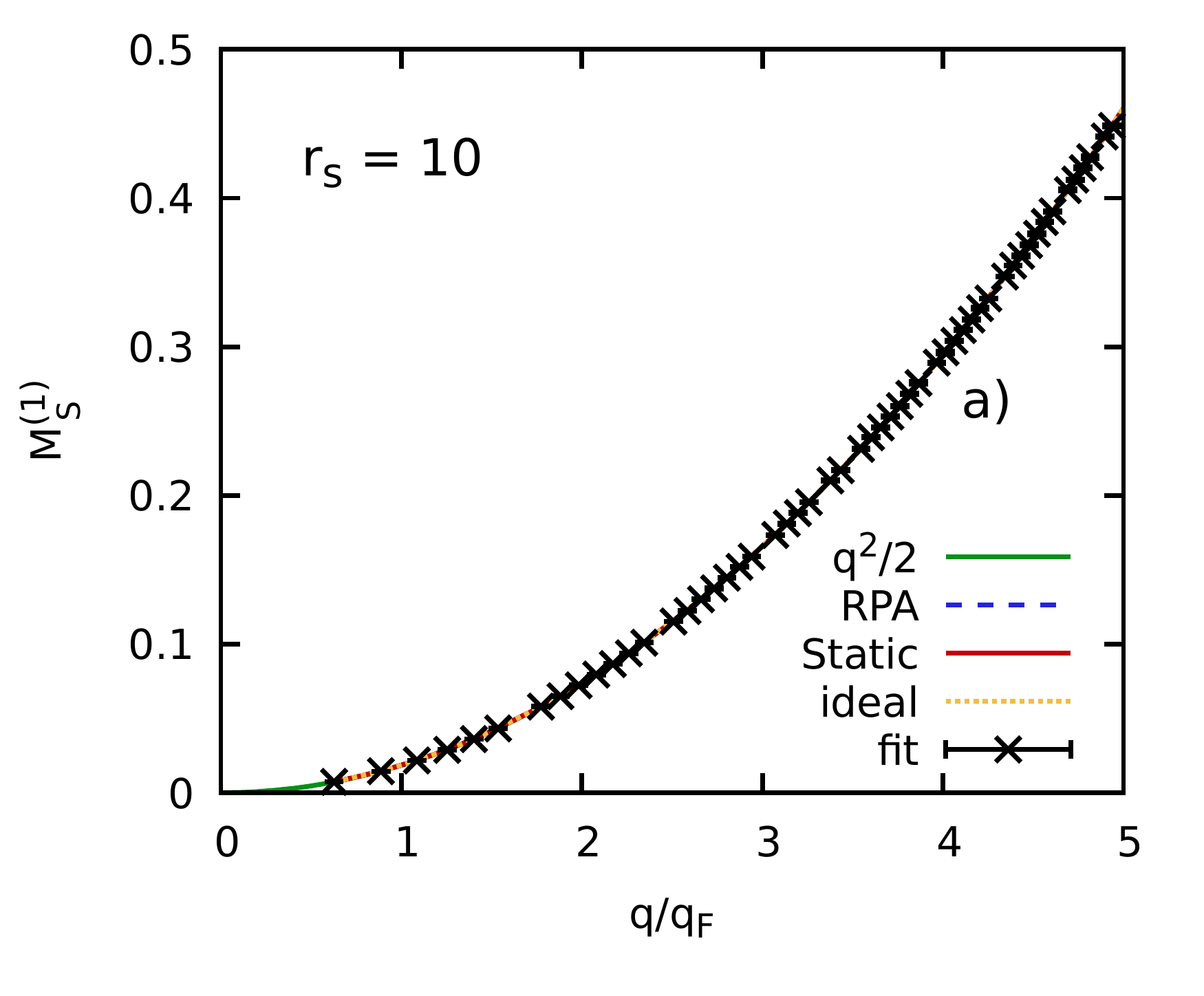}\\\vspace*{-1.3cm}\includegraphics[width=0.475\textwidth]{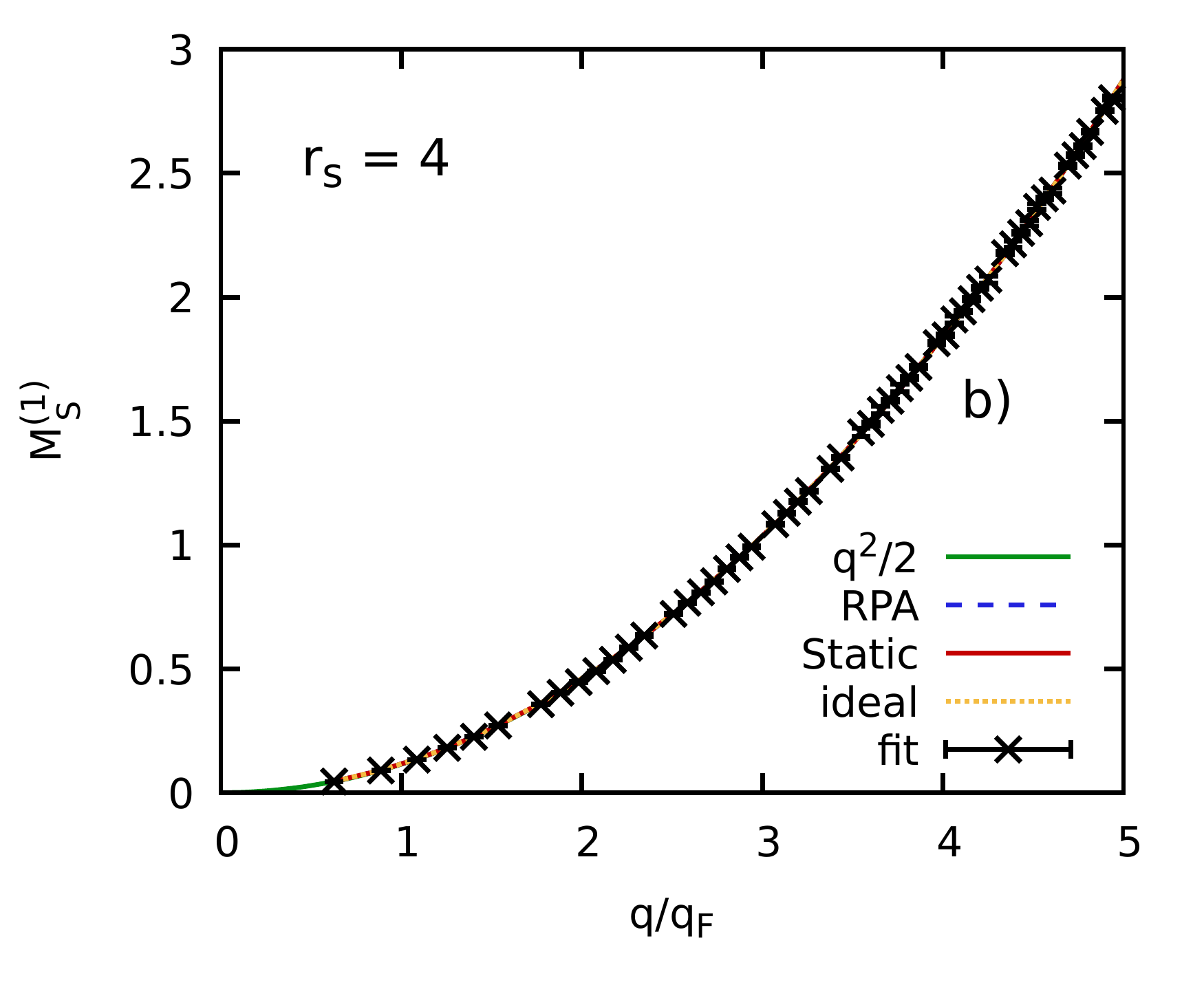}
\caption{\label{fig:Mom1}
The first frequency moment $M^{(1)}_{S}(\mathbf{q})$ for the UEG at $\Theta=1$ for a) $r_s=10$ and b) $r_s=4$. Green line: f-sum rule, Eq.~(\ref{eq:first}); black crosses: moments extracted via Eq.~(\ref{eq:final}); dashed blue, solid red, and dotted yellow: reference data within RPA, \emph{static approximation}, and for the ideal Fermi gas computed from synthetic $S(\mathbf{q},\omega)$ directly via Eq.~(\ref{eq:moments}).
}
\end{figure} 

\begin{figure}\centering
\includegraphics[width=0.475\textwidth]{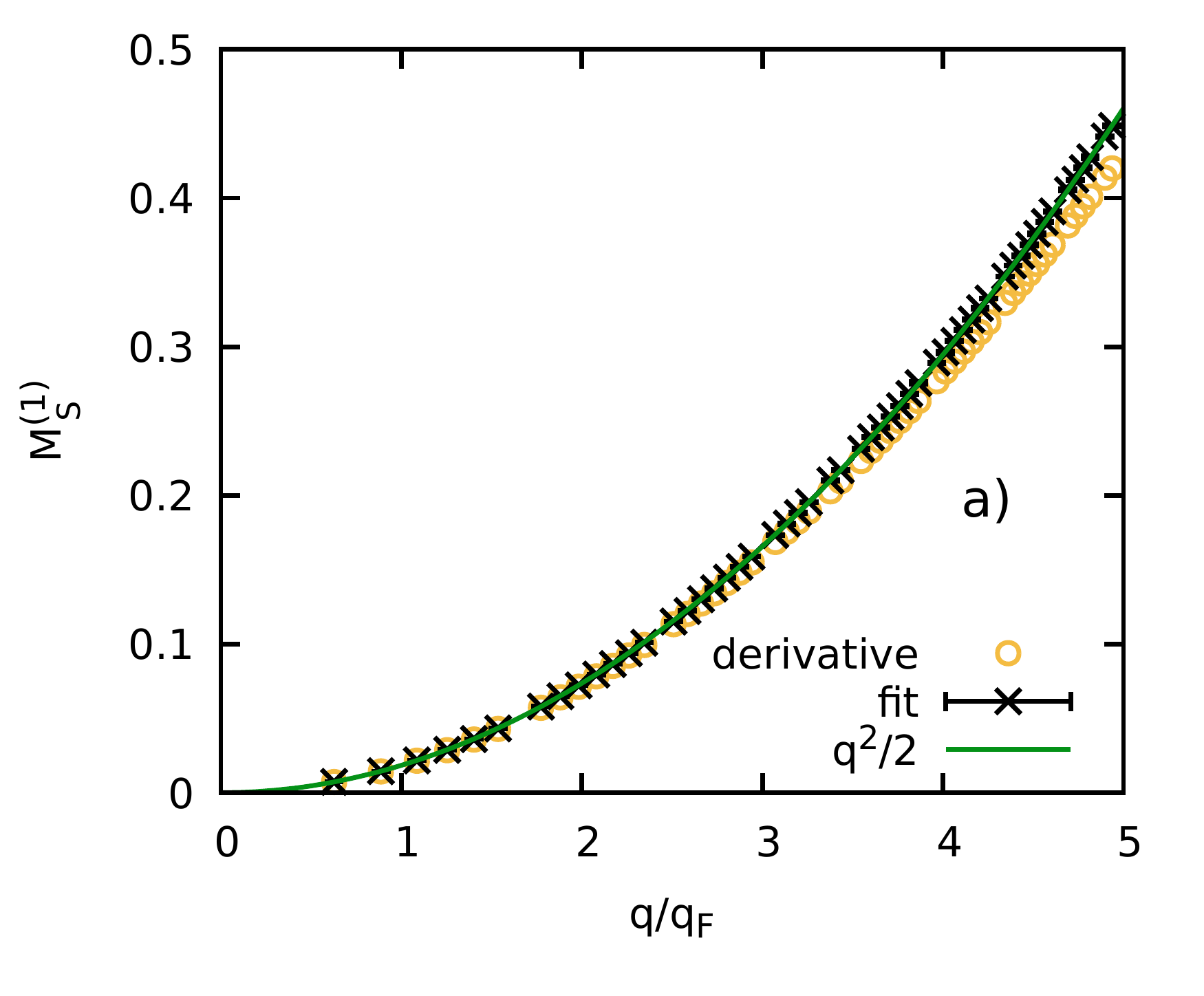}\\\includegraphics[width=0.475\textwidth]{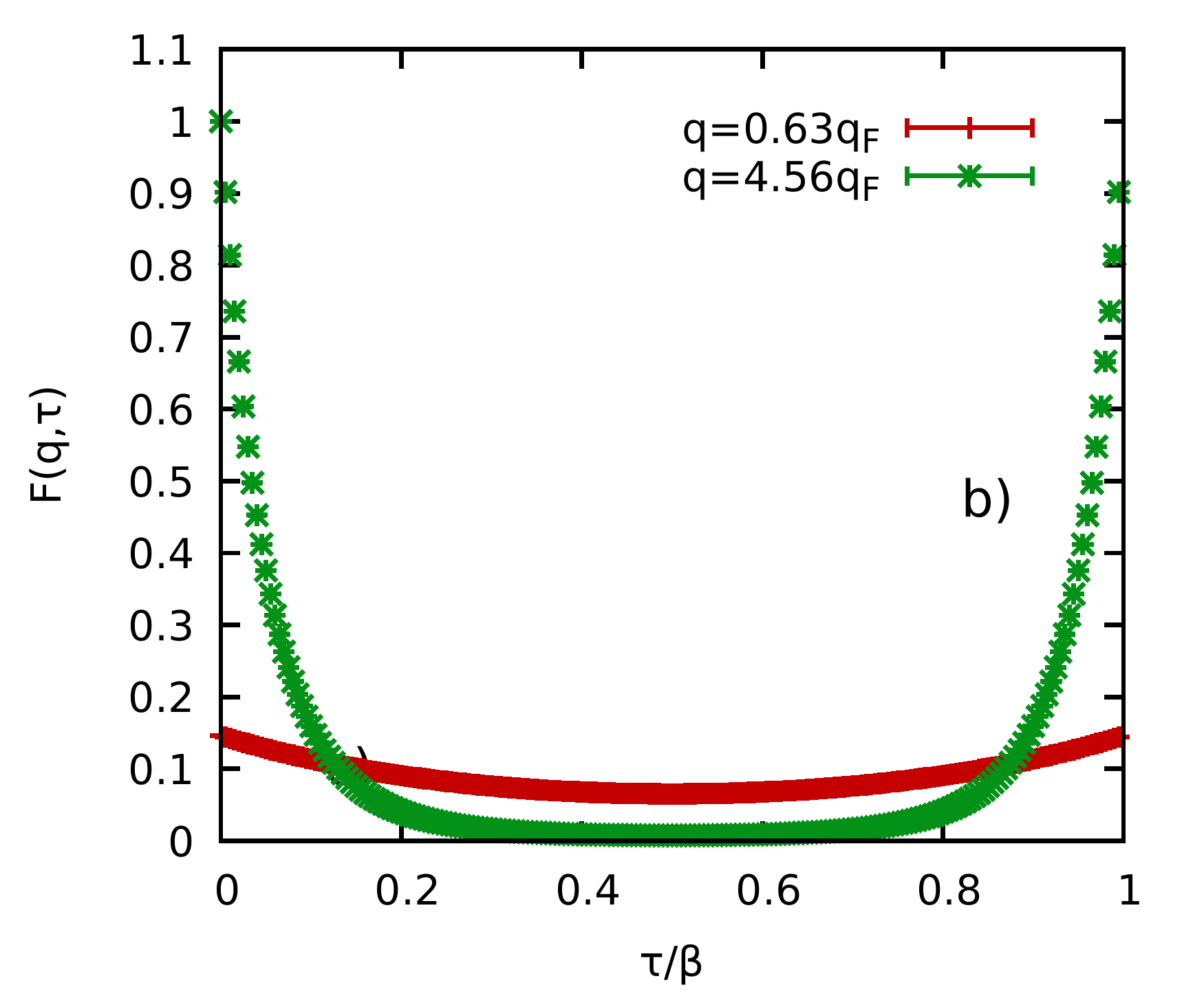}
\caption{\label{fig:Mom1_approx}
Panel a) shows the first frequency moment $M^{(1)}_{S}(\mathbf{q})$ for the UEG at $\Theta=1$ and $r_s=10$, with the yellow circles having being evaluated from the approximate derivative Eq.~(\ref{eq:approx}); b) $\tau$-dependence of PIMC results for the ITCF for $q=0.63q_\textnormal{F}$ (red crosses) and $q=4.56q_\textnormal{F}$ (green stars).
}
\end{figure} 

In Fig.~\ref{fig:Mom1}, we repeat this analysis for the first moment $M^{(1)}_{S}(\mathbf{q})$, with the solid green curve depicting the exact f-sum rule, Eq.~(\ref{eq:first}). Clearly, all data sets are in perfect agreement with the latter, including all synthetic curves. As an alternative route to the polynomial expansion Eq.~(\ref{eq:Taylor}), one might also attempt to numerically evaluate the first derivative of the ITCF with respect to $\tau$ on the given PIMC $\tau$-grid,
\begin{eqnarray}\label{eq:approx}
 \left.   \frac{\partial F(\mathbf{q},\tau)}{\partial\tau} \right|_{\tau=0} \approx \frac{F(\mathbf{q},\epsilon)-F(\mathbf{q},0)}{\epsilon}\,.
\end{eqnarray}
The results are shown as the yellow circles in Fig.~\ref{fig:Mom1_approx}a). Evidently, the numerical derivative only agrees with the exact f-sum rule for $q\lesssim3q_\textnormal{F}$, but becomes increasingly inaccurate in the limit of large $q$. This observation can be directly traced back to the behavior of the ITCF, as it is illustrated in Fig.~\ref{fig:Mom1_approx}b). Specifically, the ITCF becomes increasingly steep for large $q$. For $q=0.63q_\textnormal{F}$ (red crosses), the ITCF is comparably flat, and the corresponding evaluation of Eq.~(\ref{eq:approx}) is accurate. In contrast, we find a very sharp $\tau$-decay around $\tau=0$ for $q=4.56q_\textnormal{F}$ (green stars), and the available $\tau$-grid in the PIMC simulation is not sufficient to accurately estimate the first derivative directly.
At the same time, we stress that the proposed polynomial fit of $F(\mathbf{q},\tau)$ [Eq.~(\ref{eq:Taylor})] completely overcomes this issue and, therefore, constitutes the preferable option.

\begin{figure}\centering
\includegraphics[width=0.475\textwidth]{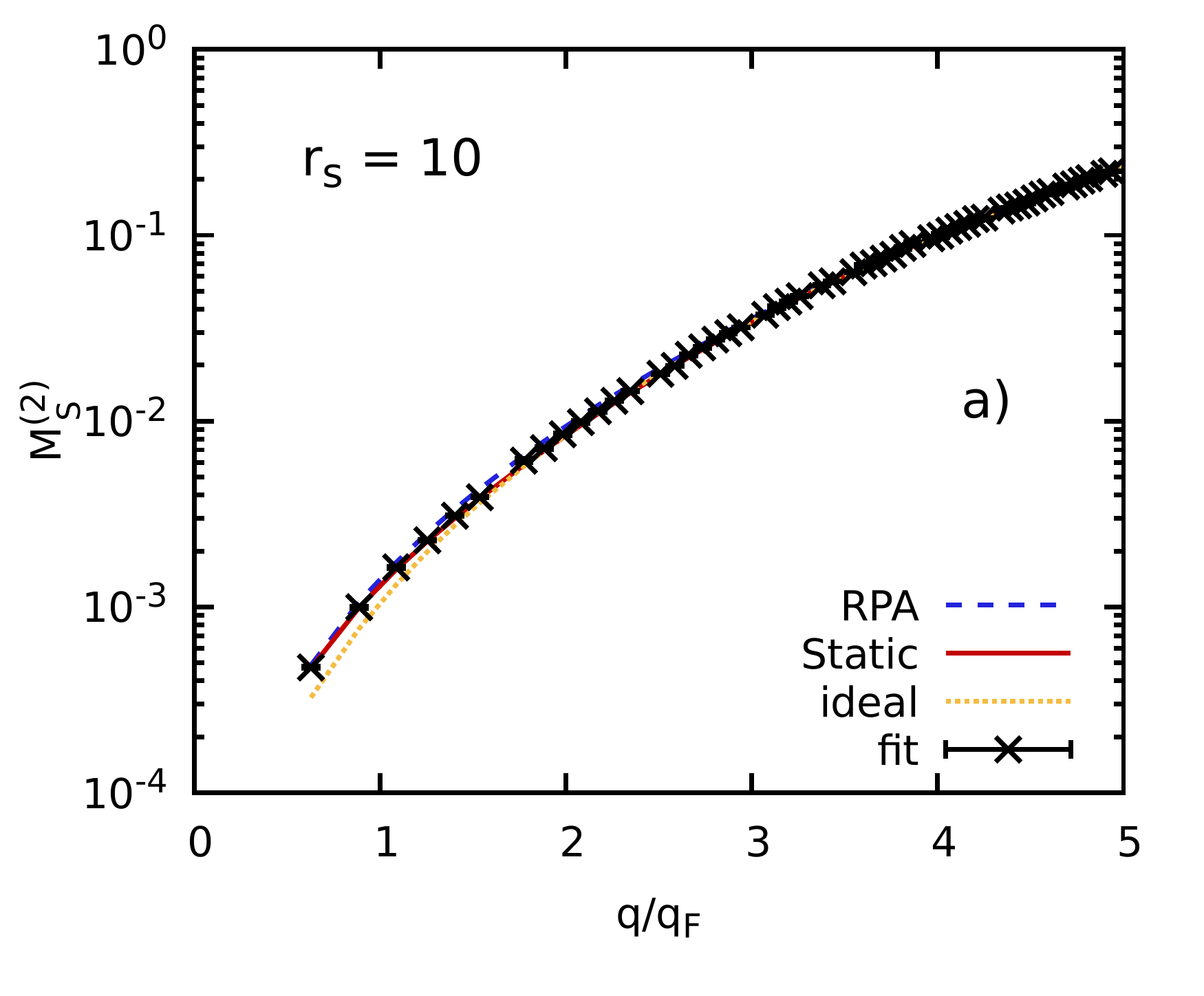}\\\vspace*{-1.3cm}\includegraphics[width=0.475\textwidth]{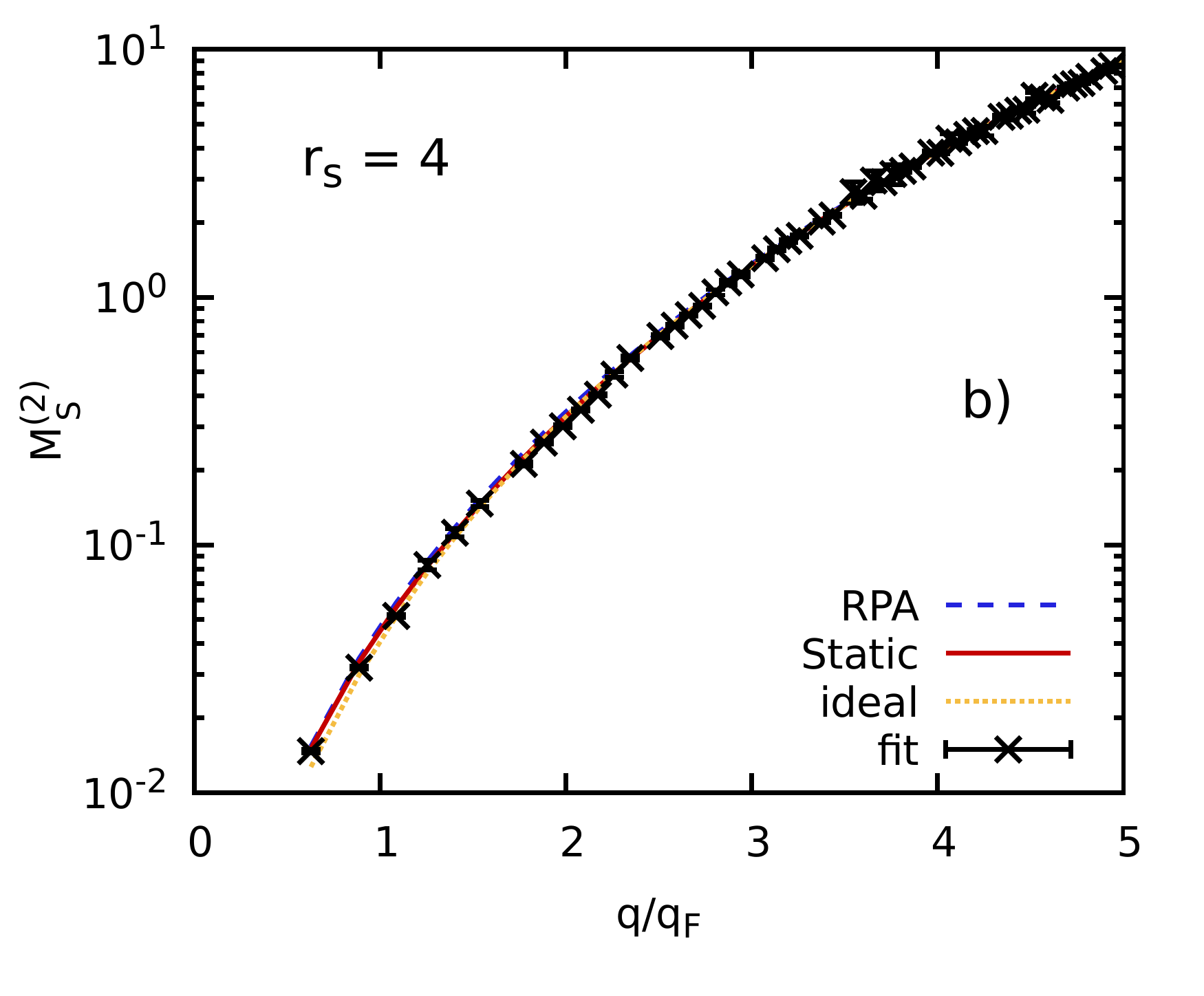}
\caption{\label{fig:Mom2}
The second frequency moment $M^{(2)}_{S}(\mathbf{q})$ for the UEG at $\Theta=1$ for a) $r_s=10$ and b) $r_s=4$. Black crosses: moments extracted via Eq.~(\ref{eq:final}); dashed blue, solid red, and dotted yellow: reference data within RPA, \emph{static approximation}, and for the ideal Fermi gas computed from synthetic $S(\mathbf{q},\omega)$ directly via Eq.~(\ref{eq:moments}).
}
\end{figure} 

Next, we consider the second moment $M^{(2)}_{S}(\mathbf{q})$ shown in Fig.~\ref{fig:Mom2}. In this case, no exact reference data are available either from a sum-rule or from another source. At the same time, the RPA and \emph{static approximation} are in close agreement with each other and also closely agree with the extracted moments for both densities. In contrast, the reference data computed from the ideal Fermi gas model exhibits significant deviations for small wave numbers. Lastly, we point out that the extracted moments for the higher density of $r_s=4$ exhibit small yet visible fluctuations for $q\gtrsim 3q_\textnormal{F}$ for $M^{(2)}_{S}(\mathbf{q})$, while no such fluctuations are visible for $r_s=10$ with the naked eye. This is a direct consequence of the increased statistical uncertainty in the PIMC data, which, in turn, is due to the more severe fermion sign problem at $r_s=4$~\cite{dornheim_sign_problem}. At the same time, we find that the observed fluctuations in $M^{(2)}_{S}(\mathbf{q})$ are well captured by the corresponding error bars, the calculation of which is explained in more detail in Appendix~\ref{sec:details}.

\begin{figure}\centering
\includegraphics[width=0.475\textwidth]{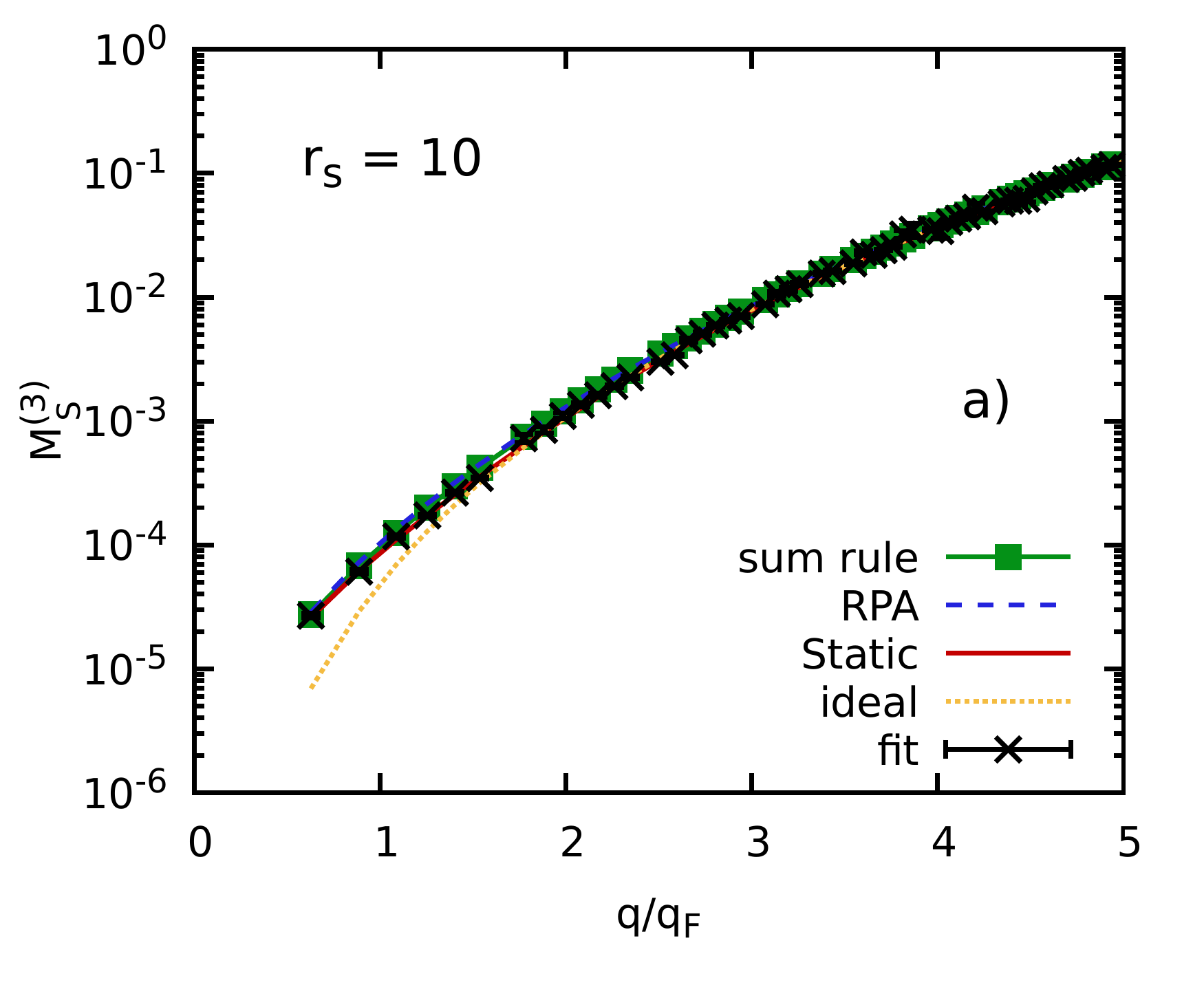}\\\vspace*{-1.3cm}\includegraphics[width=0.475\textwidth]{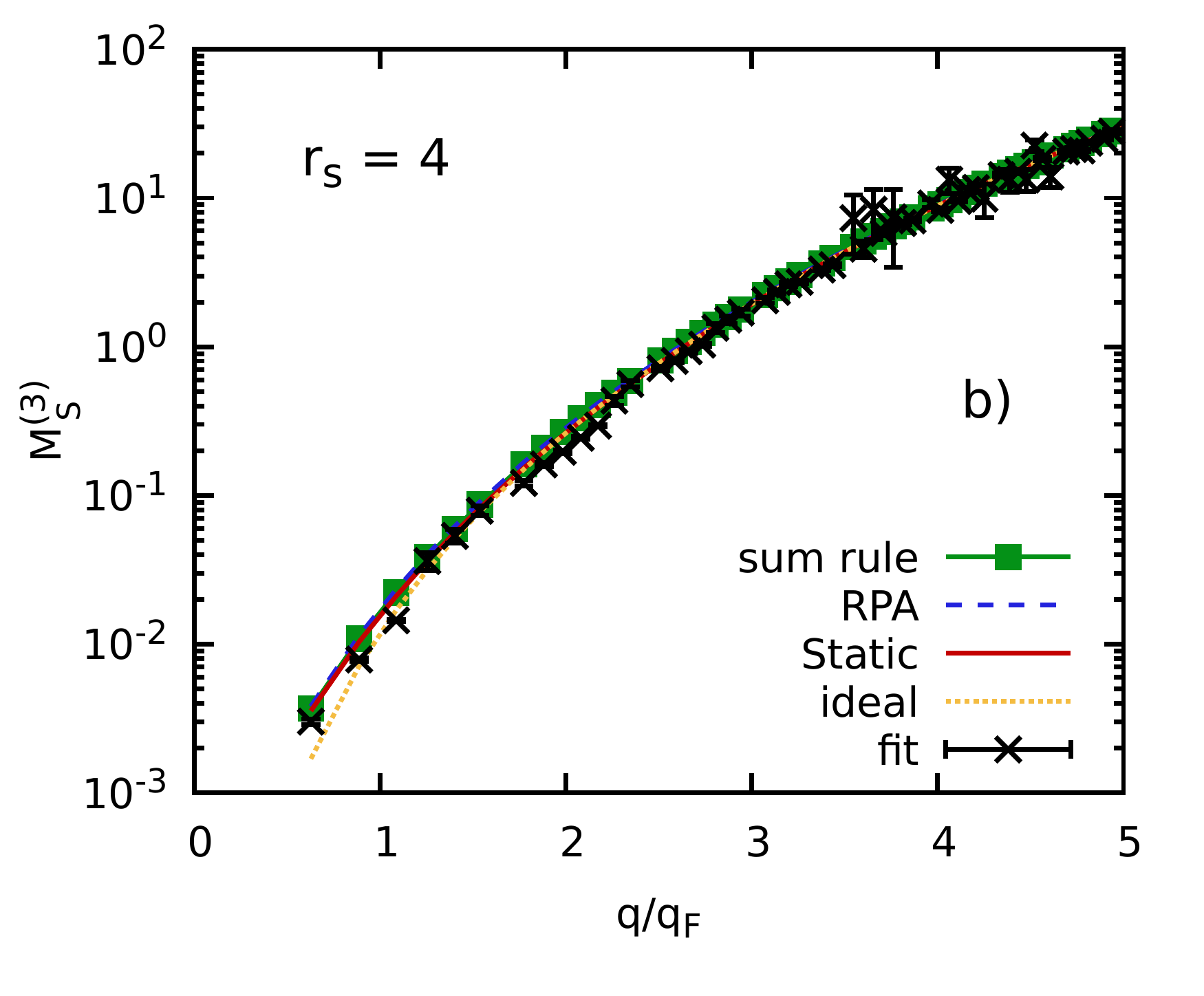}
\caption{\label{fig:Mom3}
The third frequency moment $M^{(3)}_{S}(\mathbf{q})$ for the UEG at $\Theta=1$ for a) $r_s=10$ and b) $r_s=4$. Green squares: cubic sum rule, Eq.~(\ref{eq:third}); black crosses: moments extracted via Eq.~(\ref{eq:final}); dashed blue, solid red, and dotted yellow: reference data within RPA, \emph{static approximation}, and for the ideal Fermi gas computed from synthetic $S(\mathbf{q},\omega)$ directly via Eq.~(\ref{eq:moments}).
}
\end{figure} 

A particularly interesting frequency moment of the DSF is given by $M^{(3)}_{S}(\mathbf{q})$, as it is directly connected to the high-frequency limit of the local field correction~\cite{Iwamoto_PRB_1984}. The corresponding results are shown in Fig.~\ref{fig:Mom3}, with the green squares being reference data computed from the cubic sum rule, Eq.~(\ref{eq:third}), using PIMC data for the kinetic energy $K$ and the static structure factor $S(\mathbf{q})$. Overall, the latter is in close agreement with the RPA and \emph{static approximation} data sets over the entire $q$-range for both densities; the ideal Fermi gas model again deviates for small $q$. For $r_s=10$, the proposed canonical fitting method gives accurate results over four orders of magnitude in $M^{(3)}_{S}(\mathbf{q})$ and is in good agreement with the sum-rule reference data. For $r_s=4$, the agreement is noticeably less good. This is likely a consequence of the larger statistical errors in the PIMC results for the ITCF but only captured by the error bars of the extracted moments for $q\gtrsim q_\textnormal{F}$. At the same time, it is important to note that the results for the cubic sum rule, too, are not carved in stone and might be subject to a small bias, e.g.~due to the discrete sum in Eq.~(\ref{eq:third}) that only becomes a continuous integral in the thermodynamic limit (i.e., $N\to\infty$).

\begin{figure}\centering
\includegraphics[width=0.475\textwidth]{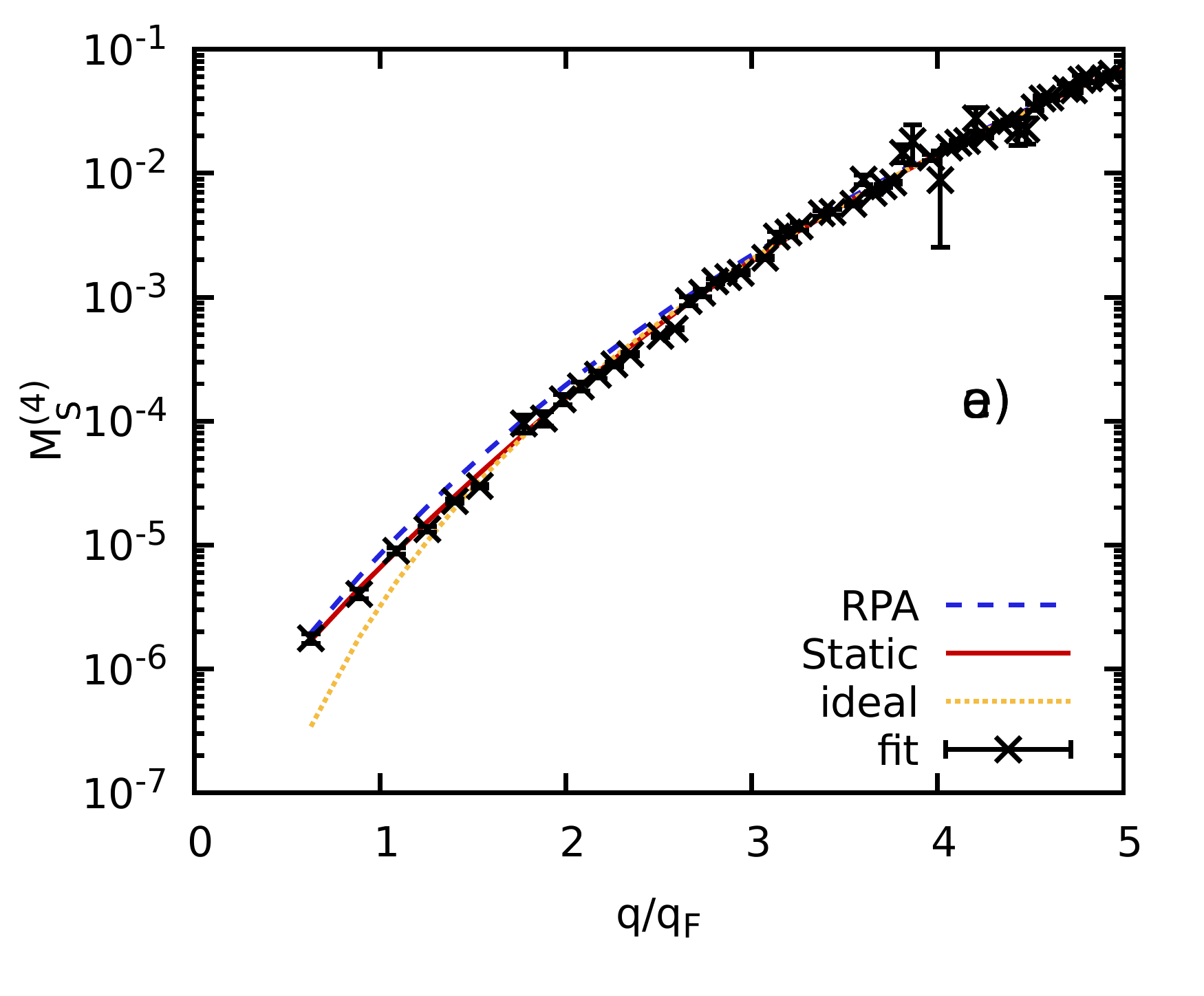}\\\vspace*{-1.3cm}\includegraphics[width=0.475\textwidth]{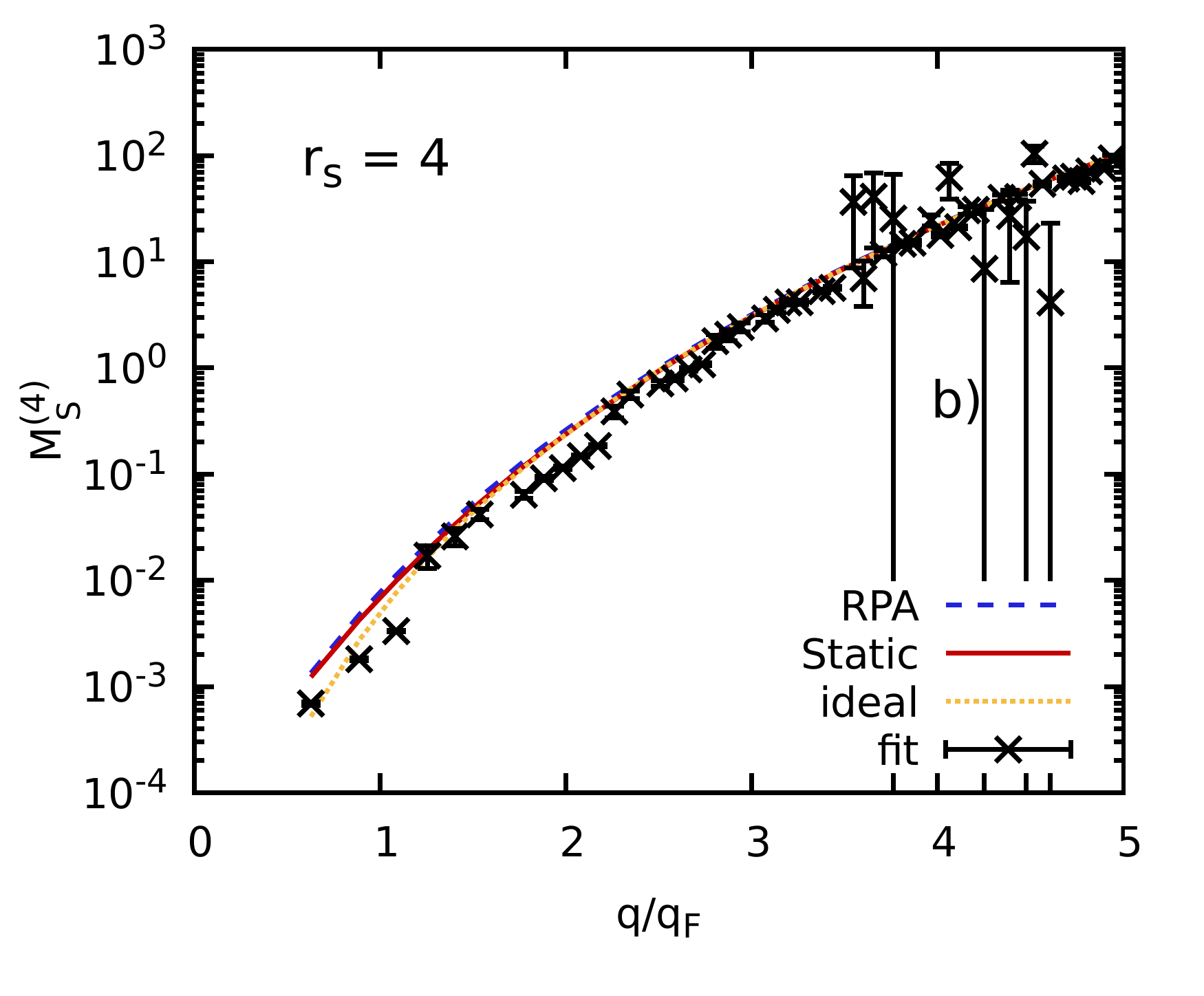}
\caption{\label{fig:Mom4}
The fourth frequency moment $M^{(4)}_{S}(\mathbf{q})$ for the UEG at $\Theta=1$ for a) $r_s=10$ and b) $r_s=4$. Black crosses: moments extracted via Eq.~(\ref{eq:final}); dashed blue, solid red, and dotted yellow: reference data within RPA, \emph{static approximation}, and for the ideal Fermi gas computed from synthetic $S(\mathbf{q},\omega)$ directly via Eq.~(\ref{eq:moments}).
}
\end{figure} 

\begin{figure}\centering
\includegraphics[width=0.475\textwidth]{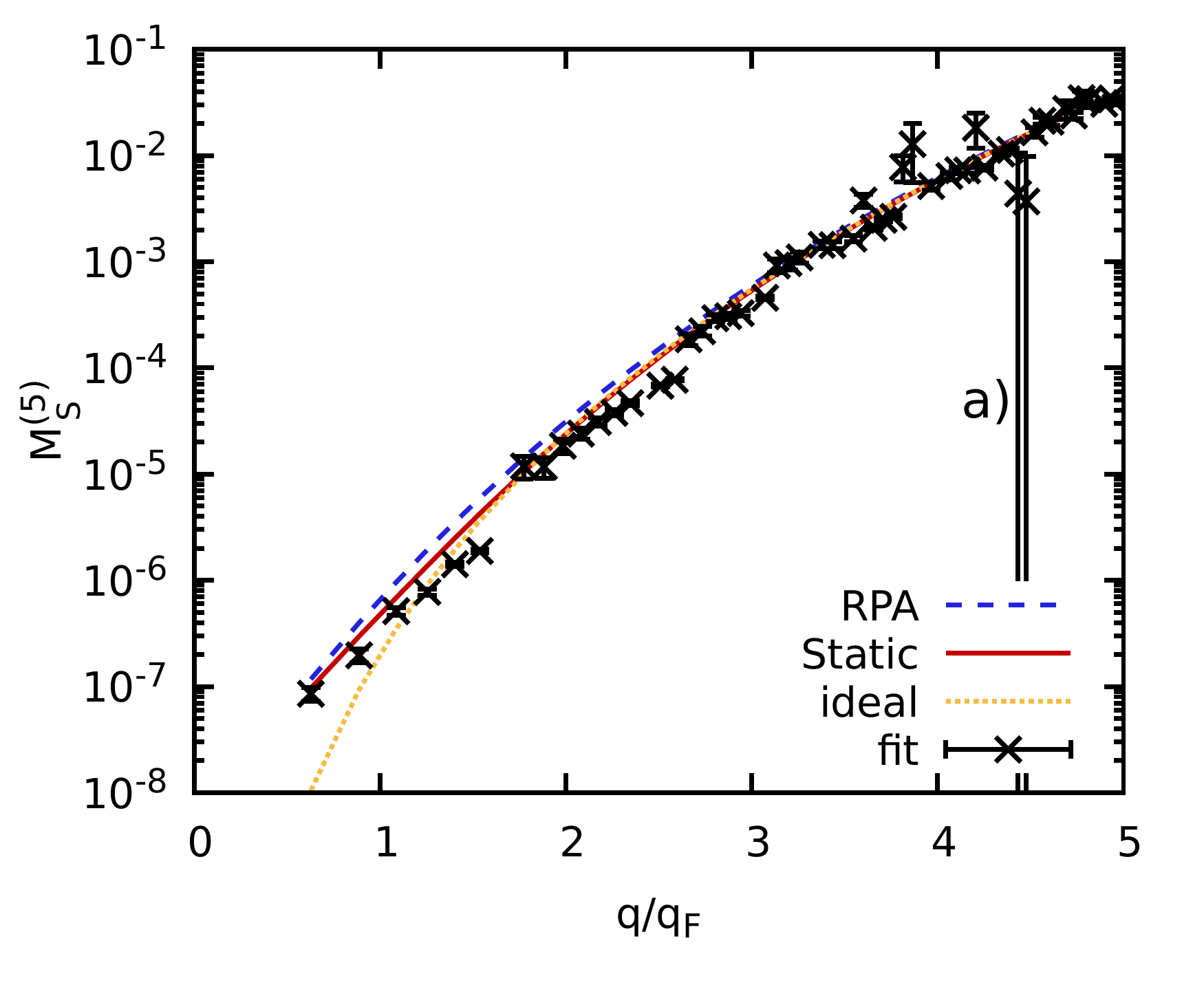}\\\vspace*{-1.3cm}\includegraphics[width=0.475\textwidth]{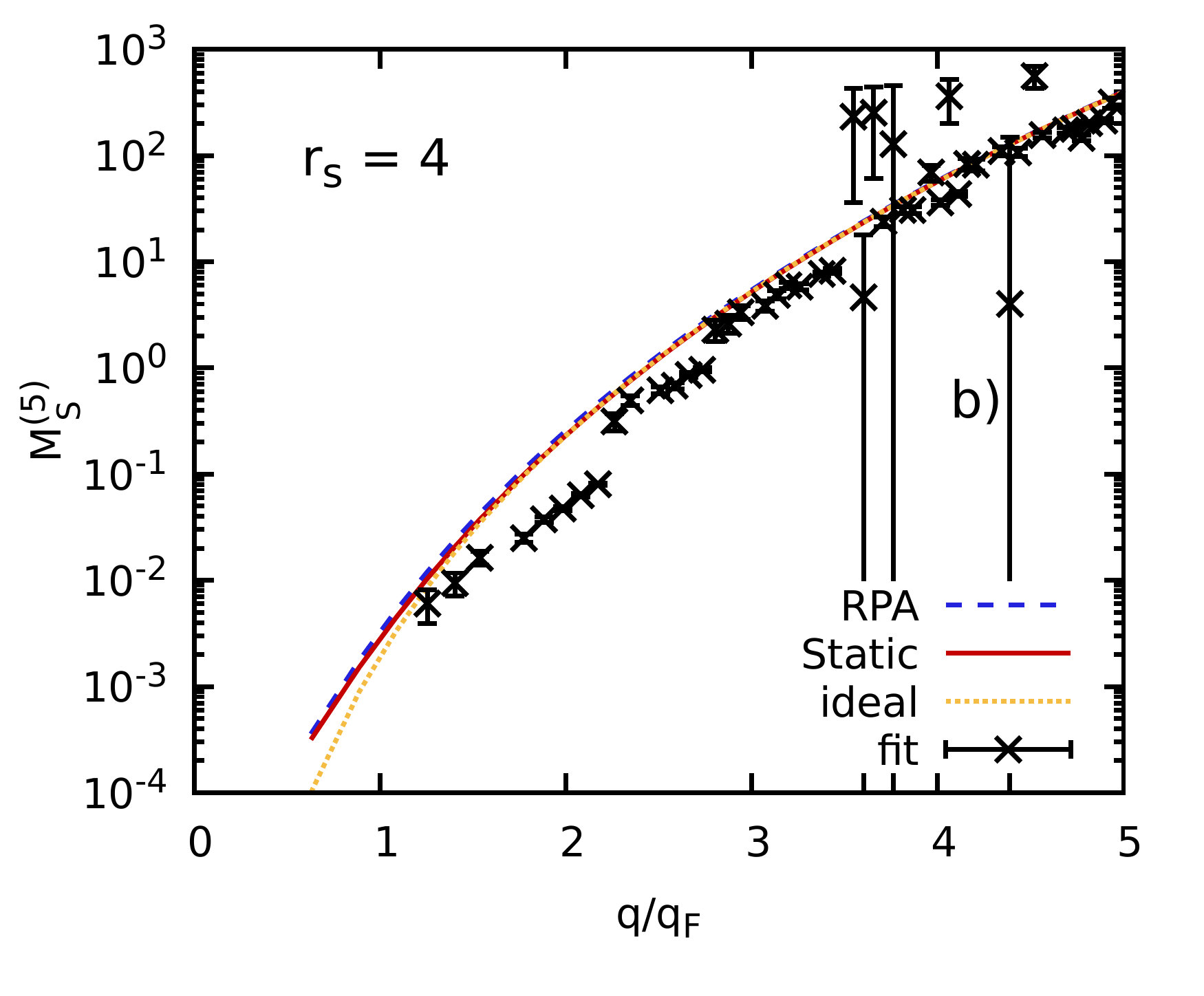}
\caption{\label{fig:Mom5}
The fifth frequency moment $M^{(5)}_{S}(\mathbf{q})$ for the UEG at $\Theta=1$ for a) $r_s=10$ and b) $r_s=4$. Black crosses: moments extracted via Eq.~(\ref{eq:final}); dashed blue, solid red, and dotted yellow: reference data within RPA, \emph{static approximation}, and for the ideal Fermi gas computed from synthetic $S(\mathbf{q},\omega)$ directly via Eq.~(\ref{eq:moments}).
}
\end{figure} 

Let us proceed to the fourth moment $M^{(4)}_{S}(\mathbf{q})$ shown in Fig.~\ref{fig:Mom4}. In this case, we find that our method is still capable of accurately resolving $M^{(4)}_{S}(\mathbf{q})$ over five orders of magnitude for $r_s=10$, whereas the quality is noticeably less good for $r_s=4$. Still, we can obtain valuable insights into the correct qualitative behavior even for the higher density.

Finally, we analyze $M^{(5)}_{S}(\mathbf{q})$ in Fig.~\ref{fig:Mom5}. In this case, both the synthetic data and our extracted values span seven orders of magnitude in the depicted relevant range of wave numbers. For $r_s=10$, we obtain reasonable results for all $q$, although there do appear noticeable fluctuations in the extracted moments. For $r_s=4$, the quality is significantly worse, as it is expected, and the fluctuations are not fully captured by the error bars.

\section{Summary and Outlook\label{sec:summary}}

In this work, we have presented a new, formally exact method to extract all positive integer frequency moments of dynamic properties from imaginary-time correlation functions. As a practical example, we have investigated the DSF $S(\mathbf{q},\omega)$ of the UEG, which is directly connected to the ITCF $F(\mathbf{q},\tau)$ via the two-sided Laplace transform Eq.~(\ref{eq:analytic_continuation}).
We have demonstrated that the frequency moments $M^{(\alpha)}_{S}(\mathbf{q})$ directly correspond to the fitting coefficients from a polynomial fit to PIMC data for the ITCF in the canonical representation. In practice, we find good agreement between our newly extracted results and the existing sum rules for $\alpha=0,1,3$. In addition, we have presented, to our knowledge, the first data for the cases of $\alpha=2,4,5$. These results are interesting in their own right and will serve as a valuable benchmark for future developments such as the derivation of the $M^{(5)}_{S}(\mathbf{q})$ sum-rule featuring static three-body correlation functions, or the construction of novel DSF approximation schemes. From a physical perspective, we observe that only $M^{(0)}_{S}(\mathbf{q})$ (and also $M^{(-1)}_{S}(\mathbf{q})$, see e.g.~Ref.~\cite{Dornheim_review}) exhibits a pronounced structure with respect to the wave number $q$, whereas the cases of $\alpha=1,\dots,5$ are strictly monotonic. This constitutes a nontrivial finding that deserves to be explored in more depth in future works.

We are convinced that our work opens up enticing opportunities for impactful future research in a gamut of research fields, including the study of ultracold atoms~\cite{Boninsegni1996,Filinov_PRA_2012,Dornheim_SciRep_2022}, warm dense matter~\cite{dornheim_dynamic,dynamic_folgepaper,Hamann_PRB_2020}, as well as condensed matter physics~\cite{Mishchenko_PRB_2000,Georges_RMP_1996}. For example, the accurate knowledge of different $M^{(\alpha)}_{S}(\mathbf{q})$ is directly useful to further constrain the analytic continuation from the imaginary-time domain to real frequencies~\cite{Vitali_PRB_2010,Filinov_PRA_2012}. Moreover, the frequency moments are the key input for the method of moments~\cite{tkachenko_book}, which constitutes a promising route for the direct calculation of dynamic properties based on static QMC simulation data without the need for an explicit numerical inversion of Eq.~(\ref{eq:analytic_continuation}). The corresponding recent results for the warm dense UEG based on the odd moments already look promising~\cite{Ara_proceeding_2022}, and it is likely that the incorporation of the hitherto unknown moments of $\alpha=2,4,5$ would lead to further improvement.

In addition to its value for quantum many-body theory, our approach for the study of the frequency moments of the DSF is also of direct practical use for the interpretation of XRTS experiments of matter under extreme conditions. In particular, the measured XRTS intensity signal is given by the convolution of the DSF with the combined probe and instrument function $R(\omega)$~\cite{siegfried_review},
\begin{eqnarray}\label{eq:convolution}
I(\mathbf{q},\omega) = S(\mathbf{q},\omega) \circledast R(\omega)\ .
\end{eqnarray}
In practice, XRTS thus does not give one direct access to the DSF (and its frequency moments $M_\alpha$) as the deconvolution is typically rendered highly unstable by the inevitable noise in the experimental measurement. This restriction does not pose an obstacle in the Laplace domain, where one can make use of the well-known convolution theorem, which, in combination with Eqs.~(\ref{eq:analytic_continuation}) and (\ref{eq:convolution}), gives~\cite{Dornheim_insight_2022,Dornheim_T_2022}
\begin{eqnarray}\label{eq:convolution_theorem}
F(\mathbf{q},\tau) = \frac{\mathcal{L}\left[I(\mathbf{q},\omega)\right]}{\mathcal{L}\left[R(\omega)\right]}\ .
\end{eqnarray}
Since, in addition to the actual intensity $I(\mathbf{q},\omega)$, the source and instrument function is often known with high accuracy e.g.~from additional source monitoring as it is employed at modern X-ray free-electron laser facilities~\cite{Tschentscher_2017}, the evaluation of the RHS of Eq.~(\ref{eq:convolution_theorem}) gives one access to the ITCF $F(\mathbf{q},\tau)$ of the probed system. Therefore, our new framework for the estimation of the frequency moments $M^{(\alpha)}_{S}(\mathbf{q})$ is also directly useful for the interpretation of XRTS experiments of real materials. 

Finally, we stress that our idea is not limited to the DSF and the corresponding ITCF $F(\mathbf{q},\tau)$ and can easily be extended to other dynamic properties. For example, the Matsubara Green function $G_\textnormal{M}(\mathbf{q},\tau)$ [see Ref.~\cite{boninsegni1} for an accessible discussion] is connected to the single-particle spectral function $A(\mathbf{q},\omega)$ via the relation~\cite{Filinov_PRA_2012,Schueler_PRB_2018}
\begin{eqnarray}\label{eq:Matsubara}
G_\textnormal{M}(\mathbf{q},\tau) = \int_{-\infty}^\infty \frac{\textnormal{d}\omega}{2\pi} \frac{e^{-\tau\omega}}{1\pm e^{-\beta\omega}} A(\mathbf{q},\omega) \ ,
\end{eqnarray}
with the $\pm$ in the denominator corresponding to fermions and bosons, respectively. 
Single particle excitations of the system are most visible in $A(\mathbf{q},\omega)$ as peaks, plasmons as well as other quasi-particles leave their signatures in the spectral function. An integration over the momenta will produce the density of states from the spectral function~\cite{quantum_theory}.
It is easy to see that the frequency moments of $A(\mathbf{q},\omega)$ ---here denoted as $M^{(\alpha)}_{A}(\mathbf{q})$ ---can be obtained from $G_\textnormal{M}(\mathbf{q},\tau)$ via
\begin{eqnarray}\label{eq:A_moments}
M^{(\alpha)}_{A}(\mathbf{q}) &=& \left(-1\right)^\alpha 2\pi \left\{\left.
\frac{\partial^\alpha G_\textnormal{M}(\mathbf{q},\tau)}{\partial\tau^\alpha}\right|_{\tau=0}\right. \\\nonumber & &\pm \left. \left.
\frac{\partial^\alpha G_\textnormal{M}(\mathbf{q},\tau)}{\partial\tau^\alpha}\right|_{\tau=\beta}
\right\} \ .
\end{eqnarray}
In contrast to the DSF, the frequency moments of the single-particle spectral function thus require evaluation of the derivatives around both $\tau=0$ and $\tau=\beta$. This makes intuitive sense as the Matsubara Green function does not have a symmetry relation around $\tau=\beta/2$ such as $F(\mathbf{q},\tau)$, for which both derivatives would be equal up to a sign change. The practical evaluation of Eq.~(\ref{eq:A_moments}) thus requires polynomial expansions around both boundary values of $\tau$, which does not pose an obstacle.

\appendix
\section{Methodology of the fitting scheme\label{sec:details}}

Classic $1$-dimensional polynomial interpolation goes back to Newton, Lagrange, and others, see, e.g., Ref.~\cite{LIP}. Its generalization to regression tasks was
mainly proposed and developed by Gau\ss, Markov, and Gergonne \cite{gergonne1974application,stigler1974gergonne}
and is omnipresent in mathematics and computing till today. Due to 
Ref.~\cite{platte:2011}, however, there are theoretical and practical limits when it comes to fitting functions sampled on equidistant data nodes or grids. Here, often the term "over-fitting" is used for pointing to Runge's phenomenon, being a classic problem in applied mathematics 
\cite{runge,hewitt1979gibbs,dimarogonas1996vibration}.

In Ref.~\cite{REG_arxiv} the problem is addressed even in multi-dimensions, and, based on the results in 
Refs.~\cite{PIP1,PIP2,MIP,IEEE}, implementations are condensed into the open source package {\sc minterpy} \cite{minterpy}. In contrast to na\"ively fitting functions with respect to the canonical polynomial basis $1,x,x^2,\dots,x^n$, {\sc minterpy} rests on Lagrange polynomials
$$l_{i} (x) = \prod_{j\neq i}^n \frac{x-q_{j}}{q_{i} - q_{j}}\,, \quad l_{i}(q_j) = \delta_{ij}\,, 0\leq i,j\leq n$$
being located in the Chebyshev-Lobatto nodes \cite{trefethen2019}
$$q_i \in \mathrm{Cheb}_n = \left\{ \cos\Big(\frac{i\pi}{n}\Big) : 0 \leq i \leq n\right\}\,.$$
Fitting a function $f: [-1,1] \longrightarrow \mathbb{R}$, 
sampled in (equidistant) data points 
$P=\{p_1,\dots,p_m\}\subseteq [-1,1]$, $F =(f(p_1),\dots,f(p_m)) \in \mathbb{R}^m$, $m\in \mathbb{N}$ is realised due to solving a classic least square problem
$$ C = \mathrm{argmin}_{X\in \mathbb{R}^{n+1}} \|RX - F\|^2\,,$$
where $R =(r_{k,i})_{1\leq k\leq m, 1\leq i\leq n+1 } \in \mathbb{R}^{m \times n+1}$, with $r_{k,i} = l_i(p_k)$, denotes the regression matrix.
Once the coefficients $C =(c_0,\dots,c_n)$ are computed, the polynomial $Q_{f,n}$ of degree $n\in \mathbb{N}$ fitting the function $f$ is given by 
$$ f(x) \approx Q_{f,n}(x) = \sum_{i=0}^n c_il_i(x)\,.$$
This Lagrange-regression scheme turns out to maintain stability for high polynomial degrees, and shows more approximation power, suppressing Runge's phenomenon than regression with respect to the canonical basis \cite{REG_arxiv,MIP,trefethen2019}.
{\sc minterpy} includes a domain re-scaling routine and a basis transformation that enables a numerically stable transformation of the Lagrange coefficients to the canonical coefficients $D\in\mathbb{R}^{n+1}$
$$Q_{f,n} = \sum_{i=0}^n c_il_i(x)  = \sum_{i=0}^n d_ix^i\,, \quad D = (d_0,\dots,d_n)\,,$$
fitting the initial data.

For determining the maximum degree $\alpha_{\textnormal{max}}$ used for the truncation in Eq.~(\ref{eq:Taylor}), we apply a Monte-Carlo cross-validation strategy.
In this strategy, a large proportion of the dataset ($90\%$) is randomly sampled and used to fit a polynomial of a given degree, while the rest of the dataset ($10\%$) is used to compute the maximum absolute error.
Furthermore, to have a more robust estimate of the maximum degree, we randomly and uniformly perturb $F$ within the range of its PIMC error bars.
This procedure is then repeated multiple times ($250$), each time giving an estimate of the maximum polynomial degree for the given dataset split.
We pick $\alpha_{\textnormal{max}}$ as the polynomial degree that both minimizes the maximum absolute error and appears the most over many repetitions.

Once $\alpha_{\textnormal{max}}$ has been determined, we estimate the error associated with the polynomial coefficients by fitting polynomials of the same degree many times ($1000$) using the whole dataset.
As before, $F$ is also randomly and uniformly perturbed within the range of its error estimate.
The standard deviation of the polynomial coefficients over many repetitions represents the error estimate of the coefficients.
\begin{figure}[!htb]\centering
\includegraphics[width=0.475\textwidth]{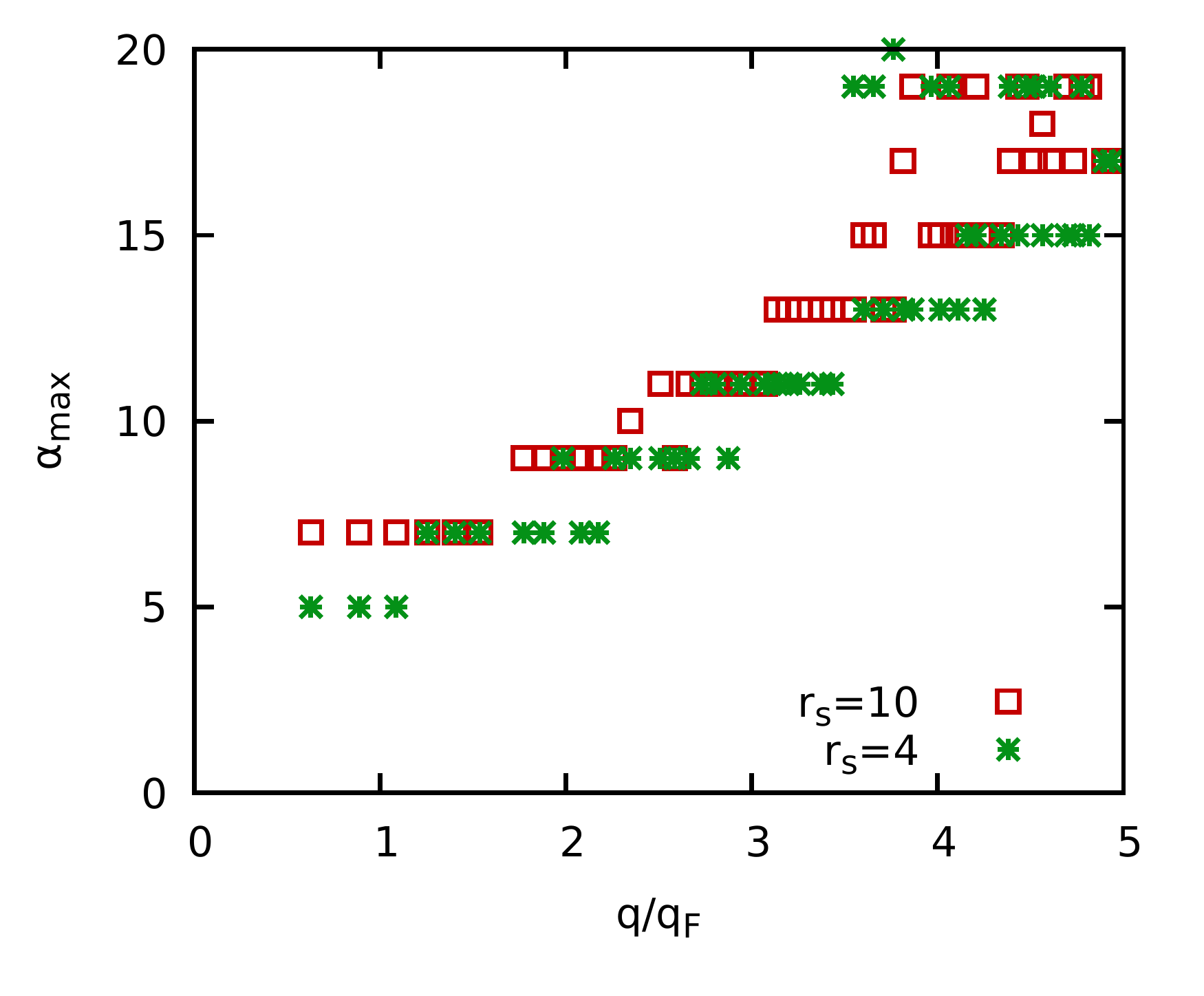}
\caption{\label{fig:AlphaMax}
Truncated polynomial degree $\alpha_\textnormal{max}$ for $r_s=10$ (red squares) and $r_s=4$ (green stars) as a function of the wave number $q$.
}
\end{figure} 

In Fig.~\ref{fig:AlphaMax}, we show the truncated polynomial degree $\alpha_\textnormal{max}$ over the entire considered range of wave numbers $q$. Overall, we find that $\alpha_\textnormal{max}$ tends to increase with $q$, as $F(\mathbf{q},\tau)$ exhibits more curvature for large wave numbers. These trends are similar for $r_s=10$ (red squares) and $r_s=4$ (green stars), although $\alpha_\textnormal{max}$ tends to be slightly lower in the latter case. 
\FloatBarrier
\section*{Acknowledgments}
This work was partially supported by the Center for Advanced Systems Understanding (CASUS) which is financed by Germany’s Federal Ministry of Education and Research (BMBF) and by the Saxon state government out of the State budget approved by the Saxon State Parliament. 
The PIMC calculations were partly carried out at the Norddeutscher Verbund f\"ur Hoch- und H\"ochstleistungsrechnen (HLRN) under grant shp00026 and on a Bull Cluster at the Center for Information Services and High Performance Computing (ZIH) at Technische Universit\"at Dresden.

\bibliography{bibliography}

\begin{thebibliography}{158}%
\makeatletter
\providecommand \@ifxundefined [1]{%
 \@ifx{#1\undefined}
}%
\providecommand \@ifnum [1]{%
 \ifnum #1\expandafter \@firstoftwo
 \else \expandafter \@secondoftwo
 \fi
}%
\providecommand \@ifx [1]{%
 \ifx #1\expandafter \@firstoftwo
 \else \expandafter \@secondoftwo
 \fi
}%
\providecommand \natexlab [1]{#1}%
\providecommand \enquote  [1]{``#1''}%
\providecommand \bibnamefont  [1]{#1}%
\providecommand \bibfnamefont [1]{#1}%
\providecommand \citenamefont [1]{#1}%
\providecommand \href@noop [0]{\@secondoftwo}%
\providecommand \href [0]{\begingroup \@sanitize@url \@href}%
\providecommand \@href[1]{\@@startlink{#1}\@@href}%
\providecommand \@@href[1]{\endgroup#1\@@endlink}%
\providecommand \@sanitize@url [0]{\catcode `\\12\catcode `\$12\catcode
  `\&12\catcode `\#12\catcode `\^12\catcode `\_12\catcode `\%12\relax}%
\providecommand \@@startlink[1]{}%
\providecommand \@@endlink[0]{}%
\providecommand \url  [0]{\begingroup\@sanitize@url \@url }%
\providecommand \@url [1]{\endgroup\@href {#1}{\urlprefix }}%
\providecommand \urlprefix  [0]{URL }%
\providecommand \Eprint [0]{\href }%
\providecommand \doibase [0]{http://dx.doi.org/}%
\providecommand \selectlanguage [0]{\@gobble}%
\providecommand \bibinfo  [0]{\@secondoftwo}%
\providecommand \bibfield  [0]{\@secondoftwo}%
\providecommand \translation [1]{[#1]}%
\providecommand \BibitemOpen [0]{}%
\providecommand \bibitemStop [0]{}%
\providecommand \bibitemNoStop [0]{.\EOS\space}%
\providecommand \EOS [0]{\spacefactor3000\relax}%
\providecommand \BibitemShut  [1]{\csname bibitem#1\endcsname}%
\let\auto@bib@innerbib\@empty
\bibitem [{\citenamefont {Nagy}\ \emph {et~al.}(1989)\citenamefont {Nagy},
  \citenamefont {Arnau},\ and\ \citenamefont {Echenique}}]{Nagy_PRA_1989}%
  \BibitemOpen
  \bibfield  {author} {\bibinfo {author} {\bibfnamefont {I.}~\bibnamefont
  {Nagy}}, \bibinfo {author} {\bibfnamefont {A.}~\bibnamefont {Arnau}}, \ and\
  \bibinfo {author} {\bibfnamefont {P.~M.}\ \bibnamefont {Echenique}},\
  }\bibfield  {title} {\enquote {\bibinfo {title} {Nonlinear stopping power and
  energy-loss straggling of an interacting electron gas for slow ions},}\
  }\href {\doibase 10.1103/PhysRevA.40.987} {\bibfield  {journal} {\bibinfo
  {journal} {Phys. Rev. A}\ }\textbf {\bibinfo {volume} {40}},\ \bibinfo
  {pages} {987--994} (\bibinfo {year} {1989})}\BibitemShut {NoStop}%
\bibitem [{\citenamefont {Balzer}\ \emph {et~al.}(2016)\citenamefont {Balzer},
  \citenamefont {Schl\"unzen},\ and\ \citenamefont {Bonitz}}]{Balzer_PRB_2016}%
  \BibitemOpen
  \bibfield  {author} {\bibinfo {author} {\bibfnamefont {Karsten}\ \bibnamefont
  {Balzer}}, \bibinfo {author} {\bibfnamefont {Niclas}\ \bibnamefont
  {Schl\"unzen}}, \ and\ \bibinfo {author} {\bibfnamefont {Michael}\
  \bibnamefont {Bonitz}},\ }\bibfield  {title} {\enquote {\bibinfo {title}
  {Stopping dynamics of ions passing through correlated honeycomb clusters},}\
  }\href {\doibase 10.1103/PhysRevB.94.245118} {\bibfield  {journal} {\bibinfo
  {journal} {Phys. Rev. B}\ }\textbf {\bibinfo {volume} {94}},\ \bibinfo
  {pages} {245118} (\bibinfo {year} {2016})}\BibitemShut {NoStop}%
\bibitem [{\citenamefont {Blaga}\ \emph {et~al.}(2009)\citenamefont {Blaga},
  \citenamefont {Catoire}, \citenamefont {Colosimo}, \citenamefont {Paulus},
  \citenamefont {Muller}, \citenamefont {Agostini},\ and\ \citenamefont
  {DiMauro}}]{Blaga2009}%
  \BibitemOpen
  \bibfield  {author} {\bibinfo {author} {\bibfnamefont {C.~I.}\ \bibnamefont
  {Blaga}}, \bibinfo {author} {\bibfnamefont {F.}~\bibnamefont {Catoire}},
  \bibinfo {author} {\bibfnamefont {P.}~\bibnamefont {Colosimo}}, \bibinfo
  {author} {\bibfnamefont {G.~G.}\ \bibnamefont {Paulus}}, \bibinfo {author}
  {\bibfnamefont {H.~G.}\ \bibnamefont {Muller}}, \bibinfo {author}
  {\bibfnamefont {P.}~\bibnamefont {Agostini}}, \ and\ \bibinfo {author}
  {\bibfnamefont {L.~F.}\ \bibnamefont {DiMauro}},\ }\bibfield  {title}
  {\enquote {\bibinfo {title} {Strong-field photoionization revisited},}\
  }\href {\doibase 10.1038/nphys1228} {\bibfield  {journal} {\bibinfo
  {journal} {Nature Physics}\ }\textbf {\bibinfo {volume} {5}},\ \bibinfo
  {pages} {335--338} (\bibinfo {year} {2009})}\BibitemShut {NoStop}%
\bibitem [{\citenamefont {Hochstuhl}\ \emph {et~al.}(2010)\citenamefont
  {Hochstuhl}, \citenamefont {Balzer}, \citenamefont {Bauch},\ and\
  \citenamefont {Bonitz}}]{HOCHSTUHL2010513}%
  \BibitemOpen
  \bibfield  {author} {\bibinfo {author} {\bibfnamefont {D.}~\bibnamefont
  {Hochstuhl}}, \bibinfo {author} {\bibfnamefont {K.}~\bibnamefont {Balzer}},
  \bibinfo {author} {\bibfnamefont {S.}~\bibnamefont {Bauch}}, \ and\ \bibinfo
  {author} {\bibfnamefont {M.}~\bibnamefont {Bonitz}},\ }\bibfield  {title}
  {\enquote {\bibinfo {title} {Nonequilibrium green function approach to
  photoionization processes in atoms},}\ }\href {\doibase
  https://doi.org/10.1016/j.physe.2009.06.044} {\bibfield  {journal} {\bibinfo
  {journal} {Physica E: Low-dimensional Systems and Nanostructures}\ }\textbf
  {\bibinfo {volume} {42}},\ \bibinfo {pages} {513--519} (\bibinfo {year}
  {2010})},\ \bibinfo {note} {proceedings of the international conference
  Frontiers of Quantum and Mesoscopic Thermodynamics FQMT '08}\BibitemShut
  {NoStop}%
\bibitem [{\citenamefont {Vorberger}\ \emph {et~al.}(2010)\citenamefont
  {Vorberger}, \citenamefont {Gericke}, \citenamefont {Bornath},\ and\
  \citenamefont {Schlanges}}]{transfer1}%
  \BibitemOpen
  \bibfield  {author} {\bibinfo {author} {\bibfnamefont {J.}~\bibnamefont
  {Vorberger}}, \bibinfo {author} {\bibfnamefont {D.~O.}\ \bibnamefont
  {Gericke}}, \bibinfo {author} {\bibfnamefont {Th.}\ \bibnamefont {Bornath}},
  \ and\ \bibinfo {author} {\bibfnamefont {M.}~\bibnamefont {Schlanges}},\
  }\bibfield  {title} {\enquote {\bibinfo {title} {Energy relaxation in dense,
  strongly coupled two-temperature plasmas},}\ }\href
  {https://journals.aps.org/pre/abstract/10.1103/PhysRevE.96.023203} {\bibfield
   {journal} {\bibinfo  {journal} {Phys. Rev. E}\ }\textbf {\bibinfo {volume}
  {81}},\ \bibinfo {pages} {046404} (\bibinfo {year} {2010})}\BibitemShut
  {NoStop}%
\bibitem [{\citenamefont {Zhang}\ \emph {et~al.}(2022)\citenamefont {Zhang},
  \citenamefont {Qin}, \citenamefont {Zhu},\ and\ \citenamefont
  {Vorberger}}]{ma15051902}%
  \BibitemOpen
  \bibfield  {author} {\bibinfo {author} {\bibfnamefont {Jia}\ \bibnamefont
  {Zhang}}, \bibinfo {author} {\bibfnamefont {Rui}\ \bibnamefont {Qin}},
  \bibinfo {author} {\bibfnamefont {Wenjun}\ \bibnamefont {Zhu}}, \ and\
  \bibinfo {author} {\bibfnamefont {Jan}\ \bibnamefont {Vorberger}},\
  }\bibfield  {title} {\enquote {\bibinfo {title} {Energy relaxation and
  electron-phonon coupling in laser-excited metals},}\ }\href {\doibase
  10.3390/ma15051902} {\bibfield  {journal} {\bibinfo  {journal} {Materials}\
  }\textbf {\bibinfo {volume} {15}} (\bibinfo {year} {2022}),\
  10.3390/ma15051902}\BibitemShut {NoStop}%
\bibitem [{\citenamefont {Bonitz}(2016)}]{bonitz_book}%
  \BibitemOpen
  \bibfield  {author} {\bibinfo {author} {\bibfnamefont {M.}~\bibnamefont
  {Bonitz}},\ }\href@noop {} {\emph {\bibinfo {title} {Quantum kinetic
  theory}}}\ (\bibinfo  {publisher} {Springer},\ \bibinfo {address}
  {Heidelberg},\ \bibinfo {year} {2016})\BibitemShut {NoStop}%
\bibitem [{\citenamefont {Stefanucci}\ and\ \citenamefont {van
  Leeuwen}(2013)}]{stefanucci2013nonequilibrium}%
  \BibitemOpen
  \bibfield  {author} {\bibinfo {author} {\bibfnamefont {G.}~\bibnamefont
  {Stefanucci}}\ and\ \bibinfo {author} {\bibfnamefont {R.}~\bibnamefont {van
  Leeuwen}},\ }\href {https://books.google.de/books?id=6GsrjPFXLDYC} {\emph
  {\bibinfo {title} {Nonequilibrium Many-Body Theory of Quantum Systems: A
  Modern Introduction}}}\ (\bibinfo  {publisher} {Cambridge University Press},\
  \bibinfo {year} {2013})\BibitemShut {NoStop}%
\bibitem [{\citenamefont {Anderson}(2007)}]{anderson2007quantum}%
  \BibitemOpen
  \bibfield  {author} {\bibinfo {author} {\bibfnamefont {J.B.}\ \bibnamefont
  {Anderson}},\ }\href {https://books.google.de/books?id=\_QUSDAAAQBAJ} {\emph
  {\bibinfo {title} {Quantum Monte Carlo: Origins, Development,
  Applications}}}\ (\bibinfo  {publisher} {Oxford University Press, USA},\
  \bibinfo {year} {2007})\BibitemShut {NoStop}%
\bibitem [{\citenamefont {Ceperley}(1995)}]{cep}%
  \BibitemOpen
  \bibfield  {author} {\bibinfo {author} {\bibfnamefont {D.~M.}\ \bibnamefont
  {Ceperley}},\ }\bibfield  {title} {\enquote {\bibinfo {title} {Path integrals
  in the theory of condensed helium},}\ }\href
  {https://journals.aps.org/rmp/abstract/10.1103/RevModPhys.67.279} {\bibfield
  {journal} {\bibinfo  {journal} {Rev. Mod. Phys}\ }\textbf {\bibinfo {volume}
  {67}},\ \bibinfo {pages} {279} (\bibinfo {year} {1995})}\BibitemShut
  {NoStop}%
\bibitem [{\citenamefont {Herman}\ \emph {et~al.}(1982)\citenamefont {Herman},
  \citenamefont {Bruskin},\ and\ \citenamefont {Berne}}]{Berne_JCP_1982}%
  \BibitemOpen
  \bibfield  {author} {\bibinfo {author} {\bibfnamefont {M.~F.}\ \bibnamefont
  {Herman}}, \bibinfo {author} {\bibfnamefont {E.~J.}\ \bibnamefont {Bruskin}},
  \ and\ \bibinfo {author} {\bibfnamefont {B.~J.}\ \bibnamefont {Berne}},\
  }\bibfield  {title} {\enquote {\bibinfo {title} {On path integral monte carlo
  simulations},}\ }\href {\doibase 10.1063/1.442815} {\bibfield  {journal}
  {\bibinfo  {journal} {The Journal of Chemical Physics}\ }\textbf {\bibinfo
  {volume} {76}},\ \bibinfo {pages} {5150--5155} (\bibinfo {year}
  {1982})}\BibitemShut {NoStop}%
\bibitem [{\citenamefont {Takahashi}\ and\ \citenamefont
  {Imada}(1984)}]{Takahashi_Imada_PIMC_1984}%
  \BibitemOpen
  \bibfield  {author} {\bibinfo {author} {\bibfnamefont {Minoru}\ \bibnamefont
  {Takahashi}}\ and\ \bibinfo {author} {\bibfnamefont {Masatoshi}\ \bibnamefont
  {Imada}},\ }\bibfield  {title} {\enquote {\bibinfo {title} {Monte carlo
  calculation of quantum systems},}\ }\href {\doibase 10.1143/JPSJ.53.963}
  {\bibfield  {journal} {\bibinfo  {journal} {Journal of the Physical Society
  of Japan}\ }\textbf {\bibinfo {volume} {53}},\ \bibinfo {pages} {963--974}
  (\bibinfo {year} {1984})}\BibitemShut {NoStop}%
\bibitem [{\citenamefont {Thirumalai}\ and\ \citenamefont
  {Berne}(1983)}]{Berne_JCP_1983}%
  \BibitemOpen
  \bibfield  {author} {\bibinfo {author} {\bibfnamefont {Devarajan}\
  \bibnamefont {Thirumalai}}\ and\ \bibinfo {author} {\bibfnamefont {Bruce~J.}\
  \bibnamefont {Berne}},\ }\bibfield  {title} {\enquote {\bibinfo {title} {On
  the calculation of time correlation functions in quantum systems: Path
  integral techniquesa)},}\ }\href {\doibase 10.1063/1.445597} {\bibfield
  {journal} {\bibinfo  {journal} {The Journal of Chemical Physics}\ }\textbf
  {\bibinfo {volume} {79}},\ \bibinfo {pages} {5029--5033} (\bibinfo {year}
  {1983})}\BibitemShut {NoStop}%
\bibitem [{\citenamefont {Dornheim}\ \emph
  {et~al.}(2021{\natexlab{a}})\citenamefont {Dornheim}, \citenamefont
  {Moldabekov},\ and\ \citenamefont {Vorberger}}]{Dornheim_JCP_ITCF_2021}%
  \BibitemOpen
  \bibfield  {author} {\bibinfo {author} {\bibfnamefont {Tobias}\ \bibnamefont
  {Dornheim}}, \bibinfo {author} {\bibfnamefont {Zhandos~A.}\ \bibnamefont
  {Moldabekov}}, \ and\ \bibinfo {author} {\bibfnamefont {Jan}\ \bibnamefont
  {Vorberger}},\ }\bibfield  {title} {\enquote {\bibinfo {title} {Nonlinear
  density response from imaginary-time correlation functions: Ab initio path
  integral monte carlo simulations of the warm dense electron gas},}\ }\href
  {\doibase 10.1063/5.0058988} {\bibfield  {journal} {\bibinfo  {journal} {The
  Journal of Chemical Physics}\ }\textbf {\bibinfo {volume} {155}},\ \bibinfo
  {pages} {054110} (\bibinfo {year} {2021}{\natexlab{a}})}\BibitemShut
  {NoStop}%
\bibitem [{\citenamefont {Vitali}\ \emph {et~al.}(2010)\citenamefont {Vitali},
  \citenamefont {Rossi}, \citenamefont {Reatto},\ and\ \citenamefont
  {Galli}}]{Vitali_PRB_2010}%
  \BibitemOpen
  \bibfield  {author} {\bibinfo {author} {\bibfnamefont {E.}~\bibnamefont
  {Vitali}}, \bibinfo {author} {\bibfnamefont {M.}~\bibnamefont {Rossi}},
  \bibinfo {author} {\bibfnamefont {L.}~\bibnamefont {Reatto}}, \ and\ \bibinfo
  {author} {\bibfnamefont {D.~E.}\ \bibnamefont {Galli}},\ }\bibfield  {title}
  {\enquote {\bibinfo {title} {Ab initio low-energy dynamics of superfluid and
  solid $^{4}\textnormal{H}\textnormal{e}$},}\ }\href {\doibase
  10.1103/PhysRevB.82.174510} {\bibfield  {journal} {\bibinfo  {journal} {Phys.
  Rev. B}\ }\textbf {\bibinfo {volume} {82}},\ \bibinfo {pages} {174510}
  (\bibinfo {year} {2010})}\BibitemShut {NoStop}%
\bibitem [{\citenamefont {Filinov}\ and\ \citenamefont
  {Bonitz}(2012)}]{Filinov_PRA_2012}%
  \BibitemOpen
  \bibfield  {author} {\bibinfo {author} {\bibfnamefont {A.}~\bibnamefont
  {Filinov}}\ and\ \bibinfo {author} {\bibfnamefont {M.}~\bibnamefont
  {Bonitz}},\ }\bibfield  {title} {\enquote {\bibinfo {title} {Collective and
  single-particle excitations in two-dimensional dipolar bose gases},}\ }\href
  {\doibase 10.1103/PhysRevA.86.043628} {\bibfield  {journal} {\bibinfo
  {journal} {Phys. Rev. A}\ }\textbf {\bibinfo {volume} {86}},\ \bibinfo
  {pages} {043628} (\bibinfo {year} {2012})}\BibitemShut {NoStop}%
\bibitem [{\citenamefont {Kora}\ and\ \citenamefont
  {Boninsegni}(2018)}]{Boninsegni_maximum_entropy}%
  \BibitemOpen
  \bibfield  {author} {\bibinfo {author} {\bibfnamefont {Youssef}\ \bibnamefont
  {Kora}}\ and\ \bibinfo {author} {\bibfnamefont {Massimo}\ \bibnamefont
  {Boninsegni}},\ }\bibfield  {title} {\enquote {\bibinfo {title} {Dynamic
  structure factor of superfluid $^{4}\mathrm{He}$ from quantum monte carlo:
  Maximum entropy revisited},}\ }\href {\doibase 10.1103/PhysRevB.98.134509}
  {\bibfield  {journal} {\bibinfo  {journal} {Phys. Rev. B}\ }\textbf {\bibinfo
  {volume} {98}},\ \bibinfo {pages} {134509} (\bibinfo {year}
  {2018})}\BibitemShut {NoStop}%
\bibitem [{\citenamefont {Dornheim}\ \emph
  {et~al.}(2018{\natexlab{a}})\citenamefont {Dornheim}, \citenamefont {Groth},
  \citenamefont {Vorberger},\ and\ \citenamefont {Bonitz}}]{dornheim_dynamic}%
  \BibitemOpen
  \bibfield  {author} {\bibinfo {author} {\bibfnamefont {T.}~\bibnamefont
  {Dornheim}}, \bibinfo {author} {\bibfnamefont {S.}~\bibnamefont {Groth}},
  \bibinfo {author} {\bibfnamefont {J.}~\bibnamefont {Vorberger}}, \ and\
  \bibinfo {author} {\bibfnamefont {M.}~\bibnamefont {Bonitz}},\ }\bibfield
  {title} {\enquote {\bibinfo {title} {Ab initio path integral {M}onte {C}arlo
  results for the dynamic structure factor of correlated electrons: From the
  electron liquid to warm dense matter},}\ }\href
  {https://journals.aps.org/prl/abstract/10.1103/PhysRevLett.121.255001}
  {\bibfield  {journal} {\bibinfo  {journal} {Phys. Rev. Lett.}\ }\textbf
  {\bibinfo {volume} {121}},\ \bibinfo {pages} {255001} (\bibinfo {year}
  {2018}{\natexlab{a}})}\BibitemShut {NoStop}%
\bibitem [{\citenamefont {Ferr\'e}\ and\ \citenamefont
  {Boronat}(2016)}]{Ferre_PRB_2016}%
  \BibitemOpen
  \bibfield  {author} {\bibinfo {author} {\bibfnamefont {G.}~\bibnamefont
  {Ferr\'e}}\ and\ \bibinfo {author} {\bibfnamefont {J.}~\bibnamefont
  {Boronat}},\ }\bibfield  {title} {\enquote {\bibinfo {title} {Dynamic
  structure factor of liquid $^{4}\mathrm{He}$ across the normal-superfluid
  transition},}\ }\href {\doibase 10.1103/PhysRevB.93.104510} {\bibfield
  {journal} {\bibinfo  {journal} {Phys. Rev. B}\ }\textbf {\bibinfo {volume}
  {93}},\ \bibinfo {pages} {104510} (\bibinfo {year} {2016})}\BibitemShut
  {NoStop}%
\bibitem [{\citenamefont {Motta}\ \emph {et~al.}(2015)\citenamefont {Motta},
  \citenamefont {Galli}, \citenamefont {Moroni},\ and\ \citenamefont
  {Vitali}}]{Motta_JCP_2015}%
  \BibitemOpen
  \bibfield  {author} {\bibinfo {author} {\bibfnamefont {M.}~\bibnamefont
  {Motta}}, \bibinfo {author} {\bibfnamefont {D.~E.}\ \bibnamefont {Galli}},
  \bibinfo {author} {\bibfnamefont {S.}~\bibnamefont {Moroni}}, \ and\ \bibinfo
  {author} {\bibfnamefont {E.}~\bibnamefont {Vitali}},\ }\bibfield  {title}
  {\enquote {\bibinfo {title} {Imaginary time density-density correlations for
  two-dimensional electron gases at high density},}\ }\href {\doibase
  10.1063/1.4934666} {\bibfield  {journal} {\bibinfo  {journal} {The Journal of
  Chemical Physics}\ }\textbf {\bibinfo {volume} {143}},\ \bibinfo {pages}
  {164108} (\bibinfo {year} {2015})}\BibitemShut {NoStop}%
\bibitem [{\citenamefont {Filinov}(2016)}]{Filinov_PRA_2016}%
  \BibitemOpen
  \bibfield  {author} {\bibinfo {author} {\bibfnamefont {A.}~\bibnamefont
  {Filinov}},\ }\bibfield  {title} {\enquote {\bibinfo {title} {Correlation
  effects and collective excitations in bosonic bilayers: Role of quantum
  statistics, superfluidity, and the dimerization transition},}\ }\href
  {\doibase 10.1103/PhysRevA.94.013603} {\bibfield  {journal} {\bibinfo
  {journal} {Phys. Rev. A}\ }\textbf {\bibinfo {volume} {94}},\ \bibinfo
  {pages} {013603} (\bibinfo {year} {2016})}\BibitemShut {NoStop}%
\bibitem [{\citenamefont {Dornheim}\ \emph
  {et~al.}(2022{\natexlab{a}})\citenamefont {Dornheim}, \citenamefont
  {Moldabekov}, \citenamefont {Vorberger},\ and\ \citenamefont
  {Militzer}}]{Dornheim_SciRep_2022}%
  \BibitemOpen
  \bibfield  {author} {\bibinfo {author} {\bibfnamefont {Tobias}\ \bibnamefont
  {Dornheim}}, \bibinfo {author} {\bibfnamefont {Zhandos~A.}\ \bibnamefont
  {Moldabekov}}, \bibinfo {author} {\bibfnamefont {Jan}\ \bibnamefont
  {Vorberger}}, \ and\ \bibinfo {author} {\bibfnamefont {Burkhard}\
  \bibnamefont {Militzer}},\ }\bibfield  {title} {\enquote {\bibinfo {title}
  {Path integral monte carlo approach to the structural properties and
  collective excitations of liquid $^3$he without fixed nodes},}\ }\href
  {\doibase 10.1038/s41598-021-04355-9} {\bibfield  {journal} {\bibinfo
  {journal} {Scientific Reports}\ }\textbf {\bibinfo {volume} {12}},\ \bibinfo
  {pages} {708} (\bibinfo {year} {2022}{\natexlab{a}})}\BibitemShut {NoStop}%
\bibitem [{\citenamefont {Dornheim}\ \emph
  {et~al.}(2022{\natexlab{b}})\citenamefont {Dornheim}, \citenamefont
  {Moldabekov}, \citenamefont {Tolias}, \citenamefont {Böhme},\ and\
  \citenamefont {Vorberger}}]{Dornheim_insight_2022}%
  \BibitemOpen
  \bibfield  {author} {\bibinfo {author} {\bibfnamefont {Tobias}\ \bibnamefont
  {Dornheim}}, \bibinfo {author} {\bibfnamefont {Zhandos}\ \bibnamefont
  {Moldabekov}}, \bibinfo {author} {\bibfnamefont {Panagiotis}\ \bibnamefont
  {Tolias}}, \bibinfo {author} {\bibfnamefont {Maximilian}\ \bibnamefont
  {Böhme}}, \ and\ \bibinfo {author} {\bibfnamefont {Jan}\ \bibnamefont
  {Vorberger}},\ }\bibfield  {title} {\enquote {\bibinfo {title} {Physical
  insights from imaginary-time density--density correlation functions},}\
  }\href {\doibase 10.48550/ARXIV.2209.02254} {\  (\bibinfo {year}
  {2022}{\natexlab{b}}),\ 10.48550/ARXIV.2209.02254}\BibitemShut {NoStop}%
\bibitem [{\citenamefont {Boninsegni}\ and\ \citenamefont
  {Ceperley}(1996)}]{Boninsegni1996}%
  \BibitemOpen
  \bibfield  {author} {\bibinfo {author} {\bibfnamefont {Massimo}\ \bibnamefont
  {Boninsegni}}\ and\ \bibinfo {author} {\bibfnamefont {David~M.}\ \bibnamefont
  {Ceperley}},\ }\bibfield  {title} {\enquote {\bibinfo {title} {Density
  fluctuations in liquid4he. path integrals and maximum entropy},}\ }\href
  {\doibase 10.1007/BF00751861} {\bibfield  {journal} {\bibinfo  {journal}
  {Journal of Low Temperature Physics}\ }\textbf {\bibinfo {volume} {104}},\
  \bibinfo {pages} {339--357} (\bibinfo {year} {1996})}\BibitemShut {NoStop}%
\bibitem [{\citenamefont {Groth}\ \emph {et~al.}(2019)\citenamefont {Groth},
  \citenamefont {Dornheim},\ and\ \citenamefont
  {Vorberger}}]{dynamic_folgepaper}%
  \BibitemOpen
  \bibfield  {author} {\bibinfo {author} {\bibfnamefont {S.}~\bibnamefont
  {Groth}}, \bibinfo {author} {\bibfnamefont {T.}~\bibnamefont {Dornheim}}, \
  and\ \bibinfo {author} {\bibfnamefont {J.}~\bibnamefont {Vorberger}},\
  }\bibfield  {title} {\enquote {\bibinfo {title} {Ab initio path integral
  {M}onte {C}arlo approach to the static and dynamic density response of the
  uniform electron gas},}\ }\href
  {https://link.aps.org/doi/10.1103/PhysRevB.99.235122} {\bibfield  {journal}
  {\bibinfo  {journal} {Phys. Rev. B}\ }\textbf {\bibinfo {volume} {99}},\
  \bibinfo {pages} {235122} (\bibinfo {year} {2019})}\BibitemShut {NoStop}%
\bibitem [{\citenamefont {Dornheim}\ and\ \citenamefont
  {Vorberger}(2020)}]{Dornheim_PRE_2020}%
  \BibitemOpen
  \bibfield  {author} {\bibinfo {author} {\bibfnamefont {Tobias}\ \bibnamefont
  {Dornheim}}\ and\ \bibinfo {author} {\bibfnamefont {Jan}\ \bibnamefont
  {Vorberger}},\ }\bibfield  {title} {\enquote {\bibinfo {title} {Finite-size
  effects in the reconstruction of dynamic properties from ab initio path
  integral monte carlo simulations},}\ }\href {\doibase
  10.1103/PhysRevE.102.063301} {\bibfield  {journal} {\bibinfo  {journal}
  {Phys. Rev. E}\ }\textbf {\bibinfo {volume} {102}},\ \bibinfo {pages}
  {063301} (\bibinfo {year} {2020})}\BibitemShut {NoStop}%
\bibitem [{\citenamefont {Mishchenko}\ \emph {et~al.}(2000)\citenamefont
  {Mishchenko}, \citenamefont {Prokof'ev}, \citenamefont {Sakamoto},\ and\
  \citenamefont {Svistunov}}]{Mishchenko_PRB_2000}%
  \BibitemOpen
  \bibfield  {author} {\bibinfo {author} {\bibfnamefont {A.~S.}\ \bibnamefont
  {Mishchenko}}, \bibinfo {author} {\bibfnamefont {N.~V.}\ \bibnamefont
  {Prokof'ev}}, \bibinfo {author} {\bibfnamefont {A.}~\bibnamefont {Sakamoto}},
  \ and\ \bibinfo {author} {\bibfnamefont {B.~V.}\ \bibnamefont {Svistunov}},\
  }\bibfield  {title} {\enquote {\bibinfo {title} {Diagrammatic quantum monte
  carlo study of the fr\"ohlich polaron},}\ }\href {\doibase
  10.1103/PhysRevB.62.6317} {\bibfield  {journal} {\bibinfo  {journal} {Phys.
  Rev. B}\ }\textbf {\bibinfo {volume} {62}},\ \bibinfo {pages} {6317--6336}
  (\bibinfo {year} {2000})}\BibitemShut {NoStop}%
\bibitem [{\citenamefont {Silver}\ \emph
  {et~al.}(1990{\natexlab{a}})\citenamefont {Silver}, \citenamefont {Sivia},\
  and\ \citenamefont {Gubernatis}}]{Silver_PRB_1990}%
  \BibitemOpen
  \bibfield  {author} {\bibinfo {author} {\bibfnamefont {R.~N.}\ \bibnamefont
  {Silver}}, \bibinfo {author} {\bibfnamefont {D.~S.}\ \bibnamefont {Sivia}}, \
  and\ \bibinfo {author} {\bibfnamefont {J.~E.}\ \bibnamefont {Gubernatis}},\
  }\bibfield  {title} {\enquote {\bibinfo {title} {Maximum-entropy method for
  analytic continuation of quantum monte carlo data},}\ }\href {\doibase
  10.1103/PhysRevB.41.2380} {\bibfield  {journal} {\bibinfo  {journal} {Phys.
  Rev. B}\ }\textbf {\bibinfo {volume} {41}},\ \bibinfo {pages} {2380--2389}
  (\bibinfo {year} {1990}{\natexlab{a}})}\BibitemShut {NoStop}%
\bibitem [{\citenamefont {Fei}\ \emph {et~al.}(2021)\citenamefont {Fei},
  \citenamefont {Yeh},\ and\ \citenamefont {Gull}}]{Gull_PRL_2021}%
  \BibitemOpen
  \bibfield  {author} {\bibinfo {author} {\bibfnamefont {Jiani}\ \bibnamefont
  {Fei}}, \bibinfo {author} {\bibfnamefont {Chia-Nan}\ \bibnamefont {Yeh}}, \
  and\ \bibinfo {author} {\bibfnamefont {Emanuel}\ \bibnamefont {Gull}},\
  }\bibfield  {title} {\enquote {\bibinfo {title} {Nevanlinna analytical
  continuation},}\ }\href {\doibase 10.1103/PhysRevLett.126.056402} {\bibfield
  {journal} {\bibinfo  {journal} {Phys. Rev. Lett.}\ }\textbf {\bibinfo
  {volume} {126}},\ \bibinfo {pages} {056402} (\bibinfo {year}
  {2021})}\BibitemShut {NoStop}%
\bibitem [{\citenamefont {Kotliar}\ \emph {et~al.}(2006)\citenamefont
  {Kotliar}, \citenamefont {Savrasov}, \citenamefont {Haule}, \citenamefont
  {Oudovenko}, \citenamefont {Parcollet},\ and\ \citenamefont
  {Marianetti}}]{RevModPhys.78.865}%
  \BibitemOpen
  \bibfield  {author} {\bibinfo {author} {\bibfnamefont {G.}~\bibnamefont
  {Kotliar}}, \bibinfo {author} {\bibfnamefont {S.~Y.}\ \bibnamefont
  {Savrasov}}, \bibinfo {author} {\bibfnamefont {K.}~\bibnamefont {Haule}},
  \bibinfo {author} {\bibfnamefont {V.~S.}\ \bibnamefont {Oudovenko}}, \bibinfo
  {author} {\bibfnamefont {O.}~\bibnamefont {Parcollet}}, \ and\ \bibinfo
  {author} {\bibfnamefont {C.~A.}\ \bibnamefont {Marianetti}},\ }\bibfield
  {title} {\enquote {\bibinfo {title} {Electronic structure calculations with
  dynamical mean-field theory},}\ }\href {\doibase 10.1103/RevModPhys.78.865}
  {\bibfield  {journal} {\bibinfo  {journal} {Rev. Mod. Phys.}\ }\textbf
  {\bibinfo {volume} {78}},\ \bibinfo {pages} {865--951} (\bibinfo {year}
  {2006})}\BibitemShut {NoStop}%
\bibitem [{\citenamefont {Georges}\ \emph {et~al.}(1996)\citenamefont
  {Georges}, \citenamefont {Kotliar}, \citenamefont {Krauth},\ and\
  \citenamefont {Rozenberg}}]{Georges_RMP_1996}%
  \BibitemOpen
  \bibfield  {author} {\bibinfo {author} {\bibfnamefont {Antoine}\ \bibnamefont
  {Georges}}, \bibinfo {author} {\bibfnamefont {Gabriel}\ \bibnamefont
  {Kotliar}}, \bibinfo {author} {\bibfnamefont {Werner}\ \bibnamefont
  {Krauth}}, \ and\ \bibinfo {author} {\bibfnamefont {Marcelo~J.}\ \bibnamefont
  {Rozenberg}},\ }\bibfield  {title} {\enquote {\bibinfo {title} {Dynamical
  mean-field theory of strongly correlated fermion systems and the limit of
  infinite dimensions},}\ }\href {\doibase 10.1103/RevModPhys.68.13} {\bibfield
   {journal} {\bibinfo  {journal} {Rev. Mod. Phys.}\ }\textbf {\bibinfo
  {volume} {68}},\ \bibinfo {pages} {13--125} (\bibinfo {year}
  {1996})}\BibitemShut {NoStop}%
\bibitem [{\citenamefont {Jarrell}\ and\ \citenamefont
  {Gubernatis}(1996)}]{JARRELL1996133}%
  \BibitemOpen
  \bibfield  {author} {\bibinfo {author} {\bibfnamefont {Mark}\ \bibnamefont
  {Jarrell}}\ and\ \bibinfo {author} {\bibfnamefont {J.E.}\ \bibnamefont
  {Gubernatis}},\ }\bibfield  {title} {\enquote {\bibinfo {title} {Bayesian
  inference and the analytic continuation of imaginary-time quantum monte carlo
  data},}\ }\href {\doibase https://doi.org/10.1016/0370-1573(95)00074-7}
  {\bibfield  {journal} {\bibinfo  {journal} {Physics Reports}\ }\textbf
  {\bibinfo {volume} {269}},\ \bibinfo {pages} {133--195} (\bibinfo {year}
  {1996})}\BibitemShut {NoStop}%
\bibitem [{\citenamefont {Goulko}\ \emph {et~al.}(2017)\citenamefont {Goulko},
  \citenamefont {Mishchenko}, \citenamefont {Pollet}, \citenamefont
  {Prokof'ev},\ and\ \citenamefont {Svistunov}}]{Goulko_PRB_2017}%
  \BibitemOpen
  \bibfield  {author} {\bibinfo {author} {\bibfnamefont {Olga}\ \bibnamefont
  {Goulko}}, \bibinfo {author} {\bibfnamefont {Andrey~S.}\ \bibnamefont
  {Mishchenko}}, \bibinfo {author} {\bibfnamefont {Lode}\ \bibnamefont
  {Pollet}}, \bibinfo {author} {\bibfnamefont {Nikolay}\ \bibnamefont
  {Prokof'ev}}, \ and\ \bibinfo {author} {\bibfnamefont {Boris}\ \bibnamefont
  {Svistunov}},\ }\bibfield  {title} {\enquote {\bibinfo {title} {Numerical
  analytic continuation: Answers to well-posed questions},}\ }\href {\doibase
  10.1103/PhysRevB.95.014102} {\bibfield  {journal} {\bibinfo  {journal} {Phys.
  Rev. B}\ }\textbf {\bibinfo {volume} {95}},\ \bibinfo {pages} {014102}
  (\bibinfo {year} {2017})}\BibitemShut {NoStop}%
\bibitem [{\citenamefont {Silver}\ \emph
  {et~al.}(1990{\natexlab{b}})\citenamefont {Silver}, \citenamefont {Sivia},\
  and\ \citenamefont {Gubernatis}}]{PhysRevB.41.2380}%
  \BibitemOpen
  \bibfield  {author} {\bibinfo {author} {\bibfnamefont {R.~N.}\ \bibnamefont
  {Silver}}, \bibinfo {author} {\bibfnamefont {D.~S.}\ \bibnamefont {Sivia}}, \
  and\ \bibinfo {author} {\bibfnamefont {J.~E.}\ \bibnamefont {Gubernatis}},\
  }\bibfield  {title} {\enquote {\bibinfo {title} {Maximum-entropy method for
  analytic continuation of quantum monte carlo data},}\ }\href {\doibase
  10.1103/PhysRevB.41.2380} {\bibfield  {journal} {\bibinfo  {journal} {Phys.
  Rev. B}\ }\textbf {\bibinfo {volume} {41}},\ \bibinfo {pages} {2380--2389}
  (\bibinfo {year} {1990}{\natexlab{b}})}\BibitemShut {NoStop}%
\bibitem [{\citenamefont {Fuchs}\ \emph {et~al.}(2010)\citenamefont {Fuchs},
  \citenamefont {Pruschke},\ and\ \citenamefont {Jarrell}}]{Fuchs_PRE_2010}%
  \BibitemOpen
  \bibfield  {author} {\bibinfo {author} {\bibfnamefont {Sebastian}\
  \bibnamefont {Fuchs}}, \bibinfo {author} {\bibfnamefont {Thomas}\
  \bibnamefont {Pruschke}}, \ and\ \bibinfo {author} {\bibfnamefont {Mark}\
  \bibnamefont {Jarrell}},\ }\bibfield  {title} {\enquote {\bibinfo {title}
  {Analytic continuation of quantum monte carlo data by stochastic analytical
  inference},}\ }\href {\doibase 10.1103/PhysRevE.81.056701} {\bibfield
  {journal} {\bibinfo  {journal} {Phys. Rev. E}\ }\textbf {\bibinfo {volume}
  {81}},\ \bibinfo {pages} {056701} (\bibinfo {year} {2010})}\BibitemShut
  {NoStop}%
\bibitem [{\citenamefont {Sandvik}(2016)}]{Sandvik}%
  \BibitemOpen
  \bibfield  {author} {\bibinfo {author} {\bibfnamefont {Anders~W.}\
  \bibnamefont {Sandvik}},\ }\bibfield  {title} {\enquote {\bibinfo {title}
  {Constrained sampling method for analytic continuation},}\ }\href {\doibase
  10.1103/PhysRevE.94.063308} {\bibfield  {journal} {\bibinfo  {journal} {Phys.
  Rev. E}\ }\textbf {\bibinfo {volume} {94}},\ \bibinfo {pages} {063308}
  (\bibinfo {year} {2016})}\BibitemShut {NoStop}%
\bibitem [{\citenamefont {Bertaina}\ \emph {et~al.}(2017)\citenamefont
  {Bertaina}, \citenamefont {Galli},\ and\ \citenamefont
  {Vitali}}]{Bertaina_GIFT_2017}%
  \BibitemOpen
  \bibfield  {author} {\bibinfo {author} {\bibfnamefont {Gianluca}\
  \bibnamefont {Bertaina}}, \bibinfo {author} {\bibfnamefont {Davide~Emilio}\
  \bibnamefont {Galli}}, \ and\ \bibinfo {author} {\bibfnamefont {Ettore}\
  \bibnamefont {Vitali}},\ }\bibfield  {title} {\enquote {\bibinfo {title}
  {Statistical and computational intelligence approach to analytic continuation
  in quantum monte carlo},}\ }\href {\doibase 10.1080/23746149.2017.1288585}
  {\bibfield  {journal} {\bibinfo  {journal} {Advances in Physics: X}\ }\textbf
  {\bibinfo {volume} {2}},\ \bibinfo {pages} {302--323} (\bibinfo {year}
  {2017})}\BibitemShut {NoStop}%
\bibitem [{\citenamefont {Krivenko}\ and\ \citenamefont
  {Harland}(2019)}]{KRIVENKO2019166}%
  \BibitemOpen
  \bibfield  {author} {\bibinfo {author} {\bibfnamefont {Igor}\ \bibnamefont
  {Krivenko}}\ and\ \bibinfo {author} {\bibfnamefont {Malte}\ \bibnamefont
  {Harland}},\ }\bibfield  {title} {\enquote {\bibinfo {title} {Triqs/som:
  Implementation of the stochastic optimization method for analytic
  continuation},}\ }\href {\doibase https://doi.org/10.1016/j.cpc.2019.01.021}
  {\bibfield  {journal} {\bibinfo  {journal} {Computer Physics Communications}\
  }\textbf {\bibinfo {volume} {239}},\ \bibinfo {pages} {166--183} (\bibinfo
  {year} {2019})}\BibitemShut {NoStop}%
\bibitem [{\citenamefont {Otsuki}\ \emph {et~al.}(2017)\citenamefont {Otsuki},
  \citenamefont {Ohzeki}, \citenamefont {Shinaoka},\ and\ \citenamefont
  {Yoshimi}}]{Otsuki_PRE_2017}%
  \BibitemOpen
  \bibfield  {author} {\bibinfo {author} {\bibfnamefont {Junya}\ \bibnamefont
  {Otsuki}}, \bibinfo {author} {\bibfnamefont {Masayuki}\ \bibnamefont
  {Ohzeki}}, \bibinfo {author} {\bibfnamefont {Hiroshi}\ \bibnamefont
  {Shinaoka}}, \ and\ \bibinfo {author} {\bibfnamefont {Kazuyoshi}\
  \bibnamefont {Yoshimi}},\ }\bibfield  {title} {\enquote {\bibinfo {title}
  {Sparse modeling approach to analytical continuation of imaginary-time
  quantum monte carlo data},}\ }\href {\doibase 10.1103/PhysRevE.95.061302}
  {\bibfield  {journal} {\bibinfo  {journal} {Phys. Rev. E}\ }\textbf {\bibinfo
  {volume} {95}},\ \bibinfo {pages} {061302} (\bibinfo {year}
  {2017})}\BibitemShut {NoStop}%
\bibitem [{\citenamefont {Otsuki}\ \emph {et~al.}(2020)\citenamefont {Otsuki},
  \citenamefont {Ohzeki}, \citenamefont {Shinaoka},\ and\ \citenamefont
  {Yoshimi}}]{Otsuki_JPSJ_2020}%
  \BibitemOpen
  \bibfield  {author} {\bibinfo {author} {\bibfnamefont {Junya}\ \bibnamefont
  {Otsuki}}, \bibinfo {author} {\bibfnamefont {Masayuki}\ \bibnamefont
  {Ohzeki}}, \bibinfo {author} {\bibfnamefont {Hiroshi}\ \bibnamefont
  {Shinaoka}}, \ and\ \bibinfo {author} {\bibfnamefont {Kazuyoshi}\
  \bibnamefont {Yoshimi}},\ }\bibfield  {title} {\enquote {\bibinfo {title}
  {Sparse modeling in quantum many-body problems},}\ }\href {\doibase
  10.7566/JPSJ.89.012001} {\bibfield  {journal} {\bibinfo  {journal} {Journal
  of the Physical Society of Japan}\ }\textbf {\bibinfo {volume} {89}},\
  \bibinfo {pages} {012001} (\bibinfo {year} {2020})}\BibitemShut {NoStop}%
\bibitem [{\citenamefont {Motoyama}\ \emph {et~al.}(2022)\citenamefont
  {Motoyama}, \citenamefont {Yoshimi},\ and\ \citenamefont
  {Otsuki}}]{PhysRevB.105.035139}%
  \BibitemOpen
  \bibfield  {author} {\bibinfo {author} {\bibfnamefont {Yuichi}\ \bibnamefont
  {Motoyama}}, \bibinfo {author} {\bibfnamefont {Kazuyoshi}\ \bibnamefont
  {Yoshimi}}, \ and\ \bibinfo {author} {\bibfnamefont {Junya}\ \bibnamefont
  {Otsuki}},\ }\bibfield  {title} {\enquote {\bibinfo {title} {Robust analytic
  continuation combining the advantages of the sparse modeling approach and the
  pad\'e approximation},}\ }\href {\doibase 10.1103/PhysRevB.105.035139}
  {\bibfield  {journal} {\bibinfo  {journal} {Phys. Rev. B}\ }\textbf {\bibinfo
  {volume} {105}},\ \bibinfo {pages} {035139} (\bibinfo {year}
  {2022})}\BibitemShut {NoStop}%
\bibitem [{\citenamefont {Mihara}\ and\ \citenamefont
  {Puff}(1968)}]{Mihara_Puff_PR_1968}%
  \BibitemOpen
  \bibfield  {author} {\bibinfo {author} {\bibfnamefont {N.}~\bibnamefont
  {Mihara}}\ and\ \bibinfo {author} {\bibfnamefont {R.~D.}\ \bibnamefont
  {Puff}},\ }\bibfield  {title} {\enquote {\bibinfo {title} {{Liquid Structure
  Factor of Ground-State ${\mathrm{He}}^{4}$}},}\ }\href {\doibase
  10.1103/PhysRev.174.221} {\bibfield  {journal} {\bibinfo  {journal} {Phys.
  Rev.}\ }\textbf {\bibinfo {volume} {174}},\ \bibinfo {pages} {221--227}
  (\bibinfo {year} {1968})}\BibitemShut {NoStop}%
\bibitem [{\citenamefont {Tkachenko}\ \emph {et~al.}(2012)\citenamefont
  {Tkachenko}, \citenamefont {Arkhipov},\ and\ \citenamefont
  {Askaruly}}]{tkachenko_book}%
  \BibitemOpen
  \bibfield  {author} {\bibinfo {author} {\bibfnamefont {Igor~M.}\ \bibnamefont
  {Tkachenko}}, \bibinfo {author} {\bibfnamefont {Yuriy~V.}\ \bibnamefont
  {Arkhipov}}, \ and\ \bibinfo {author} {\bibfnamefont {Adil}\ \bibnamefont
  {Askaruly}},\ }\href@noop {} {\emph {\bibinfo {title} {The Method of Moments
  and its Applications in Plasma Physics}}}\ (\bibinfo  {publisher}
  {Akademikerverlag, Saarbr\"ucken, Germany},\ \bibinfo {year}
  {2012})\BibitemShut {NoStop}%
\bibitem [{\citenamefont {Vorberger}\ \emph {et~al.}(2012)\citenamefont
  {Vorberger}, \citenamefont {Donko}, \citenamefont {Tkachenko},\ and\
  \citenamefont {Gericke}}]{Vorberger_PRL_2012}%
  \BibitemOpen
  \bibfield  {author} {\bibinfo {author} {\bibfnamefont {J.}~\bibnamefont
  {Vorberger}}, \bibinfo {author} {\bibfnamefont {Z.}~\bibnamefont {Donko}},
  \bibinfo {author} {\bibfnamefont {I.~M.}\ \bibnamefont {Tkachenko}}, \ and\
  \bibinfo {author} {\bibfnamefont {D.~O.}\ \bibnamefont {Gericke}},\
  }\bibfield  {title} {\enquote {\bibinfo {title} {Dynamic ion structure factor
  of warm dense matter},}\ }\href {\doibase 10.1103/PhysRevLett.109.225001}
  {\bibfield  {journal} {\bibinfo  {journal} {Phys. Rev. Lett.}\ }\textbf
  {\bibinfo {volume} {109}},\ \bibinfo {pages} {225001} (\bibinfo {year}
  {2012})}\BibitemShut {NoStop}%
\bibitem [{\citenamefont {Arkhipov}\ \emph {et~al.}(2018)\citenamefont
  {Arkhipov}, \citenamefont {Ashikbayeva}, \citenamefont {Askaruly},
  \citenamefont {Bonitz}, \citenamefont {Conde}, \citenamefont {Davletov},
  \citenamefont {Dornheim}, \citenamefont {Dubovtsev}, \citenamefont {Groth},
  \citenamefont {Santybayev}, \citenamefont {Syzganbayeva},\ and\ \citenamefont
  {Tkachenko}}]{Arkhipov_CPP_2018}%
  \BibitemOpen
  \bibfield  {author} {\bibinfo {author} {\bibfnamefont {Yu.V.}\ \bibnamefont
  {Arkhipov}}, \bibinfo {author} {\bibfnamefont {A.B.}\ \bibnamefont
  {Ashikbayeva}}, \bibinfo {author} {\bibfnamefont {A.}~\bibnamefont
  {Askaruly}}, \bibinfo {author} {\bibfnamefont {M.}~\bibnamefont {Bonitz}},
  \bibinfo {author} {\bibfnamefont {L.}~\bibnamefont {Conde}}, \bibinfo
  {author} {\bibfnamefont {A.E.}\ \bibnamefont {Davletov}}, \bibinfo {author}
  {\bibfnamefont {T.}~\bibnamefont {Dornheim}}, \bibinfo {author}
  {\bibfnamefont {D.Yu.}\ \bibnamefont {Dubovtsev}}, \bibinfo {author}
  {\bibfnamefont {S.}~\bibnamefont {Groth}}, \bibinfo {author} {\bibfnamefont
  {Kh.}\ \bibnamefont {Santybayev}}, \bibinfo {author} {\bibfnamefont {S.A.}\
  \bibnamefont {Syzganbayeva}}, \ and\ \bibinfo {author} {\bibfnamefont {I.M}\
  \bibnamefont {Tkachenko}},\ }\bibfield  {title} {\enquote {\bibinfo {title}
  {Sum rules and exact inequalities for strongly coupled one-component
  plasmas},}\ }\href {\doibase https://doi.org/10.1002/ctpp.201700136}
  {\bibfield  {journal} {\bibinfo  {journal} {Contributions to Plasma Physics}\
  }\textbf {\bibinfo {volume} {58}},\ \bibinfo {pages} {967--975} (\bibinfo
  {year} {2018})}\BibitemShut {NoStop}%
\bibitem [{\citenamefont {Ara}\ \emph {et~al.}(2022)\citenamefont {Ara},
  \citenamefont {Filinov},\ and\ \citenamefont
  {Tkachenko}}]{Ara_proceeding_2022}%
  \BibitemOpen
  \bibfield  {author} {\bibinfo {author} {\bibfnamefont {J.}~\bibnamefont
  {Ara}}, \bibinfo {author} {\bibfnamefont {A.V.}\ \bibnamefont {Filinov}}, \
  and\ \bibinfo {author} {\bibfnamefont {I.M.}\ \bibnamefont {Tkachenko}},\
  }\bibfield  {title} {\enquote {\bibinfo {title} {Classical and quantum warm
  dense electron gas dynamic characteristics: analytic predictions},}\ }\href
  {\doibase 10.1088/1742-6596/2270/1/012041} {\bibfield  {journal} {\bibinfo
  {journal} {Journal of Physics: Conference Series}\ }\textbf {\bibinfo
  {volume} {2270}},\ \bibinfo {pages} {012041} (\bibinfo {year}
  {2022})}\BibitemShut {NoStop}%
\bibitem [{\citenamefont {Dornheim}\ \emph
  {et~al.}(2018{\natexlab{b}})\citenamefont {Dornheim}, \citenamefont {Groth},\
  and\ \citenamefont {Bonitz}}]{review}%
  \BibitemOpen
  \bibfield  {author} {\bibinfo {author} {\bibfnamefont {T.}~\bibnamefont
  {Dornheim}}, \bibinfo {author} {\bibfnamefont {S.}~\bibnamefont {Groth}}, \
  and\ \bibinfo {author} {\bibfnamefont {M.}~\bibnamefont {Bonitz}},\
  }\bibfield  {title} {\enquote {\bibinfo {title} {The uniform electron gas at
  warm dense matter conditions},}\ }\href
  {https://www.sciencedirect.com/science/article/abs/pii/S0370157318300516}
  {\bibfield  {journal} {\bibinfo  {journal} {Phys. Reports}\ }\textbf
  {\bibinfo {volume} {744}},\ \bibinfo {pages} {1--86} (\bibinfo {year}
  {2018}{\natexlab{b}})}\BibitemShut {NoStop}%
\bibitem [{\citenamefont {Loos}\ and\ \citenamefont {Gill}(2016)}]{loos}%
  \BibitemOpen
  \bibfield  {author} {\bibinfo {author} {\bibfnamefont {P.-F.}\ \bibnamefont
  {Loos}}\ and\ \bibinfo {author} {\bibfnamefont {P.~M.~W.}\ \bibnamefont
  {Gill}},\ }\bibfield  {title} {\enquote {\bibinfo {title} {The uniform
  electron gas},}\ }\href
  {http://onlinelibrary.wiley.com/doi/10.1002/wcms.1257/abstract} {\bibfield
  {journal} {\bibinfo  {journal} {Comput. Mol. Sci}\ }\textbf {\bibinfo
  {volume} {6}},\ \bibinfo {pages} {410--429} (\bibinfo {year}
  {2016})}\BibitemShut {NoStop}%
\bibitem [{\citenamefont {Giuliani}\ and\ \citenamefont
  {Vignale}(2008)}]{quantum_theory}%
  \BibitemOpen
  \bibfield  {author} {\bibinfo {author} {\bibfnamefont {G.}~\bibnamefont
  {Giuliani}}\ and\ \bibinfo {author} {\bibfnamefont {G.}~\bibnamefont
  {Vignale}},\ }\href@noop {} {\emph {\bibinfo {title} {Quantum Theory of the
  Electron Liquid}}}\ (\bibinfo  {publisher} {Cambridge University Press},\
  \bibinfo {address} {Cambridge},\ \bibinfo {year} {2008})\BibitemShut
  {NoStop}%
\bibitem [{\citenamefont {Ott}\ \emph {et~al.}(2018)\citenamefont {Ott},
  \citenamefont {Thomsen}, \citenamefont {Abraham}, \citenamefont {Dornheim},\
  and\ \citenamefont {Bonitz}}]{Ott2018}%
  \BibitemOpen
  \bibfield  {author} {\bibinfo {author} {\bibfnamefont {Torben}\ \bibnamefont
  {Ott}}, \bibinfo {author} {\bibfnamefont {Hauke}\ \bibnamefont {Thomsen}},
  \bibinfo {author} {\bibfnamefont {Jan~Willem}\ \bibnamefont {Abraham}},
  \bibinfo {author} {\bibfnamefont {Tobias}\ \bibnamefont {Dornheim}}, \ and\
  \bibinfo {author} {\bibfnamefont {Michael}\ \bibnamefont {Bonitz}},\
  }\bibfield  {title} {\enquote {\bibinfo {title} {Recent progress in the
  theory and simulation of strongly correlated plasmas: phase transitions,
  transport, quantum, and magnetic field effects},}\ }\href {\doibase
  10.1140/epjd/e2018-80385-7} {\bibfield  {journal} {\bibinfo  {journal} {The
  European Physical Journal D}\ }\textbf {\bibinfo {volume} {72}},\ \bibinfo
  {pages} {84} (\bibinfo {year} {2018})}\BibitemShut {NoStop}%
\bibitem [{\citenamefont {Brown}\ \emph
  {et~al.}(2013{\natexlab{a}})\citenamefont {Brown}, \citenamefont {DuBois},
  \citenamefont {Holzmann},\ and\ \citenamefont {Ceperley}}]{Brown_PRB_2013}%
  \BibitemOpen
  \bibfield  {author} {\bibinfo {author} {\bibfnamefont {Ethan~W.}\
  \bibnamefont {Brown}}, \bibinfo {author} {\bibfnamefont {Jonathan~L.}\
  \bibnamefont {DuBois}}, \bibinfo {author} {\bibfnamefont {Markus}\
  \bibnamefont {Holzmann}}, \ and\ \bibinfo {author} {\bibfnamefont {David~M.}\
  \bibnamefont {Ceperley}},\ }\bibfield  {title} {\enquote {\bibinfo {title}
  {Exchange-correlation energy for the three-dimensional homogeneous electron
  gas at arbitrary temperature},}\ }\href {\doibase 10.1103/PhysRevB.88.081102}
  {\bibfield  {journal} {\bibinfo  {journal} {Phys. Rev. B}\ }\textbf {\bibinfo
  {volume} {88}},\ \bibinfo {pages} {081102} (\bibinfo {year}
  {2013}{\natexlab{a}})}\BibitemShut {NoStop}%
\bibitem [{\citenamefont {Brown}\ \emph
  {et~al.}(2013{\natexlab{b}})\citenamefont {Brown}, \citenamefont {Clark},
  \citenamefont {DuBois},\ and\ \citenamefont {Ceperley}}]{Brown_PRL_2013}%
  \BibitemOpen
  \bibfield  {author} {\bibinfo {author} {\bibfnamefont {Ethan~W.}\
  \bibnamefont {Brown}}, \bibinfo {author} {\bibfnamefont {Bryan~K.}\
  \bibnamefont {Clark}}, \bibinfo {author} {\bibfnamefont {Jonathan~L.}\
  \bibnamefont {DuBois}}, \ and\ \bibinfo {author} {\bibfnamefont {David~M.}\
  \bibnamefont {Ceperley}},\ }\bibfield  {title} {\enquote {\bibinfo {title}
  {Path-integral monte carlo simulation of the warm dense homogeneous electron
  gas},}\ }\href {\doibase 10.1103/PhysRevLett.110.146405} {\bibfield
  {journal} {\bibinfo  {journal} {Phys. Rev. Lett.}\ }\textbf {\bibinfo
  {volume} {110}},\ \bibinfo {pages} {146405} (\bibinfo {year}
  {2013}{\natexlab{b}})}\BibitemShut {NoStop}%
\bibitem [{\citenamefont {Dutta}\ and\ \citenamefont
  {Dufty}(2013)}]{Dutta_2013}%
  \BibitemOpen
  \bibfield  {author} {\bibinfo {author} {\bibfnamefont {Sandipan}\
  \bibnamefont {Dutta}}\ and\ \bibinfo {author} {\bibfnamefont {James}\
  \bibnamefont {Dufty}},\ }\bibfield  {title} {\enquote {\bibinfo {title}
  {Uniform electron gas at warm, dense matter conditions},}\ }\href {\doibase
  10.1209/0295-5075/102/67005} {\bibfield  {journal} {\bibinfo  {journal}
  {{EPL} (Europhysics Letters)}\ }\textbf {\bibinfo {volume} {102}},\ \bibinfo
  {pages} {67005} (\bibinfo {year} {2013})}\BibitemShut {NoStop}%
\bibitem [{\citenamefont {Sjostrom}\ and\ \citenamefont {Dufty}(2013)}]{stls2}%
  \BibitemOpen
  \bibfield  {author} {\bibinfo {author} {\bibfnamefont {T.}~\bibnamefont
  {Sjostrom}}\ and\ \bibinfo {author} {\bibfnamefont {J.}~\bibnamefont
  {Dufty}},\ }\bibfield  {title} {\enquote {\bibinfo {title} {Uniform electron
  gas at finite temperatures},}\ }\href
  {http://link.aps.org/doi/10.1103/PhysRevB.88.115123} {\bibfield  {journal}
  {\bibinfo  {journal} {Phys. Rev. B}\ }\textbf {\bibinfo {volume} {88}},\
  \bibinfo {pages} {115123} (\bibinfo {year} {2013})}\BibitemShut {NoStop}%
\bibitem [{\citenamefont {Dornheim}\ \emph {et~al.}(2016)\citenamefont
  {Dornheim}, \citenamefont {Groth}, \citenamefont {Sjostrom}, \citenamefont
  {Malone}, \citenamefont {Foulkes},\ and\ \citenamefont
  {Bonitz}}]{dornheim_prl}%
  \BibitemOpen
  \bibfield  {author} {\bibinfo {author} {\bibfnamefont {T.}~\bibnamefont
  {Dornheim}}, \bibinfo {author} {\bibfnamefont {S.}~\bibnamefont {Groth}},
  \bibinfo {author} {\bibfnamefont {T.}~\bibnamefont {Sjostrom}}, \bibinfo
  {author} {\bibfnamefont {F.~D.}\ \bibnamefont {Malone}}, \bibinfo {author}
  {\bibfnamefont {W.~M.~C.}\ \bibnamefont {Foulkes}}, \ and\ \bibinfo {author}
  {\bibfnamefont {M.}~\bibnamefont {Bonitz}},\ }\bibfield  {title} {\enquote
  {\bibinfo {title} {Ab initio quantum {M}onte {C}arlo simulation of the warm
  dense electron gas in the thermodynamic limit},}\ }\href
  {http://link.aps.org/doi/10.1103/PhysRevLett.117.156403} {\bibfield
  {journal} {\bibinfo  {journal} {Phys. Rev. Lett.}\ }\textbf {\bibinfo
  {volume} {117}},\ \bibinfo {pages} {156403} (\bibinfo {year}
  {2016})}\BibitemShut {NoStop}%
\bibitem [{\citenamefont {Groth}\ \emph {et~al.}(2017)\citenamefont {Groth},
  \citenamefont {Dornheim}, \citenamefont {Sjostrom}, \citenamefont {Malone},
  \citenamefont {Foulkes},\ and\ \citenamefont {Bonitz}}]{groth_prl}%
  \BibitemOpen
  \bibfield  {author} {\bibinfo {author} {\bibfnamefont {S.}~\bibnamefont
  {Groth}}, \bibinfo {author} {\bibfnamefont {T.}~\bibnamefont {Dornheim}},
  \bibinfo {author} {\bibfnamefont {T.}~\bibnamefont {Sjostrom}}, \bibinfo
  {author} {\bibfnamefont {F.~D.}\ \bibnamefont {Malone}}, \bibinfo {author}
  {\bibfnamefont {W.~M.~C.}\ \bibnamefont {Foulkes}}, \ and\ \bibinfo {author}
  {\bibfnamefont {M.}~\bibnamefont {Bonitz}},\ }\bibfield  {title} {\enquote
  {\bibinfo {title} {Ab initio exchange--correlation free energy of the uniform
  electron gas at warm dense matter conditions},}\ }\href
  {https://journals.aps.org/prl/abstract/10.1103/PhysRevLett.119.135001}
  {\bibfield  {journal} {\bibinfo  {journal} {Phys. Rev. Lett.}\ }\textbf
  {\bibinfo {volume} {119}},\ \bibinfo {pages} {135001} (\bibinfo {year}
  {2017})}\BibitemShut {NoStop}%
\bibitem [{\citenamefont {Malone}\ \emph {et~al.}(2016)\citenamefont {Malone},
  \citenamefont {Blunt}, \citenamefont {Brown}, \citenamefont {Lee},
  \citenamefont {Spencer}, \citenamefont {Foulkes},\ and\ \citenamefont
  {Shepherd}}]{Malone_PRL_2016}%
  \BibitemOpen
  \bibfield  {author} {\bibinfo {author} {\bibfnamefont {Fionn~D.}\
  \bibnamefont {Malone}}, \bibinfo {author} {\bibfnamefont {N.~S.}\
  \bibnamefont {Blunt}}, \bibinfo {author} {\bibfnamefont {Ethan~W.}\
  \bibnamefont {Brown}}, \bibinfo {author} {\bibfnamefont {D.~K.~K.}\
  \bibnamefont {Lee}}, \bibinfo {author} {\bibfnamefont {J.~S.}\ \bibnamefont
  {Spencer}}, \bibinfo {author} {\bibfnamefont {W.~M.~C.}\ \bibnamefont
  {Foulkes}}, \ and\ \bibinfo {author} {\bibfnamefont {James~J.}\ \bibnamefont
  {Shepherd}},\ }\bibfield  {title} {\enquote {\bibinfo {title} {Accurate
  exchange-correlation energies for the warm dense electron gas},}\ }\href
  {\doibase 10.1103/PhysRevLett.117.115701} {\bibfield  {journal} {\bibinfo
  {journal} {Phys. Rev. Lett.}\ }\textbf {\bibinfo {volume} {117}},\ \bibinfo
  {pages} {115701} (\bibinfo {year} {2016})}\BibitemShut {NoStop}%
\bibitem [{\citenamefont {Karasiev}\ \emph {et~al.}(2014)\citenamefont
  {Karasiev}, \citenamefont {Sjostrom}, \citenamefont {Dufty},\ and\
  \citenamefont {Trickey}}]{ksdt}%
  \BibitemOpen
  \bibfield  {author} {\bibinfo {author} {\bibfnamefont {Valentin~V.}\
  \bibnamefont {Karasiev}}, \bibinfo {author} {\bibfnamefont {Travis}\
  \bibnamefont {Sjostrom}}, \bibinfo {author} {\bibfnamefont {James}\
  \bibnamefont {Dufty}}, \ and\ \bibinfo {author} {\bibfnamefont {S.~B.}\
  \bibnamefont {Trickey}},\ }\bibfield  {title} {\enquote {\bibinfo {title}
  {Accurate homogeneous electron gas exchange-correlation free energy for local
  spin-density calculations},}\ }\href {\doibase
  10.1103/PhysRevLett.112.076403} {\bibfield  {journal} {\bibinfo  {journal}
  {Phys. Rev. Lett.}\ }\textbf {\bibinfo {volume} {112}},\ \bibinfo {pages}
  {076403} (\bibinfo {year} {2014})}\BibitemShut {NoStop}%
\bibitem [{\citenamefont {Karasiev}\ \emph
  {et~al.}(2019{\natexlab{a}})\citenamefont {Karasiev}, \citenamefont
  {Trickey},\ and\ \citenamefont {Dufty}}]{Karasiev_status_2019}%
  \BibitemOpen
  \bibfield  {author} {\bibinfo {author} {\bibfnamefont {Valentin~V.}\
  \bibnamefont {Karasiev}}, \bibinfo {author} {\bibfnamefont {S.~B.}\
  \bibnamefont {Trickey}}, \ and\ \bibinfo {author} {\bibfnamefont {James~W.}\
  \bibnamefont {Dufty}},\ }\bibfield  {title} {\enquote {\bibinfo {title}
  {Status of free-energy representations for the homogeneous electron gas},}\
  }\href {\doibase 10.1103/PhysRevB.99.195134} {\bibfield  {journal} {\bibinfo
  {journal} {Phys. Rev. B}\ }\textbf {\bibinfo {volume} {99}},\ \bibinfo
  {pages} {195134} (\bibinfo {year} {2019}{\natexlab{a}})}\BibitemShut
  {NoStop}%
\bibitem [{\citenamefont {Dornheim}\ \emph
  {et~al.}(2020{\natexlab{a}})\citenamefont {Dornheim}, \citenamefont
  {Vorberger},\ and\ \citenamefont {Bonitz}}]{Dornheim_PRL_2020}%
  \BibitemOpen
  \bibfield  {author} {\bibinfo {author} {\bibfnamefont {Tobias}\ \bibnamefont
  {Dornheim}}, \bibinfo {author} {\bibfnamefont {Jan}\ \bibnamefont
  {Vorberger}}, \ and\ \bibinfo {author} {\bibfnamefont {Michael}\ \bibnamefont
  {Bonitz}},\ }\bibfield  {title} {\enquote {\bibinfo {title} {Nonlinear
  electronic density response in warm dense matter},}\ }\href {\doibase
  10.1103/PhysRevLett.125.085001} {\bibfield  {journal} {\bibinfo  {journal}
  {Phys. Rev. Lett.}\ }\textbf {\bibinfo {volume} {125}},\ \bibinfo {pages}
  {085001} (\bibinfo {year} {2020}{\natexlab{a}})}\BibitemShut {NoStop}%
\bibitem [{\citenamefont {Dornheim}\ \emph
  {et~al.}(2020{\natexlab{b}})\citenamefont {Dornheim}, \citenamefont
  {Sjostrom}, \citenamefont {Tanaka},\ and\ \citenamefont
  {Vorberger}}]{dornheim_electron_liquid}%
  \BibitemOpen
  \bibfield  {author} {\bibinfo {author} {\bibfnamefont {Tobias}\ \bibnamefont
  {Dornheim}}, \bibinfo {author} {\bibfnamefont {Travis}\ \bibnamefont
  {Sjostrom}}, \bibinfo {author} {\bibfnamefont {Shigenori}\ \bibnamefont
  {Tanaka}}, \ and\ \bibinfo {author} {\bibfnamefont {Jan}\ \bibnamefont
  {Vorberger}},\ }\bibfield  {title} {\enquote {\bibinfo {title} {Strongly
  coupled electron liquid: Ab initio path integral monte carlo simulations and
  dielectric theories},}\ }\href {\doibase 10.1103/PhysRevB.101.045129}
  {\bibfield  {journal} {\bibinfo  {journal} {Phys. Rev. B}\ }\textbf {\bibinfo
  {volume} {101}},\ \bibinfo {pages} {045129} (\bibinfo {year}
  {2020}{\natexlab{b}})}\BibitemShut {NoStop}%
\bibitem [{\citenamefont {Dornheim}\ \emph
  {et~al.}(2020{\natexlab{c}})\citenamefont {Dornheim}, \citenamefont
  {Moldabekov}, \citenamefont {Vorberger},\ and\ \citenamefont
  {Groth}}]{dornheim_HEDP}%
  \BibitemOpen
  \bibfield  {author} {\bibinfo {author} {\bibfnamefont {Tobias}\ \bibnamefont
  {Dornheim}}, \bibinfo {author} {\bibfnamefont {Zhandos~A}\ \bibnamefont
  {Moldabekov}}, \bibinfo {author} {\bibfnamefont {Jan}\ \bibnamefont
  {Vorberger}}, \ and\ \bibinfo {author} {\bibfnamefont {Simon}\ \bibnamefont
  {Groth}},\ }\bibfield  {title} {\enquote {\bibinfo {title} {Ab initio path
  integral monte carlo simulation of the uniform electron gas in the high
  energy density regime},}\ }\href {\doibase 10.1088/1361-6587/ab8bb4}
  {\bibfield  {journal} {\bibinfo  {journal} {Plasma Physics and Controlled
  Fusion}\ }\textbf {\bibinfo {volume} {62}},\ \bibinfo {pages} {075003}
  (\bibinfo {year} {2020}{\natexlab{c}})}\BibitemShut {NoStop}%
\bibitem [{\citenamefont {Dornheim}\ \emph
  {et~al.}(2022{\natexlab{c}})\citenamefont {Dornheim}, \citenamefont
  {Vorberger}, \citenamefont {Moldabekov}, \citenamefont {Röpke},\ and\
  \citenamefont {Kraeft}}]{Dornheim_HEDP_2022}%
  \BibitemOpen
  \bibfield  {author} {\bibinfo {author} {\bibfnamefont {Tobias}\ \bibnamefont
  {Dornheim}}, \bibinfo {author} {\bibfnamefont {Jan}\ \bibnamefont
  {Vorberger}}, \bibinfo {author} {\bibfnamefont {Zhandos}\ \bibnamefont
  {Moldabekov}}, \bibinfo {author} {\bibfnamefont {Gerd}\ \bibnamefont
  {Röpke}}, \ and\ \bibinfo {author} {\bibfnamefont {Wolf-Dietrich}\
  \bibnamefont {Kraeft}},\ }\bibfield  {title} {\enquote {\bibinfo {title} {The
  uniform electron gas at high temperatures: ab initio path integral monte
  carlo simulations and analytical theory},}\ }\href {\doibase
  https://doi.org/10.1016/j.hedp.2022.101015} {\bibfield  {journal} {\bibinfo
  {journal} {High Energy Density Physics}\ }\textbf {\bibinfo {volume} {45}},\
  \bibinfo {pages} {101015} (\bibinfo {year} {2022}{\natexlab{c}})}\BibitemShut
  {NoStop}%
\bibitem [{\citenamefont {Castello}\ \emph {et~al.}(2022)\citenamefont
  {Castello}, \citenamefont {Tolias},\ and\ \citenamefont
  {Dornheim}}]{castello2021classical}%
  \BibitemOpen
  \bibfield  {author} {\bibinfo {author} {\bibfnamefont {F.~Lucco}\
  \bibnamefont {Castello}}, \bibinfo {author} {\bibfnamefont {P.}~\bibnamefont
  {Tolias}}, \ and\ \bibinfo {author} {\bibfnamefont {T.}~\bibnamefont
  {Dornheim}},\ }\bibfield  {title} {\enquote {\bibinfo {title} {Classical
  bridge functions in classical and quantum plasma liquids},}\ }\href {\doibase
  10.1209/0295-5075/ac7166} {\bibfield  {journal} {\bibinfo  {journal}
  {Europhysics Letters}\ }\textbf {\bibinfo {volume} {138}},\ \bibinfo {pages}
  {44003} (\bibinfo {year} {2022})}\BibitemShut {NoStop}%
\bibitem [{\citenamefont {Tolias}\ \emph {et~al.}(2021)\citenamefont {Tolias},
  \citenamefont {Lucco~Castello},\ and\ \citenamefont
  {Dornheim}}]{Tolias_JCP_2021}%
  \BibitemOpen
  \bibfield  {author} {\bibinfo {author} {\bibfnamefont {P.}~\bibnamefont
  {Tolias}}, \bibinfo {author} {\bibfnamefont {F.}~\bibnamefont
  {Lucco~Castello}}, \ and\ \bibinfo {author} {\bibfnamefont {T.}~\bibnamefont
  {Dornheim}},\ }\bibfield  {title} {\enquote {\bibinfo {title} {Integral
  equation theory based dielectric scheme for strongly coupled electron
  liquids},}\ }\href {\doibase 10.1063/5.0065988} {\bibfield  {journal}
  {\bibinfo  {journal} {The Journal of Chemical Physics}\ }\textbf {\bibinfo
  {volume} {155}},\ \bibinfo {pages} {134115} (\bibinfo {year}
  {2021})}\BibitemShut {NoStop}%
\bibitem [{\citenamefont {Dornheim}\ \emph
  {et~al.}(2022{\natexlab{d}})\citenamefont {Dornheim}, \citenamefont
  {Moldabekov}, \citenamefont {Vorberger}, \citenamefont {K{\"a}hlert},\ and\
  \citenamefont {Bonitz}}]{Dornheim_Nature_2022}%
  \BibitemOpen
  \bibfield  {author} {\bibinfo {author} {\bibfnamefont {Tobias}\ \bibnamefont
  {Dornheim}}, \bibinfo {author} {\bibfnamefont {Zhandos}\ \bibnamefont
  {Moldabekov}}, \bibinfo {author} {\bibfnamefont {Jan}\ \bibnamefont
  {Vorberger}}, \bibinfo {author} {\bibfnamefont {Hanno}\ \bibnamefont
  {K{\"a}hlert}}, \ and\ \bibinfo {author} {\bibfnamefont {Michael}\
  \bibnamefont {Bonitz}},\ }\bibfield  {title} {\enquote {\bibinfo {title}
  {Electronic pair alignment and roton feature in the warm dense electron
  gas},}\ }\href {\doibase 10.1038/s42005-022-01078-9} {\bibfield  {journal}
  {\bibinfo  {journal} {Communications Physics}\ }\textbf {\bibinfo {volume}
  {5}},\ \bibinfo {pages} {304} (\bibinfo {year}
  {2022}{\natexlab{d}})}\BibitemShut {NoStop}%
\bibitem [{\citenamefont {Graziani}\ \emph {et~al.}(2014)\citenamefont
  {Graziani}, \citenamefont {Desjarlais}, \citenamefont {Redmer},\ and\
  \citenamefont {Trickey}}]{wdm_book}%
  \BibitemOpen
  \bibinfo {editor} {\bibfnamefont {F.}~\bibnamefont {Graziani}}, \bibinfo
  {editor} {\bibfnamefont {M.~P.}\ \bibnamefont {Desjarlais}}, \bibinfo
  {editor} {\bibfnamefont {R.}~\bibnamefont {Redmer}}, \ and\ \bibinfo {editor}
  {\bibfnamefont {S.~B.}\ \bibnamefont {Trickey}},\ eds.,\ \href@noop {} {\emph
  {\bibinfo {title} {Frontiers and Challenges in Warm Dense Matter}}}\
  (\bibinfo  {publisher} {Springer},\ \bibinfo {address} {International
  Publishing},\ \bibinfo {year} {2014})\BibitemShut {NoStop}%
\bibitem [{\citenamefont {Bonitz}\ \emph {et~al.}(2020)\citenamefont {Bonitz},
  \citenamefont {Dornheim}, \citenamefont {Moldabekov}, \citenamefont {Zhang},
  \citenamefont {Hamann}, \citenamefont {Kählert}, \citenamefont {Filinov},
  \citenamefont {Ramakrishna},\ and\ \citenamefont {Vorberger}}]{new_POP}%
  \BibitemOpen
  \bibfield  {author} {\bibinfo {author} {\bibfnamefont {M.}~\bibnamefont
  {Bonitz}}, \bibinfo {author} {\bibfnamefont {T.}~\bibnamefont {Dornheim}},
  \bibinfo {author} {\bibfnamefont {Zh.~A.}\ \bibnamefont {Moldabekov}},
  \bibinfo {author} {\bibfnamefont {S.}~\bibnamefont {Zhang}}, \bibinfo
  {author} {\bibfnamefont {P.}~\bibnamefont {Hamann}}, \bibinfo {author}
  {\bibfnamefont {H.}~\bibnamefont {Kählert}}, \bibinfo {author}
  {\bibfnamefont {A.}~\bibnamefont {Filinov}}, \bibinfo {author} {\bibfnamefont
  {K.}~\bibnamefont {Ramakrishna}}, \ and\ \bibinfo {author} {\bibfnamefont
  {J.}~\bibnamefont {Vorberger}},\ }\bibfield  {title} {\enquote {\bibinfo
  {title} {Ab initio simulation of warm dense matter},}\ }\href {\doibase
  10.1063/1.5143225} {\bibfield  {journal} {\bibinfo  {journal} {Physics of
  Plasmas}\ }\textbf {\bibinfo {volume} {27}},\ \bibinfo {pages} {042710}
  (\bibinfo {year} {2020})}\BibitemShut {NoStop}%
\bibitem [{\citenamefont {Fortov}(2009)}]{fortov_review}%
  \BibitemOpen
  \bibfield  {author} {\bibinfo {author} {\bibfnamefont {V.~E.}\ \bibnamefont
  {Fortov}},\ }\bibfield  {title} {\enquote {\bibinfo {title} {Extreme states
  of matter on earth and in space},}\ }\href
  {https://www.turpion.org/php/paper.phtml?journal_id=pu&paper_id=6821}
  {\bibfield  {journal} {\bibinfo  {journal} {Phys.-Usp}\ }\textbf {\bibinfo
  {volume} {52}},\ \bibinfo {pages} {615--647} (\bibinfo {year}
  {2009})}\BibitemShut {NoStop}%
\bibitem [{\citenamefont {Drake}(2018)}]{drake2018high}%
  \BibitemOpen
  \bibfield  {author} {\bibinfo {author} {\bibfnamefont {R.P.}\ \bibnamefont
  {Drake}},\ }\href {https://books.google.de/books?id=1AoZtAEACAAJ} {\emph
  {\bibinfo {title} {High-Energy-Density Physics: Foundation of Inertial Fusion
  and Experimental Astrophysics}}},\ Graduate Texts in Physics\ (\bibinfo
  {publisher} {Springer International Publishing},\ \bibinfo {year}
  {2018})\BibitemShut {NoStop}%
\bibitem [{\citenamefont {Militzer}\ \emph {et~al.}(2008)\citenamefont
  {Militzer}, \citenamefont {Hubbard}, \citenamefont {Vorberger}, \citenamefont
  {Tamblyn},\ and\ \citenamefont {Bonev}}]{Militzer_2008}%
  \BibitemOpen
  \bibfield  {author} {\bibinfo {author} {\bibfnamefont {B.}~\bibnamefont
  {Militzer}}, \bibinfo {author} {\bibfnamefont {W.~B.}\ \bibnamefont
  {Hubbard}}, \bibinfo {author} {\bibfnamefont {J.}~\bibnamefont {Vorberger}},
  \bibinfo {author} {\bibfnamefont {I.}~\bibnamefont {Tamblyn}}, \ and\
  \bibinfo {author} {\bibfnamefont {S.~A.}\ \bibnamefont {Bonev}},\ }\bibfield
  {title} {\enquote {\bibinfo {title} {A massive core in jupiter predicted from
  first-principles simulations},}\ }\href {\doibase 10.1086/594364} {\bibfield
  {journal} {\bibinfo  {journal} {The Astrophysical Journal}\ }\textbf
  {\bibinfo {volume} {688}},\ \bibinfo {pages} {L45--L48} (\bibinfo {year}
  {2008})}\BibitemShut {NoStop}%
\bibitem [{\citenamefont {Benuzzi-Mounaix}\ \emph {et~al.}(2014)\citenamefont
  {Benuzzi-Mounaix}, \citenamefont {Mazevet}, \citenamefont {Ravasio},
  \citenamefont {Vinci}, \citenamefont {Denoeud}, \citenamefont {Koenig},
  \citenamefont {Amadou}, \citenamefont {Brambrink}, \citenamefont {Festa},
  \citenamefont {Levy}, \citenamefont {Harmand}, \citenamefont {Brygoo},
  \citenamefont {Huser}, \citenamefont {Recoules}, \citenamefont {Bouchet},
  \citenamefont {Morard}, \citenamefont {Guyot}, \citenamefont {de~Resseguier},
  \citenamefont {Myanishi}, \citenamefont {Ozaki}, \citenamefont {Dorchies},
  \citenamefont {Gaudin}, \citenamefont {Leguay}, \citenamefont {Peyrusse},
  \citenamefont {Henry}, \citenamefont {Raffestin}, \citenamefont {Pape},
  \citenamefont {Smith},\ and\ \citenamefont {Musella}}]{Benuzzi_Mounaix_2014}%
  \BibitemOpen
  \bibfield  {author} {\bibinfo {author} {\bibfnamefont {Alessandra}\
  \bibnamefont {Benuzzi-Mounaix}}, \bibinfo {author} {\bibfnamefont
  {St{\'{e}}phane}\ \bibnamefont {Mazevet}}, \bibinfo {author} {\bibfnamefont
  {Alessandra}\ \bibnamefont {Ravasio}}, \bibinfo {author} {\bibfnamefont
  {Tommaso}\ \bibnamefont {Vinci}}, \bibinfo {author} {\bibfnamefont {Adrien}\
  \bibnamefont {Denoeud}}, \bibinfo {author} {\bibfnamefont {Michel}\
  \bibnamefont {Koenig}}, \bibinfo {author} {\bibfnamefont {Nourou}\
  \bibnamefont {Amadou}}, \bibinfo {author} {\bibfnamefont {Erik}\ \bibnamefont
  {Brambrink}}, \bibinfo {author} {\bibfnamefont {Floriane}\ \bibnamefont
  {Festa}}, \bibinfo {author} {\bibfnamefont {Anna}\ \bibnamefont {Levy}},
  \bibinfo {author} {\bibfnamefont {Marion}\ \bibnamefont {Harmand}}, \bibinfo
  {author} {\bibfnamefont {St{\'{e}}phanie}\ \bibnamefont {Brygoo}}, \bibinfo
  {author} {\bibfnamefont {Gael}\ \bibnamefont {Huser}}, \bibinfo {author}
  {\bibfnamefont {Vanina}\ \bibnamefont {Recoules}}, \bibinfo {author}
  {\bibfnamefont {Johan}\ \bibnamefont {Bouchet}}, \bibinfo {author}
  {\bibfnamefont {Guillaume}\ \bibnamefont {Morard}}, \bibinfo {author}
  {\bibfnamefont {Fran{\c{c}}ois}\ \bibnamefont {Guyot}}, \bibinfo {author}
  {\bibfnamefont {Thibaut}\ \bibnamefont {de~Resseguier}}, \bibinfo {author}
  {\bibfnamefont {Kohei}\ \bibnamefont {Myanishi}}, \bibinfo {author}
  {\bibfnamefont {Norimasa}\ \bibnamefont {Ozaki}}, \bibinfo {author}
  {\bibfnamefont {Fabien}\ \bibnamefont {Dorchies}}, \bibinfo {author}
  {\bibfnamefont {Jer{\^{o}}me}\ \bibnamefont {Gaudin}}, \bibinfo {author}
  {\bibfnamefont {Pierre~Marie}\ \bibnamefont {Leguay}}, \bibinfo {author}
  {\bibfnamefont {Olivier}\ \bibnamefont {Peyrusse}}, \bibinfo {author}
  {\bibfnamefont {Olivier}\ \bibnamefont {Henry}}, \bibinfo {author}
  {\bibfnamefont {Didier}\ \bibnamefont {Raffestin}}, \bibinfo {author}
  {\bibfnamefont {Sebastien~Le}\ \bibnamefont {Pape}}, \bibinfo {author}
  {\bibfnamefont {Ray}\ \bibnamefont {Smith}}, \ and\ \bibinfo {author}
  {\bibfnamefont {Riccardo}\ \bibnamefont {Musella}},\ }\bibfield  {title}
  {\enquote {\bibinfo {title} {Progress in warm dense matter study with
  applications to planetology},}\ }\href {\doibase
  10.1088/0031-8949/2014/t161/014060} {\bibfield  {journal} {\bibinfo
  {journal} {Phys. Scripta}\ }\textbf {\bibinfo {volume} {T161}},\ \bibinfo
  {pages} {014060} (\bibinfo {year} {2014})}\BibitemShut {NoStop}%
\bibitem [{\citenamefont {Hu}\ \emph {et~al.}(2011)\citenamefont {Hu},
  \citenamefont {Militzer}, \citenamefont {Goncharov},\ and\ \citenamefont
  {Skupsky}}]{hu_ICF}%
  \BibitemOpen
  \bibfield  {author} {\bibinfo {author} {\bibfnamefont {S.~X.}\ \bibnamefont
  {Hu}}, \bibinfo {author} {\bibfnamefont {B.}~\bibnamefont {Militzer}},
  \bibinfo {author} {\bibfnamefont {V.~N.}\ \bibnamefont {Goncharov}}, \ and\
  \bibinfo {author} {\bibfnamefont {S.}~\bibnamefont {Skupsky}},\ }\bibfield
  {title} {\enquote {\bibinfo {title} {First-principles equation-of-state table
  of deuterium for inertial confinement fusion applications},}\ }\href
  {https://journals.aps.org/prb/abstract/10.1103/PhysRevB.84.224109} {\bibfield
   {journal} {\bibinfo  {journal} {Phys. Rev. B}\ }\textbf {\bibinfo {volume}
  {84}},\ \bibinfo {pages} {224109} (\bibinfo {year} {2011})}\BibitemShut
  {NoStop}%
\bibitem [{\citenamefont {Betti}\ and\ \citenamefont
  {Hurricane}(2016)}]{Betti2016}%
  \BibitemOpen
  \bibfield  {author} {\bibinfo {author} {\bibfnamefont {R.}~\bibnamefont
  {Betti}}\ and\ \bibinfo {author} {\bibfnamefont {O.~A.}\ \bibnamefont
  {Hurricane}},\ }\bibfield  {title} {\enquote {\bibinfo {title}
  {Inertial-confinement fusion with lasers},}\ }\href {\doibase
  10.1038/nphys3736} {\bibfield  {journal} {\bibinfo  {journal} {Nature
  Physics}\ }\textbf {\bibinfo {volume} {12}},\ \bibinfo {pages} {435--448}
  (\bibinfo {year} {2016})}\BibitemShut {NoStop}%
\bibitem [{\citenamefont {Dornheim}\ \emph
  {et~al.}(2022{\natexlab{e}})\citenamefont {Dornheim}, \citenamefont
  {Vorberger}, \citenamefont {Moldabekov},\ and\ \citenamefont
  {Böhme}}]{Dornheim_PTR_2022}%
  \BibitemOpen
  \bibfield  {author} {\bibinfo {author} {\bibfnamefont {Tobias}\ \bibnamefont
  {Dornheim}}, \bibinfo {author} {\bibfnamefont {Jan}\ \bibnamefont
  {Vorberger}}, \bibinfo {author} {\bibfnamefont {Zhandos}\ \bibnamefont
  {Moldabekov}}, \ and\ \bibinfo {author} {\bibfnamefont {Maximilian}\
  \bibnamefont {Böhme}},\ }\bibfield  {title} {\enquote {\bibinfo {title}
  {Analyzing x-ray thomson scattering experiments of warm dense matter in the
  imaginary-time domain: theoretical models and simulations},}\ }\href
  {\doibase 10.48550/ARXIV.2211.00579} {\  (\bibinfo {year}
  {2022}{\natexlab{e}}),\ 10.48550/ARXIV.2211.00579}\BibitemShut {NoStop}%
\bibitem [{\citenamefont {Dornheim}\ \emph
  {et~al.}(2022{\natexlab{f}})\citenamefont {Dornheim}, \citenamefont
  {B{\"o}hme}, \citenamefont {Kraus}, \citenamefont {D{\"o}ppner},
  \citenamefont {Preston}, \citenamefont {Moldabekov},\ and\ \citenamefont
  {Vorberger}}]{Dornheim_T_2022}%
  \BibitemOpen
  \bibfield  {author} {\bibinfo {author} {\bibfnamefont {Tobias}\ \bibnamefont
  {Dornheim}}, \bibinfo {author} {\bibfnamefont {Maximilian}\ \bibnamefont
  {B{\"o}hme}}, \bibinfo {author} {\bibfnamefont {Dominik}\ \bibnamefont
  {Kraus}}, \bibinfo {author} {\bibfnamefont {Tilo}\ \bibnamefont
  {D{\"o}ppner}}, \bibinfo {author} {\bibfnamefont {Thomas~R.}\ \bibnamefont
  {Preston}}, \bibinfo {author} {\bibfnamefont {Zhandos~A.}\ \bibnamefont
  {Moldabekov}}, \ and\ \bibinfo {author} {\bibfnamefont {Jan}\ \bibnamefont
  {Vorberger}},\ }\bibfield  {title} {\enquote {\bibinfo {title} {Accurate
  temperature diagnostics for matter under extreme conditions},}\ }\href
  {\doibase 10.1038/s41467-022-35578-7} {\bibfield  {journal} {\bibinfo
  {journal} {Nature Communications}\ }\textbf {\bibinfo {volume} {13}},\
  \bibinfo {pages} {7911} (\bibinfo {year} {2022}{\natexlab{f}})}\BibitemShut
  {NoStop}%
\bibitem [{\citenamefont {Dornheim}\ \emph
  {et~al.}(2022{\natexlab{g}})\citenamefont {Dornheim}, \citenamefont
  {B{\"o}hme}, \citenamefont {Chapman}, \citenamefont {Kraus}, \citenamefont
  {D{\"o}ppner}, \citenamefont {Preston}, \citenamefont {Moldabekov},\ and\
  \citenamefont {Vorberger}}]{Dornheim_T2_2022}%
  \BibitemOpen
  \bibfield  {author} {\bibinfo {author} {\bibfnamefont {Tobias}\ \bibnamefont
  {Dornheim}}, \bibinfo {author} {\bibfnamefont {Maximilian}\ \bibnamefont
  {B{\"o}hme}}, \bibinfo {author} {\bibfnamefont {Dave}\ \bibnamefont
  {Chapman}}, \bibinfo {author} {\bibfnamefont {Dominik}\ \bibnamefont
  {Kraus}}, \bibinfo {author} {\bibfnamefont {Tilo}\ \bibnamefont
  {D{\"o}ppner}}, \bibinfo {author} {\bibfnamefont {Thomas~R.}\ \bibnamefont
  {Preston}}, \bibinfo {author} {\bibfnamefont {Zhandos~A.}\ \bibnamefont
  {Moldabekov}}, \ and\ \bibinfo {author} {\bibfnamefont {Jan}\ \bibnamefont
  {Vorberger}},\ }\bibfield  {title} {\enquote {\bibinfo {title} {Temperature
  analysis of x-ray thomson scattering data},}\ }\href {\doibase
  10.48550/ARXIV.2212.10510} {\  (\bibinfo {year} {2022}{\natexlab{g}}),\
  10.48550/ARXIV.2212.10510}\BibitemShut {NoStop}%
\bibitem [{\citenamefont {Boninsegni}\ \emph
  {et~al.}(2006{\natexlab{a}})\citenamefont {Boninsegni}, \citenamefont
  {Prokofev},\ and\ \citenamefont {Svistunov}}]{boninsegni1}%
  \BibitemOpen
  \bibfield  {author} {\bibinfo {author} {\bibfnamefont {M.}~\bibnamefont
  {Boninsegni}}, \bibinfo {author} {\bibfnamefont {N.~V.}\ \bibnamefont
  {Prokofev}}, \ and\ \bibinfo {author} {\bibfnamefont {B.~V.}\ \bibnamefont
  {Svistunov}},\ }\bibfield  {title} {\enquote {\bibinfo {title} {Worm
  algorithm and diagrammatic {M}onte {C}arlo: A new approach to
  continuous-space path integral {M}onte {C}arlo simulations},}\ }\href
  {https://journals.aps.org/pre/abstract/10.1103/PhysRevE.74.036701} {\bibfield
   {journal} {\bibinfo  {journal} {Phys. Rev. E}\ }\textbf {\bibinfo {volume}
  {74}},\ \bibinfo {pages} {036701} (\bibinfo {year}
  {2006}{\natexlab{a}})}\BibitemShut {NoStop}%
\bibitem [{\citenamefont {Chandler}\ and\ \citenamefont
  {Wolynes}(1981)}]{Chandler_JCP_1981}%
  \BibitemOpen
  \bibfield  {author} {\bibinfo {author} {\bibfnamefont {David}\ \bibnamefont
  {Chandler}}\ and\ \bibinfo {author} {\bibfnamefont {Peter~G.}\ \bibnamefont
  {Wolynes}},\ }\bibfield  {title} {\enquote {\bibinfo {title} {Exploiting the
  isomorphism between quantum theory and classical statistical mechanics of
  polyatomic fluids},}\ }\href {\doibase 10.1063/1.441588} {\bibfield
  {journal} {\bibinfo  {journal} {The Journal of Chemical Physics}\ }\textbf
  {\bibinfo {volume} {74}},\ \bibinfo {pages} {4078--4095} (\bibinfo {year}
  {1981})}\BibitemShut {NoStop}%
\bibitem [{\citenamefont {Dornheim}(2019)}]{dornheim_sign_problem}%
  \BibitemOpen
  \bibfield  {author} {\bibinfo {author} {\bibfnamefont {T.}~\bibnamefont
  {Dornheim}},\ }\bibfield  {title} {\enquote {\bibinfo {title} {Fermion sign
  problem in path integral {M}onte {C}arlo simulations: Quantum dots, ultracold
  atoms, and warm dense matter},}\ }\href
  {https://journals.aps.org/pre/abstract/10.1103/PhysRevE.100.023307}
  {\bibfield  {journal} {\bibinfo  {journal} {Phys. Rev. E}\ }\textbf {\bibinfo
  {volume} {100}},\ \bibinfo {pages} {023307} (\bibinfo {year}
  {2019})}\BibitemShut {NoStop}%
\bibitem [{\citenamefont {Metropolis}\ \emph {et~al.}(1953)\citenamefont
  {Metropolis}, \citenamefont {Rosenbluth}, \citenamefont {Rosenbluth},
  \citenamefont {Teller},\ and\ \citenamefont {Teller}}]{metropolis}%
  \BibitemOpen
  \bibfield  {author} {\bibinfo {author} {\bibfnamefont {Nicholas}\
  \bibnamefont {Metropolis}}, \bibinfo {author} {\bibfnamefont {Arianna~W.}\
  \bibnamefont {Rosenbluth}}, \bibinfo {author} {\bibfnamefont {Marshall~N.}\
  \bibnamefont {Rosenbluth}}, \bibinfo {author} {\bibfnamefont {Augusta~H.}\
  \bibnamefont {Teller}}, \ and\ \bibinfo {author} {\bibfnamefont {Edward}\
  \bibnamefont {Teller}},\ }\bibfield  {title} {\enquote {\bibinfo {title}
  {Equation of state calculations by fast computing machines},}\ }\href
  {\doibase 10.1063/1.1699114} {\bibfield  {journal} {\bibinfo  {journal} {The
  Journal of Chemical Physics}\ }\textbf {\bibinfo {volume} {21}},\ \bibinfo
  {pages} {1087--1092} (\bibinfo {year} {1953})}\BibitemShut {NoStop}%
\bibitem [{\citenamefont {Dornheim}\ \emph
  {et~al.}(2019{\natexlab{a}})\citenamefont {Dornheim}, \citenamefont {Groth},
  \citenamefont {Filinov},\ and\ \citenamefont
  {Bonitz}}]{Dornheim_permutation_cycles}%
  \BibitemOpen
  \bibfield  {author} {\bibinfo {author} {\bibfnamefont {T.}~\bibnamefont
  {Dornheim}}, \bibinfo {author} {\bibfnamefont {S.}~\bibnamefont {Groth}},
  \bibinfo {author} {\bibfnamefont {A.~V.}\ \bibnamefont {Filinov}}, \ and\
  \bibinfo {author} {\bibfnamefont {M.}~\bibnamefont {Bonitz}},\ }\bibfield
  {title} {\enquote {\bibinfo {title} {Path integral monte carlo simulation of
  degenerate electrons: Permutation-cycle properties},}\ }\href {\doibase
  10.1063/1.5093171} {\bibfield  {journal} {\bibinfo  {journal} {The Journal of
  Chemical Physics}\ }\textbf {\bibinfo {volume} {151}},\ \bibinfo {pages}
  {014108} (\bibinfo {year} {2019}{\natexlab{a}})}\BibitemShut {NoStop}%
\bibitem [{\citenamefont {Boninsegni}\ \emph
  {et~al.}(2006{\natexlab{b}})\citenamefont {Boninsegni}, \citenamefont
  {Prokofev},\ and\ \citenamefont {Svistunov}}]{boninsegni2}%
  \BibitemOpen
  \bibfield  {author} {\bibinfo {author} {\bibfnamefont {M.}~\bibnamefont
  {Boninsegni}}, \bibinfo {author} {\bibfnamefont {N.~V.}\ \bibnamefont
  {Prokofev}}, \ and\ \bibinfo {author} {\bibfnamefont {B.~V.}\ \bibnamefont
  {Svistunov}},\ }\bibfield  {title} {\enquote {\bibinfo {title} {Worm
  algorithm for continuous-space path integral {M}onte {C}arlo simulations},}\
  }\href {https://journals.aps.org/prl/abstract/10.1103/PhysRevLett.96.070601}
  {\bibfield  {journal} {\bibinfo  {journal} {Phys. Rev. Lett}\ }\textbf
  {\bibinfo {volume} {96}},\ \bibinfo {pages} {070601} (\bibinfo {year}
  {2006}{\natexlab{b}})}\BibitemShut {NoStop}%
\bibitem [{\citenamefont {Dornheim}(2021)}]{Dornheim_2021}%
  \BibitemOpen
  \bibfield  {author} {\bibinfo {author} {\bibfnamefont {Tobias}\ \bibnamefont
  {Dornheim}},\ }\bibfield  {title} {\enquote {\bibinfo {title} {Fermion sign
  problem in path integral monte carlo simulations: grand-canonical
  ensemble},}\ }\href {\doibase 10.1088/1751-8121/ac1481} {\bibfield  {journal}
  {\bibinfo  {journal} {Journal of Physics A: Mathematical and Theoretical}\
  }\textbf {\bibinfo {volume} {54}},\ \bibinfo {pages} {335001} (\bibinfo
  {year} {2021})}\BibitemShut {NoStop}%
\bibitem [{\citenamefont {Schoof}\ \emph {et~al.}(2015)\citenamefont {Schoof},
  \citenamefont {Groth}, \citenamefont {Vorberger},\ and\ \citenamefont
  {Bonitz}}]{Schoof_PRL_2015}%
  \BibitemOpen
  \bibfield  {author} {\bibinfo {author} {\bibfnamefont {T.}~\bibnamefont
  {Schoof}}, \bibinfo {author} {\bibfnamefont {S.}~\bibnamefont {Groth}},
  \bibinfo {author} {\bibfnamefont {J.}~\bibnamefont {Vorberger}}, \ and\
  \bibinfo {author} {\bibfnamefont {M.}~\bibnamefont {Bonitz}},\ }\bibfield
  {title} {\enquote {\bibinfo {title} {Ab initio thermodynamic results for the
  degenerate electron gas at finite temperature},}\ }\href {\doibase
  10.1103/PhysRevLett.115.130402} {\bibfield  {journal} {\bibinfo  {journal}
  {Phys. Rev. Lett.}\ }\textbf {\bibinfo {volume} {115}},\ \bibinfo {pages}
  {130402} (\bibinfo {year} {2015})}\BibitemShut {NoStop}%
\bibitem [{\citenamefont {Sheffield}\ \emph {et~al.}(2010)\citenamefont
  {Sheffield}, \citenamefont {Froula}, \citenamefont {Glenzer},\ and\
  \citenamefont {Luhmann}}]{sheffield2010plasma}%
  \BibitemOpen
  \bibfield  {author} {\bibinfo {author} {\bibfnamefont {J.}~\bibnamefont
  {Sheffield}}, \bibinfo {author} {\bibfnamefont {D.}~\bibnamefont {Froula}},
  \bibinfo {author} {\bibfnamefont {S.H.}\ \bibnamefont {Glenzer}}, \ and\
  \bibinfo {author} {\bibfnamefont {N.C.}\ \bibnamefont {Luhmann}},\ }\href
  {https://books.google.de/books?id=1NS5Fxam1lkC} {\emph {\bibinfo {title}
  {Plasma Scattering of Electromagnetic Radiation: Theory and Measurement
  Techniques}}}\ (\bibinfo  {publisher} {Elsevier Science},\ \bibinfo {year}
  {2010})\BibitemShut {NoStop}%
\bibitem [{\citenamefont {Baus}\ and\ \citenamefont
  {Hansen}(1980)}]{Baus_Hansen_OCP}%
  \BibitemOpen
  \bibfield  {author} {\bibinfo {author} {\bibfnamefont {Marc}\ \bibnamefont
  {Baus}}\ and\ \bibinfo {author} {\bibfnamefont {Jean-Pierre}\ \bibnamefont
  {Hansen}},\ }\bibfield  {title} {\enquote {\bibinfo {title} {Statistical
  mechanics of simple coulomb systems},}\ }\href
  {https://www.sciencedirect.com/science/article/pii/0370157380900228}
  {\bibfield  {journal} {\bibinfo  {journal} {Phys. Rep.}\ }\textbf {\bibinfo
  {volume} {59}},\ \bibinfo {pages} {1--94} (\bibinfo {year}
  {1980})}\BibitemShut {NoStop}%
\bibitem [{\citenamefont {Hansen}\ and\ \citenamefont
  {McDonald}(2013)}]{hansen2013theory}%
  \BibitemOpen
  \bibfield  {author} {\bibinfo {author} {\bibfnamefont {J.P.}\ \bibnamefont
  {Hansen}}\ and\ \bibinfo {author} {\bibfnamefont {I.R.}\ \bibnamefont
  {McDonald}},\ }\href {https://books.google.de/books?id=agbEswEACAAJ} {\emph
  {\bibinfo {title} {Theory of simple liquids : with applications to soft
  matter}}}\ (\bibinfo  {publisher} {Academic Press},\ \bibinfo {year}
  {2013})\BibitemShut {NoStop}%
\bibitem [{\citenamefont {Kugler}(1970)}]{kugler_bounds}%
  \BibitemOpen
  \bibfield  {author} {\bibinfo {author} {\bibfnamefont {A.~A.}\ \bibnamefont
  {Kugler}},\ }\bibfield  {title} {\enquote {\bibinfo {title} {Bounds for some
  equilibrium properties of an electron gas},}\ }\href
  {https://journals.aps.org/pra/abstract/10.1103/PhysRevA.1.1688} {\bibfield
  {journal} {\bibinfo  {journal} {Phys. Rev. A}\ }\textbf {\bibinfo {volume}
  {1}},\ \bibinfo {pages} {1688} (\bibinfo {year} {1970})}\BibitemShut
  {NoStop}%
\bibitem [{\citenamefont {Mahan}(1990)}]{mahan1990many}%
  \BibitemOpen
  \bibfield  {author} {\bibinfo {author} {\bibfnamefont {G.D.}\ \bibnamefont
  {Mahan}},\ }\href {https://books.google.de/books?id=v8du6cp0vUAC} {\emph
  {\bibinfo {title} {Many-Particle Physics}}},\ Physics of Solids and Liquids\
  (\bibinfo  {publisher} {Springer US},\ \bibinfo {year} {1990})\BibitemShut
  {NoStop}%
\bibitem [{\citenamefont {Bowen}\ \emph {et~al.}(1994)\citenamefont {Bowen},
  \citenamefont {Sugiyama},\ and\ \citenamefont {Alder}}]{bowen2}%
  \BibitemOpen
  \bibfield  {author} {\bibinfo {author} {\bibfnamefont {C.}~\bibnamefont
  {Bowen}}, \bibinfo {author} {\bibfnamefont {G.}~\bibnamefont {Sugiyama}}, \
  and\ \bibinfo {author} {\bibfnamefont {B.~J.}\ \bibnamefont {Alder}},\
  }\bibfield  {title} {\enquote {\bibinfo {title} {Static dielectric response
  of the electron gas},}\ }\href
  {http://link.aps.org/doi/10.1103/PhysRevB.50.14838} {\bibfield  {journal}
  {\bibinfo  {journal} {Phys. Rev. B}\ }\textbf {\bibinfo {volume} {50}},\
  \bibinfo {pages} {14838} (\bibinfo {year} {1994})}\BibitemShut {NoStop}%
\bibitem [{\citenamefont {Ichimaru}(2018)}]{ichimaru_bookII}%
  \BibitemOpen
  \bibfield  {author} {\bibinfo {author} {\bibfnamefont {S.}~\bibnamefont
  {Ichimaru}},\ }\href
  {https://books.google.se/books?id=Sr_vAAAAMAAJ&q=Statistical+plasma+physics}
  {\emph {\bibinfo {title} {Statistical plasma physics {Vol. II}: {C}ondensed
  plasmas}}}\ (\bibinfo  {publisher} {CRC Press, Boca Raton, USA},\ \bibinfo
  {year} {2018})\BibitemShut {NoStop}%
\bibitem [{\citenamefont {Singwi}\ \emph {et~al.}(1968)\citenamefont {Singwi},
  \citenamefont {Tosi}, \citenamefont {Land},\ and\ \citenamefont
  {Sj\"olander}}]{stls_original}%
  \BibitemOpen
  \bibfield  {author} {\bibinfo {author} {\bibfnamefont {K.~S.}\ \bibnamefont
  {Singwi}}, \bibinfo {author} {\bibfnamefont {M.~P.}\ \bibnamefont {Tosi}},
  \bibinfo {author} {\bibfnamefont {R.~H.}\ \bibnamefont {Land}}, \ and\
  \bibinfo {author} {\bibfnamefont {A.}~\bibnamefont {Sj\"olander}},\
  }\bibfield  {title} {\enquote {\bibinfo {title} {Electron correlations at
  metallic densities},}\ }\href
  {http://link.aps.org/doi/10.1103/PhysRev.176.589} {\bibfield  {journal}
  {\bibinfo  {journal} {Phys. Rev}\ }\textbf {\bibinfo {volume} {176}},\
  \bibinfo {pages} {589} (\bibinfo {year} {1968})}\BibitemShut {NoStop}%
\bibitem [{\citenamefont {Singwi}\ and\ \citenamefont
  {Tosi}(1981)}]{SingwiTosi_Review}%
  \BibitemOpen
  \bibfield  {author} {\bibinfo {author} {\bibfnamefont {K.~S.}\ \bibnamefont
  {Singwi}}\ and\ \bibinfo {author} {\bibfnamefont {M.~P.}\ \bibnamefont
  {Tosi}},\ }\bibfield  {title} {\enquote {\bibinfo {title} {Correlations in
  electron liquids},}\ }\href {\doibase 10.1016/S0081-1947(08)60116-2}
  {\bibfield  {journal} {\bibinfo  {journal} {Solid State Physics}\ }\textbf
  {\bibinfo {volume} {36}},\ \bibinfo {pages} {177--266} (\bibinfo {year}
  {1981})}\BibitemShut {NoStop}%
\bibitem [{\citenamefont {Ichimaru}\ \emph {et~al.}(1987)\citenamefont
  {Ichimaru}, \citenamefont {Iyetomi},\ and\ \citenamefont {Tanaka}}]{IIT}%
  \BibitemOpen
  \bibfield  {author} {\bibinfo {author} {\bibfnamefont {Setsuo}\ \bibnamefont
  {Ichimaru}}, \bibinfo {author} {\bibfnamefont {Hiroshi}\ \bibnamefont
  {Iyetomi}}, \ and\ \bibinfo {author} {\bibfnamefont {Shigenori}\ \bibnamefont
  {Tanaka}},\ }\bibfield  {title} {\enquote {\bibinfo {title} {Statistical
  physics of dense plasmas: Thermodynamics, transport coefficients and dynamic
  correlations},}\ }\href {\doibase
  https://doi.org/10.1016/0370-1573(87)90125-6} {\bibfield  {journal} {\bibinfo
   {journal} {Physics Reports}\ }\textbf {\bibinfo {volume} {149}},\ \bibinfo
  {pages} {91--205} (\bibinfo {year} {1987})}\BibitemShut {NoStop}%
\bibitem [{\citenamefont {Kugler}(1975)}]{kugler1}%
  \BibitemOpen
  \bibfield  {author} {\bibinfo {author} {\bibfnamefont {A.~A.}\ \bibnamefont
  {Kugler}},\ }\bibfield  {title} {\enquote {\bibinfo {title} {Theory of the
  local field correction in an electron gas},}\ }\href
  {http://link.springer.com/article/10.1007/BF01024183} {\bibfield  {journal}
  {\bibinfo  {journal} {J. Stat. Phys}\ }\textbf {\bibinfo {volume} {12}},\
  \bibinfo {pages} {35} (\bibinfo {year} {1975})}\BibitemShut {NoStop}%
\bibitem [{\citenamefont {Pines}\ and\ \citenamefont
  {Nozieres}(2018)}]{pines_nozieres_bookI}%
  \BibitemOpen
  \bibfield  {author} {\bibinfo {author} {\bibfnamefont {David}\ \bibnamefont
  {Pines}}\ and\ \bibinfo {author} {\bibfnamefont {Philippe}\ \bibnamefont
  {Nozieres}},\ }\href
  {https://books.google.se/books?id=lF0PEAAAQBAJ&printsec=frontcover&dq=The+theory+of+quantum+liquids}
  {\emph {\bibinfo {title} {The theory of quantum liquids {Vol. I}: {N}ormal
  {F}ermi liquids}}}\ (\bibinfo  {publisher} {CRC Press, Boca Raton, USA},\
  \bibinfo {year} {2018})\BibitemShut {NoStop}%
\bibitem [{\citenamefont {Placzek}(1952)}]{PlaczekSumRule}%
  \BibitemOpen
  \bibfield  {author} {\bibinfo {author} {\bibfnamefont {G.}~\bibnamefont
  {Placzek}},\ }\bibfield  {title} {\enquote {\bibinfo {title} {The scattering
  of neutrons by systems of heavy nuclei},}\ }\href {\doibase
  10.1103/PhysRev.86.377} {\bibfield  {journal} {\bibinfo  {journal} {Phys.
  Rev.}\ }\textbf {\bibinfo {volume} {86}},\ \bibinfo {pages} {377--388}
  (\bibinfo {year} {1952})}\BibitemShut {NoStop}%
\bibitem [{\citenamefont {Puff}(1965)}]{PuffSumRule}%
  \BibitemOpen
  \bibfield  {author} {\bibinfo {author} {\bibfnamefont {R.~D.}\ \bibnamefont
  {Puff}},\ }\bibfield  {title} {\enquote {\bibinfo {title} {Application of sum
  rules to the low-temperature interacting boson system},}\ }\href {\doibase
  10.1103/PhysRev.137.A406} {\bibfield  {journal} {\bibinfo  {journal} {Phys.
  Rev.}\ }\textbf {\bibinfo {volume} {137}},\ \bibinfo {pages} {A406--A416}
  (\bibinfo {year} {1965})}\BibitemShut {NoStop}%
\bibitem [{\citenamefont {Ichimaru}(1982)}]{plasma2}%
  \BibitemOpen
  \bibfield  {author} {\bibinfo {author} {\bibfnamefont {S.}~\bibnamefont
  {Ichimaru}},\ }\bibfield  {title} {\enquote {\bibinfo {title} {Strongly
  coupled plasmas: high-density classical plasmas and degenerate electron
  liquids},}\ }\href
  {https://journals.aps.org/rmp/abstract/10.1103/RevModPhys.54.1017} {\bibfield
   {journal} {\bibinfo  {journal} {Rev. Mod. Phys}\ }\textbf {\bibinfo {volume}
  {54}},\ \bibinfo {pages} {1017} (\bibinfo {year} {1982})}\BibitemShut
  {NoStop}%
\bibitem [{\citenamefont {Iwamoto}\ \emph {et~al.}(1984)\citenamefont
  {Iwamoto}, \citenamefont {Krotscheck},\ and\ \citenamefont
  {Pines}}]{Iwamoto_PRB_1984}%
  \BibitemOpen
  \bibfield  {author} {\bibinfo {author} {\bibfnamefont {Naoki}\ \bibnamefont
  {Iwamoto}}, \bibinfo {author} {\bibfnamefont {Eckhard}\ \bibnamefont
  {Krotscheck}}, \ and\ \bibinfo {author} {\bibfnamefont {David}\ \bibnamefont
  {Pines}},\ }\bibfield  {title} {\enquote {\bibinfo {title} {Theory of
  electron liquids. ii. static and dynamic form factors, correlation energy,
  and plasmon dispersion},}\ }\href {\doibase 10.1103/PhysRevB.29.3936}
  {\bibfield  {journal} {\bibinfo  {journal} {Phys. Rev. B}\ }\textbf {\bibinfo
  {volume} {29}},\ \bibinfo {pages} {3936--3951} (\bibinfo {year}
  {1984})}\BibitemShut {NoStop}%
\bibitem [{\citenamefont {Khodel}\ \emph {et~al.}(1994)\citenamefont {Khodel},
  \citenamefont {Shaginyan},\ and\ \citenamefont {Khodel}}]{KhodelReview}%
  \BibitemOpen
  \bibfield  {author} {\bibinfo {author} {\bibfnamefont {V.A.}\ \bibnamefont
  {Khodel}}, \bibinfo {author} {\bibfnamefont {V.R.}\ \bibnamefont
  {Shaginyan}}, \ and\ \bibinfo {author} {\bibfnamefont {V.V.}\ \bibnamefont
  {Khodel}},\ }\bibfield  {title} {\enquote {\bibinfo {title} {New approach in
  the microscopic fermi systems theory},}\ }\href {\doibase
  https://doi.org/10.1016/0370-1573(94)00059-X} {\bibfield  {journal} {\bibinfo
   {journal} {Physics Reports}\ }\textbf {\bibinfo {volume} {249}},\ \bibinfo
  {pages} {1--134} (\bibinfo {year} {1994})}\BibitemShut {NoStop}%
\bibitem [{\citenamefont {Niklasson}(1974)}]{NiklassonLimit}%
  \BibitemOpen
  \bibfield  {author} {\bibinfo {author} {\bibfnamefont {G\"oran}\ \bibnamefont
  {Niklasson}},\ }\bibfield  {title} {\enquote {\bibinfo {title} {Dielectric
  function of the uniform electron gas for large frequencies or wave
  vectors},}\ }\href {\doibase 10.1103/PhysRevB.10.3052} {\bibfield  {journal}
  {\bibinfo  {journal} {Phys. Rev. B}\ }\textbf {\bibinfo {volume} {10}},\
  \bibinfo {pages} {3052--3061} (\bibinfo {year} {1974})}\BibitemShut {NoStop}%
\bibitem [{\citenamefont {Pathak}\ and\ \citenamefont
  {Vashishta}(1973)}]{PathakVashishtaScheme}%
  \BibitemOpen
  \bibfield  {author} {\bibinfo {author} {\bibfnamefont {K.~N.}\ \bibnamefont
  {Pathak}}\ and\ \bibinfo {author} {\bibfnamefont {P.}~\bibnamefont
  {Vashishta}},\ }\bibfield  {title} {\enquote {\bibinfo {title} {Electron
  correlations and moment sum rules},}\ }\href {\doibase
  10.1103/PhysRevB.7.3649} {\bibfield  {journal} {\bibinfo  {journal} {Phys.
  Rev. B}\ }\textbf {\bibinfo {volume} {7}},\ \bibinfo {pages} {3649--3656}
  (\bibinfo {year} {1973})}\BibitemShut {NoStop}%
\bibitem [{\citenamefont {Utsumi}\ and\ \citenamefont
  {Ichimaru}(1980)}]{UtsumiIchimaruI}%
  \BibitemOpen
  \bibfield  {author} {\bibinfo {author} {\bibfnamefont {Kenichi}\ \bibnamefont
  {Utsumi}}\ and\ \bibinfo {author} {\bibfnamefont {Setsuo}\ \bibnamefont
  {Ichimaru}},\ }\bibfield  {title} {\enquote {\bibinfo {title} {Dielectric
  formulation of strongly coupled electron liquid at metallic densities:
  Longitudinal response},}\ }\href {\doibase 10.1103/PhysRevB.22.1522}
  {\bibfield  {journal} {\bibinfo  {journal} {Phys. Rev. B}\ }\textbf {\bibinfo
  {volume} {22}},\ \bibinfo {pages} {1522--1533} (\bibinfo {year}
  {1980})}\BibitemShut {NoStop}%
\bibitem [{\citenamefont {Forster}\ \emph {et~al.}(1968)\citenamefont
  {Forster}, \citenamefont {Martin},\ and\ \citenamefont
  {Yip}}]{SixthMomentClassical}%
  \BibitemOpen
  \bibfield  {author} {\bibinfo {author} {\bibfnamefont {Dieter}\ \bibnamefont
  {Forster}}, \bibinfo {author} {\bibfnamefont {Paul~C.}\ \bibnamefont
  {Martin}}, \ and\ \bibinfo {author} {\bibfnamefont {Sidney}\ \bibnamefont
  {Yip}},\ }\bibfield  {title} {\enquote {\bibinfo {title} {Moments of the
  momentum density correlation functions in simple liquids},}\ }\href {\doibase
  10.1103/PhysRev.170.155} {\bibfield  {journal} {\bibinfo  {journal} {Phys.
  Rev.}\ }\textbf {\bibinfo {volume} {170}},\ \bibinfo {pages} {155--159}
  (\bibinfo {year} {1968})}\BibitemShut {NoStop}%
\bibitem [{\citenamefont {Bansal}\ and\ \citenamefont
  {Pathak}(1974)}]{EighthMomentClassical}%
  \BibitemOpen
  \bibfield  {author} {\bibinfo {author} {\bibfnamefont {Ravinder}\
  \bibnamefont {Bansal}}\ and\ \bibinfo {author} {\bibfnamefont {K.~N.}\
  \bibnamefont {Pathak}},\ }\bibfield  {title} {\enquote {\bibinfo {title} {Sum
  rules and atomic correlations in classical liquids},}\ }\href {\doibase
  10.1103/PhysRevA.9.2773} {\bibfield  {journal} {\bibinfo  {journal} {Phys.
  Rev. A}\ }\textbf {\bibinfo {volume} {9}},\ \bibinfo {pages} {2773--2782}
  (\bibinfo {year} {1974})}\BibitemShut {NoStop}%
\bibitem [{\citenamefont {Ichimaru}\ \emph {et~al.}(1975)\citenamefont
  {Ichimaru}, \citenamefont {Totsuji}, \citenamefont {Tange},\ and\
  \citenamefont {Pines}}]{IchimaruClassicalSumrule}%
  \BibitemOpen
  \bibfield  {author} {\bibinfo {author} {\bibfnamefont {Setsuo}\ \bibnamefont
  {Ichimaru}}, \bibinfo {author} {\bibfnamefont {Hiroo}\ \bibnamefont
  {Totsuji}}, \bibinfo {author} {\bibfnamefont {Toshio}\ \bibnamefont {Tange}},
  \ and\ \bibinfo {author} {\bibfnamefont {David}\ \bibnamefont {Pines}},\
  }\bibfield  {title} {\enquote {\bibinfo {title} {{Sum-Rule Analysis of
  Long-Wavelength Excitations in Electron Liquids: }},}\ }\href {\doibase
  10.1143/PTP.54.1077} {\bibfield  {journal} {\bibinfo  {journal} {Progress of
  Theoretical Physics}\ }\textbf {\bibinfo {volume} {54}},\ \bibinfo {pages}
  {1077--1092} (\bibinfo {year} {1975})}\BibitemShut {NoStop}%
\bibitem [{\citenamefont {Ailawadi}(1980)}]{AilawadiReview}%
  \BibitemOpen
  \bibfield  {author} {\bibinfo {author} {\bibfnamefont {N.K.}\ \bibnamefont
  {Ailawadi}},\ }\bibfield  {title} {\enquote {\bibinfo {title} {Equilibrium
  theories of simple liquids},}\ }\href {\doibase
  https://doi.org/10.1016/0370-1573(80)90063-0} {\bibfield  {journal} {\bibinfo
   {journal} {Physics Reports}\ }\textbf {\bibinfo {volume} {57}},\ \bibinfo
  {pages} {241--306} (\bibinfo {year} {1980})}\BibitemShut {NoStop}%
\bibitem [{\citenamefont {McQuarrie}(1976)}]{mcquarrie76a}%
  \BibitemOpen
  \bibfield  {author} {\bibinfo {author} {\bibfnamefont {Donald~A.}\
  \bibnamefont {McQuarrie}},\ }\href
  {https://books.google.se/books?id=itcpPnDnJM0C&dq=mcquarrie+statistical+mechanics}
  {\emph {\bibinfo {title} {Statistical Mechanics}}},\ Harper's chemistry
  series\ (\bibinfo  {publisher} {Harper Collins},\ \bibinfo {address} {New
  York},\ \bibinfo {year} {1976})\BibitemShut {NoStop}%
\bibitem [{\citenamefont {Kugler}(1973)}]{kugler_classical}%
  \BibitemOpen
  \bibfield  {author} {\bibinfo {author} {\bibfnamefont {A.~A.}\ \bibnamefont
  {Kugler}},\ }\bibfield  {title} {\enquote {\bibinfo {title} {Collective
  modes, damping, and the scattering function in classical liquids},}\ }\href
  {https://doi.org/10.1007/BF01008535} {\bibfield  {journal} {\bibinfo
  {journal} {J. Stat. Phys.}\ }\textbf {\bibinfo {volume} {8}},\ \bibinfo
  {pages} {107--153} (\bibinfo {year} {1973})}\BibitemShut {NoStop}%
\bibitem [{\citenamefont {Dornheim}\ \emph
  {et~al.}(2022{\natexlab{h}})\citenamefont {Dornheim}, \citenamefont
  {Moldabekov}, \citenamefont {Ramakrishna}, \citenamefont {Tolias},
  \citenamefont {Baczewski}, \citenamefont {Kraus}, \citenamefont {Preston},
  \citenamefont {Chapman}, \citenamefont {Böhme}, \citenamefont {Döppner},
  \citenamefont {Graziani}, \citenamefont {Bonitz}, \citenamefont {Cangi},\
  and\ \citenamefont {Vorberger}}]{Dornheim_review}%
  \BibitemOpen
  \bibfield  {author} {\bibinfo {author} {\bibfnamefont {Tobias}\ \bibnamefont
  {Dornheim}}, \bibinfo {author} {\bibfnamefont {Zhandos~A.}\ \bibnamefont
  {Moldabekov}}, \bibinfo {author} {\bibfnamefont {Kushal}\ \bibnamefont
  {Ramakrishna}}, \bibinfo {author} {\bibfnamefont {Panagiotis}\ \bibnamefont
  {Tolias}}, \bibinfo {author} {\bibfnamefont {Andrew~D.}\ \bibnamefont
  {Baczewski}}, \bibinfo {author} {\bibfnamefont {Dominik}\ \bibnamefont
  {Kraus}}, \bibinfo {author} {\bibfnamefont {Thomas~R.}\ \bibnamefont
  {Preston}}, \bibinfo {author} {\bibfnamefont {David~A.}\ \bibnamefont
  {Chapman}}, \bibinfo {author} {\bibfnamefont {Maximilian~P.}\ \bibnamefont
  {Böhme}}, \bibinfo {author} {\bibfnamefont {Tilo}\ \bibnamefont {Döppner}},
  \bibinfo {author} {\bibfnamefont {Frank}\ \bibnamefont {Graziani}}, \bibinfo
  {author} {\bibfnamefont {Michael}\ \bibnamefont {Bonitz}}, \bibinfo {author}
  {\bibfnamefont {Attila}\ \bibnamefont {Cangi}}, \ and\ \bibinfo {author}
  {\bibfnamefont {Jan}\ \bibnamefont {Vorberger}},\ }\bibfield  {title}
  {\enquote {\bibinfo {title} {Electronic density response of warm dense
  matter},}\ }\href {\doibase 10.48550/ARXIV.2212.08326} {\  (\bibinfo {year}
  {2022}{\natexlab{h}}),\ 10.48550/ARXIV.2212.08326}\BibitemShut {NoStop}%
\bibitem [{\citenamefont {Marques}\ \emph {et~al.}(2012)\citenamefont
  {Marques}, \citenamefont {Maitra}, \citenamefont {Nogueira}, \citenamefont
  {Gross},\ and\ \citenamefont {Rubio}}]{marques2012fundamentals}%
  \BibitemOpen
  \bibfield  {author} {\bibinfo {author} {\bibfnamefont {M.A.L.}\ \bibnamefont
  {Marques}}, \bibinfo {author} {\bibfnamefont {N.T.}\ \bibnamefont {Maitra}},
  \bibinfo {author} {\bibfnamefont {F.M.S.}\ \bibnamefont {Nogueira}}, \bibinfo
  {author} {\bibfnamefont {E.K.U.}\ \bibnamefont {Gross}}, \ and\ \bibinfo
  {author} {\bibfnamefont {A.}~\bibnamefont {Rubio}},\ }\href
  {https://books.google.de/books?id=bvRZhGR3BnwC} {\emph {\bibinfo {title}
  {Fundamentals of Time-Dependent Density Functional Theory}}},\ Lecture Notes
  in Physics\ (\bibinfo  {publisher} {Springer Berlin Heidelberg},\ \bibinfo
  {year} {2012})\BibitemShut {NoStop}%
\bibitem [{\citenamefont {Dornheim}\ \emph
  {et~al.}(2019{\natexlab{b}})\citenamefont {Dornheim}, \citenamefont
  {Vorberger}, \citenamefont {Groth}, \citenamefont {Hoffmann}, \citenamefont
  {Moldabekov},\ and\ \citenamefont {Bonitz}}]{dornheim_ML}%
  \BibitemOpen
  \bibfield  {author} {\bibinfo {author} {\bibfnamefont {T.}~\bibnamefont
  {Dornheim}}, \bibinfo {author} {\bibfnamefont {J.}~\bibnamefont {Vorberger}},
  \bibinfo {author} {\bibfnamefont {S.}~\bibnamefont {Groth}}, \bibinfo
  {author} {\bibfnamefont {N.}~\bibnamefont {Hoffmann}}, \bibinfo {author}
  {\bibfnamefont {Zh.A.}\ \bibnamefont {Moldabekov}}, \ and\ \bibinfo {author}
  {\bibfnamefont {M.}~\bibnamefont {Bonitz}},\ }\bibfield  {title} {\enquote
  {\bibinfo {title} {The static local field correction of the warm dense
  electron gas: An ab initio path integral {M}onte {C}arlo study and machine
  learning representation},}\ }\href
  {https://aip.scitation.org/doi/full/10.1063/1.5123013} {\bibfield  {journal}
  {\bibinfo  {journal} {J. Chem. Phys}\ }\textbf {\bibinfo {volume} {151}},\
  \bibinfo {pages} {194104} (\bibinfo {year} {2019}{\natexlab{b}})}\BibitemShut
  {NoStop}%
\bibitem [{\citenamefont {Dornheim}\ \emph
  {et~al.}(2021{\natexlab{b}})\citenamefont {Dornheim}, \citenamefont
  {Moldabekov},\ and\ \citenamefont {Tolias}}]{Dornheim_PRB_ESA_2021}%
  \BibitemOpen
  \bibfield  {author} {\bibinfo {author} {\bibfnamefont {Tobias}\ \bibnamefont
  {Dornheim}}, \bibinfo {author} {\bibfnamefont {Zhandos~A.}\ \bibnamefont
  {Moldabekov}}, \ and\ \bibinfo {author} {\bibfnamefont {Panagiotis}\
  \bibnamefont {Tolias}},\ }\bibfield  {title} {\enquote {\bibinfo {title}
  {Analytical representation of the local field correction of the uniform
  electron gas within the effective static approximation},}\ }\href {\doibase
  10.1103/PhysRevB.103.165102} {\bibfield  {journal} {\bibinfo  {journal}
  {Phys. Rev. B}\ }\textbf {\bibinfo {volume} {103}},\ \bibinfo {pages}
  {165102} (\bibinfo {year} {2021}{\natexlab{b}})}\BibitemShut {NoStop}%
\bibitem [{\citenamefont {Dornheim}\ \emph
  {et~al.}(2021{\natexlab{c}})\citenamefont {Dornheim}, \citenamefont
  {B\"ohme}, \citenamefont {Militzer},\ and\ \citenamefont
  {Vorberger}}]{Dornheim_PRB_nk_2021}%
  \BibitemOpen
  \bibfield  {author} {\bibinfo {author} {\bibfnamefont {Tobias}\ \bibnamefont
  {Dornheim}}, \bibinfo {author} {\bibfnamefont {Maximilian}\ \bibnamefont
  {B\"ohme}}, \bibinfo {author} {\bibfnamefont {Burkhard}\ \bibnamefont
  {Militzer}}, \ and\ \bibinfo {author} {\bibfnamefont {Jan}\ \bibnamefont
  {Vorberger}},\ }\bibfield  {title} {\enquote {\bibinfo {title} {Ab initio
  path integral monte carlo approach to the momentum distribution of the
  uniform electron gas at finite temperature without fixed nodes},}\ }\href
  {\doibase 10.1103/PhysRevB.103.205142} {\bibfield  {journal} {\bibinfo
  {journal} {Phys. Rev. B}\ }\textbf {\bibinfo {volume} {103}},\ \bibinfo
  {pages} {205142} (\bibinfo {year} {2021}{\natexlab{c}})}\BibitemShut
  {NoStop}%
\bibitem [{\citenamefont {Dornheim}\ and\ \citenamefont
  {Vorberger}(2021)}]{Dornheim_JCP_2021}%
  \BibitemOpen
  \bibfield  {author} {\bibinfo {author} {\bibfnamefont {Tobias}\ \bibnamefont
  {Dornheim}}\ and\ \bibinfo {author} {\bibfnamefont {Jan}\ \bibnamefont
  {Vorberger}},\ }\bibfield  {title} {\enquote {\bibinfo {title} {Overcoming
  finite-size effects in electronic structure simulations at extreme
  conditions},}\ }\href {\doibase 10.1063/5.0045634} {\bibfield  {journal}
  {\bibinfo  {journal} {The Journal of Chemical Physics}\ }\textbf {\bibinfo
  {volume} {154}},\ \bibinfo {pages} {144103} (\bibinfo {year}
  {2021})}\BibitemShut {NoStop}%
\bibitem [{\citenamefont {Lucco~Castello}\ and\ \citenamefont
  {Tolias}(2022)}]{OCP_bridge_2022}%
  \BibitemOpen
  \bibfield  {author} {\bibinfo {author} {\bibfnamefont {F.}~\bibnamefont
  {Lucco~Castello}}\ and\ \bibinfo {author} {\bibfnamefont {P.}~\bibnamefont
  {Tolias}},\ }\bibfield  {title} {\enquote {\bibinfo {title} {Bridge functions
  of classical one-component plasmas},}\ }\href {\doibase
  10.1103/PhysRevE.105.015208} {\bibfield  {journal} {\bibinfo  {journal}
  {Phys. Rev. E}\ }\textbf {\bibinfo {volume} {105}},\ \bibinfo {pages}
  {015208} (\bibinfo {year} {2022})}\BibitemShut {NoStop}%
\bibitem [{\citenamefont {Takada}(2016)}]{Takada_PRB_2016}%
  \BibitemOpen
  \bibfield  {author} {\bibinfo {author} {\bibfnamefont {Yasutami}\
  \bibnamefont {Takada}},\ }\bibfield  {title} {\enquote {\bibinfo {title}
  {Emergence of an excitonic collective mode in the dilute electron gas},}\
  }\href {\doibase 10.1103/PhysRevB.94.245106} {\bibfield  {journal} {\bibinfo
  {journal} {Phys. Rev. B}\ }\textbf {\bibinfo {volume} {94}},\ \bibinfo
  {pages} {245106} (\bibinfo {year} {2016})}\BibitemShut {NoStop}%
\bibitem [{\citenamefont {Arora}\ \emph {et~al.}(2017)\citenamefont {Arora},
  \citenamefont {Kumar},\ and\ \citenamefont {Moudgil}}]{arora}%
  \BibitemOpen
  \bibfield  {author} {\bibinfo {author} {\bibfnamefont {P.}~\bibnamefont
  {Arora}}, \bibinfo {author} {\bibfnamefont {K.}~\bibnamefont {Kumar}}, \ and\
  \bibinfo {author} {\bibfnamefont {R.~K.}\ \bibnamefont {Moudgil}},\
  }\bibfield  {title} {\enquote {\bibinfo {title} {Spin-resolved correlations
  in the warm-dense homogeneous electron gas},}\ }\href
  {https://link.springer.com/article/10.1140/epjb/e2017-70532-y} {\bibfield
  {journal} {\bibinfo  {journal} {Eur. Phys. J. B}\ }\textbf {\bibinfo {volume}
  {90}},\ \bibinfo {pages} {76} (\bibinfo {year} {2017})}\BibitemShut {NoStop}%
\bibitem [{\citenamefont {Dornheim}\ \emph
  {et~al.}(2022{\natexlab{i}})\citenamefont {Dornheim}, \citenamefont
  {Vorberger}, \citenamefont {Moldabekov},\ and\ \citenamefont
  {Tolias}}]{Dornheim_PRR_2022}%
  \BibitemOpen
  \bibfield  {author} {\bibinfo {author} {\bibfnamefont {Tobias}\ \bibnamefont
  {Dornheim}}, \bibinfo {author} {\bibfnamefont {Jan}\ \bibnamefont
  {Vorberger}}, \bibinfo {author} {\bibfnamefont {Zhandos~A.}\ \bibnamefont
  {Moldabekov}}, \ and\ \bibinfo {author} {\bibfnamefont {Panagiotis}\
  \bibnamefont {Tolias}},\ }\bibfield  {title} {\enquote {\bibinfo {title}
  {Spin-resolved density response of the warm dense electron gas},}\ }\href
  {\doibase 10.1103/PhysRevResearch.4.033018} {\bibfield  {journal} {\bibinfo
  {journal} {Phys. Rev. Research}\ }\textbf {\bibinfo {volume} {4}},\ \bibinfo
  {pages} {033018} (\bibinfo {year} {2022}{\natexlab{i}})}\BibitemShut
  {NoStop}%
\bibitem [{\citenamefont {Tanaka}(2017)}]{Tanaka_CPP_2017}%
  \BibitemOpen
  \bibfield  {author} {\bibinfo {author} {\bibfnamefont {Shigenori}\
  \bibnamefont {Tanaka}},\ }\bibfield  {title} {\enquote {\bibinfo {title}
  {Improved equation of state for finite-temperature spin-polarized electron
  liquids on the basis of singwi–tosi–land–sjölander approximation},}\
  }\href {\doibase https://doi.org/10.1002/ctpp.201600096} {\bibfield
  {journal} {\bibinfo  {journal} {Contributions to Plasma Physics}\ }\textbf
  {\bibinfo {volume} {57}},\ \bibinfo {pages} {126--136} (\bibinfo {year}
  {2017})}\BibitemShut {NoStop}%
\bibitem [{\citenamefont {Ceperley}\ and\ \citenamefont
  {Alder}(1980)}]{Ceperley_Alder_PRL_1980}%
  \BibitemOpen
  \bibfield  {author} {\bibinfo {author} {\bibfnamefont {D.~M.}\ \bibnamefont
  {Ceperley}}\ and\ \bibinfo {author} {\bibfnamefont {B.~J.}\ \bibnamefont
  {Alder}},\ }\bibfield  {title} {\enquote {\bibinfo {title} {Ground state of
  the electron gas by a stochastic method},}\ }\href {\doibase
  10.1103/PhysRevLett.45.566} {\bibfield  {journal} {\bibinfo  {journal} {Phys.
  Rev. Lett.}\ }\textbf {\bibinfo {volume} {45}},\ \bibinfo {pages} {566--569}
  (\bibinfo {year} {1980})}\BibitemShut {NoStop}%
\bibitem [{\citenamefont {Vosko}\ \emph {et~al.}(1980)\citenamefont {Vosko},
  \citenamefont {Wilk},\ and\ \citenamefont {Nusair}}]{vwn}%
  \BibitemOpen
  \bibfield  {author} {\bibinfo {author} {\bibfnamefont {S.~H.}\ \bibnamefont
  {Vosko}}, \bibinfo {author} {\bibfnamefont {L.}~\bibnamefont {Wilk}}, \ and\
  \bibinfo {author} {\bibfnamefont {M.}~\bibnamefont {Nusair}},\ }\bibfield
  {title} {\enquote {\bibinfo {title} {Accurate spin-dependent electron liquid
  correlation energies for local spin density calculations: a critical
  analysis},}\ }\href {\doibase 10.1139/p80-159} {\bibfield  {journal}
  {\bibinfo  {journal} {Canadian Journal of Physics}\ }\textbf {\bibinfo
  {volume} {58}},\ \bibinfo {pages} {1200--1211} (\bibinfo {year}
  {1980})}\BibitemShut {NoStop}%
\bibitem [{\citenamefont {Fiolhais}\ \emph {et~al.}(2008)\citenamefont
  {Fiolhais}, \citenamefont {Nogueira},\ and\ \citenamefont
  {Marques}}]{fiolhais2008primer}%
  \BibitemOpen
  \bibfield  {author} {\bibinfo {author} {\bibfnamefont {C.}~\bibnamefont
  {Fiolhais}}, \bibinfo {author} {\bibfnamefont {F.}~\bibnamefont {Nogueira}},
  \ and\ \bibinfo {author} {\bibfnamefont {M.A.L.}\ \bibnamefont {Marques}},\
  }\href {https://books.google.de/books?id=kjlsCQAAQBAJ} {\emph {\bibinfo
  {title} {A Primer in Density Functional Theory}}},\ Lecture Notes in Physics\
  (\bibinfo  {publisher} {Springer Berlin Heidelberg},\ \bibinfo {year}
  {2008})\BibitemShut {NoStop}%
\bibitem [{\citenamefont {Perdew}\ and\ \citenamefont
  {Zunger}(1981)}]{Perdew_Zunger_PRB_1981}%
  \BibitemOpen
  \bibfield  {author} {\bibinfo {author} {\bibfnamefont {J.~P.}\ \bibnamefont
  {Perdew}}\ and\ \bibinfo {author} {\bibfnamefont {Alex}\ \bibnamefont
  {Zunger}},\ }\bibfield  {title} {\enquote {\bibinfo {title} {Self-interaction
  correction to density-functional approximations for many-electron systems},}\
  }\href {\doibase 10.1103/PhysRevB.23.5048} {\bibfield  {journal} {\bibinfo
  {journal} {Phys. Rev. B}\ }\textbf {\bibinfo {volume} {23}},\ \bibinfo
  {pages} {5048--5079} (\bibinfo {year} {1981})}\BibitemShut {NoStop}%
\bibitem [{\citenamefont {Gori-Giorgi}\ \emph {et~al.}(2000)\citenamefont
  {Gori-Giorgi}, \citenamefont {Sacchetti},\ and\ \citenamefont
  {Bachelet}}]{Gori-Giorgi_PRB_2000}%
  \BibitemOpen
  \bibfield  {author} {\bibinfo {author} {\bibfnamefont {Paola}\ \bibnamefont
  {Gori-Giorgi}}, \bibinfo {author} {\bibfnamefont {Francesco}\ \bibnamefont
  {Sacchetti}}, \ and\ \bibinfo {author} {\bibfnamefont {Giovanni~B.}\
  \bibnamefont {Bachelet}},\ }\bibfield  {title} {\enquote {\bibinfo {title}
  {Analytic static structure factors and pair-correlation functions for the
  unpolarized homogeneous electron gas},}\ }\href {\doibase
  10.1103/PhysRevB.61.7353} {\bibfield  {journal} {\bibinfo  {journal} {Phys.
  Rev. B}\ }\textbf {\bibinfo {volume} {61}},\ \bibinfo {pages} {7353--7363}
  (\bibinfo {year} {2000})}\BibitemShut {NoStop}%
\bibitem [{\citenamefont {Corradini}\ \emph {et~al.}(1998)\citenamefont
  {Corradini}, \citenamefont {Sole}, \citenamefont {Onida},\ and\ \citenamefont
  {Palummo}}]{cdop}%
  \BibitemOpen
  \bibfield  {author} {\bibinfo {author} {\bibfnamefont {M.}~\bibnamefont
  {Corradini}}, \bibinfo {author} {\bibfnamefont {R.~Del}\ \bibnamefont
  {Sole}}, \bibinfo {author} {\bibfnamefont {G.}~\bibnamefont {Onida}}, \ and\
  \bibinfo {author} {\bibfnamefont {M.}~\bibnamefont {Palummo}},\ }\bibfield
  {title} {\enquote {\bibinfo {title} {Analytical expressions for the
  local-field factor $g(q)$ and the exchange-correlation kernel
  ${K}_{\mathrm{xc}}(r)$ of the homogeneous electron gas},}\ }\href
  {http://link.aps.org/doi/10.1103/PhysRevB.57.14569} {\bibfield  {journal}
  {\bibinfo  {journal} {Phys. Rev. B}\ }\textbf {\bibinfo {volume} {57}},\
  \bibinfo {pages} {14569} (\bibinfo {year} {1998})}\BibitemShut {NoStop}%
\bibitem [{\citenamefont {Ortiz}\ and\ \citenamefont
  {Ballone}(1994)}]{Ortiz_PRB_1994}%
  \BibitemOpen
  \bibfield  {author} {\bibinfo {author} {\bibfnamefont {G.}~\bibnamefont
  {Ortiz}}\ and\ \bibinfo {author} {\bibfnamefont {P.}~\bibnamefont
  {Ballone}},\ }\bibfield  {title} {\enquote {\bibinfo {title} {Correlation
  energy, structure factor, radial distribution function, and momentum
  distribution of the spin-polarized uniform electron gas},}\ }\href {\doibase
  10.1103/PhysRevB.50.1391} {\bibfield  {journal} {\bibinfo  {journal} {Phys.
  Rev. B}\ }\textbf {\bibinfo {volume} {50}},\ \bibinfo {pages} {1391--1405}
  (\bibinfo {year} {1994})}\BibitemShut {NoStop}%
\bibitem [{\citenamefont {Ortiz}\ \emph {et~al.}(1999)\citenamefont {Ortiz},
  \citenamefont {Harris},\ and\ \citenamefont {Ballone}}]{Ortiz_PRL_1999}%
  \BibitemOpen
  \bibfield  {author} {\bibinfo {author} {\bibfnamefont {G.}~\bibnamefont
  {Ortiz}}, \bibinfo {author} {\bibfnamefont {M.}~\bibnamefont {Harris}}, \
  and\ \bibinfo {author} {\bibfnamefont {P.}~\bibnamefont {Ballone}},\
  }\bibfield  {title} {\enquote {\bibinfo {title} {Zero temperature phases of
  the electron gas},}\ }\href {\doibase 10.1103/PhysRevLett.82.5317} {\bibfield
   {journal} {\bibinfo  {journal} {Phys. Rev. Lett.}\ }\textbf {\bibinfo
  {volume} {82}},\ \bibinfo {pages} {5317--5320} (\bibinfo {year}
  {1999})}\BibitemShut {NoStop}%
\bibitem [{\citenamefont {Spink}\ \emph {et~al.}(2013)\citenamefont {Spink},
  \citenamefont {Needs},\ and\ \citenamefont {Drummond}}]{Spink_PRB_2013}%
  \BibitemOpen
  \bibfield  {author} {\bibinfo {author} {\bibfnamefont {G.~G.}\ \bibnamefont
  {Spink}}, \bibinfo {author} {\bibfnamefont {R.~J.}\ \bibnamefont {Needs}}, \
  and\ \bibinfo {author} {\bibfnamefont {N.~D.}\ \bibnamefont {Drummond}},\
  }\bibfield  {title} {\enquote {\bibinfo {title} {Quantum monte carlo study of
  the three-dimensional spin-polarized homogeneous electron gas},}\ }\href
  {\doibase 10.1103/PhysRevB.88.085121} {\bibfield  {journal} {\bibinfo
  {journal} {Phys. Rev. B}\ }\textbf {\bibinfo {volume} {88}},\ \bibinfo
  {pages} {085121} (\bibinfo {year} {2013})}\BibitemShut {NoStop}%
\bibitem [{\citenamefont {Jones}(2015)}]{Jones_RMP_2015}%
  \BibitemOpen
  \bibfield  {author} {\bibinfo {author} {\bibfnamefont {R.~O.}\ \bibnamefont
  {Jones}},\ }\bibfield  {title} {\enquote {\bibinfo {title} {Density
  functional theory: Its origins, rise to prominence, and future},}\ }\href
  {\doibase 10.1103/RevModPhys.87.897} {\bibfield  {journal} {\bibinfo
  {journal} {Rev. Mod. Phys.}\ }\textbf {\bibinfo {volume} {87}},\ \bibinfo
  {pages} {897--923} (\bibinfo {year} {2015})}\BibitemShut {NoStop}%
\bibitem [{\citenamefont {Karasiev}\ \emph
  {et~al.}(2019{\natexlab{b}})\citenamefont {Karasiev}, \citenamefont
  {Trickey},\ and\ \citenamefont {Dufty}}]{status}%
  \BibitemOpen
  \bibfield  {author} {\bibinfo {author} {\bibfnamefont {V.~V.}\ \bibnamefont
  {Karasiev}}, \bibinfo {author} {\bibfnamefont {S.~B.}\ \bibnamefont
  {Trickey}}, \ and\ \bibinfo {author} {\bibfnamefont {J.~W.}\ \bibnamefont
  {Dufty}},\ }\bibfield  {title} {\enquote {\bibinfo {title} {Status of
  free-energy representations for the homogeneous electron gas},}\ }\href
  {https://journals.aps.org/prb/abstract/10.1103/PhysRevB.99.195134} {\bibfield
   {journal} {\bibinfo  {journal} {Phys. Rev. B}\ }\textbf {\bibinfo {volume}
  {99}},\ \bibinfo {pages} {195134} (\bibinfo {year}
  {2019}{\natexlab{b}})}\BibitemShut {NoStop}%
\bibitem [{\citenamefont {Malone}\ \emph {et~al.}(2015)\citenamefont {Malone},
  \citenamefont {Blunt}, \citenamefont {Shepherd}, \citenamefont {Lee},
  \citenamefont {Spencer},\ and\ \citenamefont {Foulkes}}]{Malone_JCP_2015}%
  \BibitemOpen
  \bibfield  {author} {\bibinfo {author} {\bibfnamefont {Fionn~D.}\
  \bibnamefont {Malone}}, \bibinfo {author} {\bibfnamefont {N.~S.}\
  \bibnamefont {Blunt}}, \bibinfo {author} {\bibfnamefont {James~J.}\
  \bibnamefont {Shepherd}}, \bibinfo {author} {\bibfnamefont {D.~K.~K.}\
  \bibnamefont {Lee}}, \bibinfo {author} {\bibfnamefont {J.~S.}\ \bibnamefont
  {Spencer}}, \ and\ \bibinfo {author} {\bibfnamefont {W.~M.~C.}\ \bibnamefont
  {Foulkes}},\ }\bibfield  {title} {\enquote {\bibinfo {title} {Interaction
  picture density matrix quantum monte carlo},}\ }\href {\doibase
  10.1063/1.4927434} {\bibfield  {journal} {\bibinfo  {journal} {The Journal of
  Chemical Physics}\ }\textbf {\bibinfo {volume} {143}},\ \bibinfo {pages}
  {044116} (\bibinfo {year} {2015})}\BibitemShut {NoStop}%
\bibitem [{\citenamefont {Dornheim}\ \emph
  {et~al.}(2015{\natexlab{a}})\citenamefont {Dornheim}, \citenamefont {Groth},
  \citenamefont {Filinov},\ and\ \citenamefont {Bonitz}}]{Dornheim_NJP_2015}%
  \BibitemOpen
  \bibfield  {author} {\bibinfo {author} {\bibfnamefont {Tobias}\ \bibnamefont
  {Dornheim}}, \bibinfo {author} {\bibfnamefont {Simon}\ \bibnamefont {Groth}},
  \bibinfo {author} {\bibfnamefont {Alexey}\ \bibnamefont {Filinov}}, \ and\
  \bibinfo {author} {\bibfnamefont {Michael}\ \bibnamefont {Bonitz}},\
  }\bibfield  {title} {\enquote {\bibinfo {title} {Permutation blocking path
  integral monte carlo: a highly efficient approach to the simulation of
  strongly degenerate non-ideal fermions},}\ }\href {\doibase
  10.1088/1367-2630/17/7/073017} {\bibfield  {journal} {\bibinfo  {journal}
  {New Journal of Physics}\ }\textbf {\bibinfo {volume} {17}},\ \bibinfo
  {pages} {073017} (\bibinfo {year} {2015}{\natexlab{a}})}\BibitemShut
  {NoStop}%
\bibitem [{\citenamefont {Dornheim}\ \emph
  {et~al.}(2015{\natexlab{b}})\citenamefont {Dornheim}, \citenamefont {Schoof},
  \citenamefont {Groth}, \citenamefont {Filinov},\ and\ \citenamefont
  {Bonitz}}]{Dornheim_JCP_2015}%
  \BibitemOpen
  \bibfield  {author} {\bibinfo {author} {\bibfnamefont {Tobias}\ \bibnamefont
  {Dornheim}}, \bibinfo {author} {\bibfnamefont {Tim}\ \bibnamefont {Schoof}},
  \bibinfo {author} {\bibfnamefont {Simon}\ \bibnamefont {Groth}}, \bibinfo
  {author} {\bibfnamefont {Alexey}\ \bibnamefont {Filinov}}, \ and\ \bibinfo
  {author} {\bibfnamefont {Michael}\ \bibnamefont {Bonitz}},\ }\bibfield
  {title} {\enquote {\bibinfo {title} {Permutation blocking path integral monte
  carlo approach to the uniform electron gas at finite temperature},}\ }\href
  {\doibase 10.1063/1.4936145} {\bibfield  {journal} {\bibinfo  {journal} {The
  Journal of Chemical Physics}\ }\textbf {\bibinfo {volume} {143}},\ \bibinfo
  {pages} {204101} (\bibinfo {year} {2015}{\natexlab{b}})}\BibitemShut
  {NoStop}%
\bibitem [{\citenamefont {Lee}\ \emph {et~al.}(2021)\citenamefont {Lee},
  \citenamefont {Morales},\ and\ \citenamefont {Malone}}]{Joonho_JCP_2021}%
  \BibitemOpen
  \bibfield  {author} {\bibinfo {author} {\bibfnamefont {Joonho}\ \bibnamefont
  {Lee}}, \bibinfo {author} {\bibfnamefont {Miguel~A.}\ \bibnamefont
  {Morales}}, \ and\ \bibinfo {author} {\bibfnamefont {Fionn~D.}\ \bibnamefont
  {Malone}},\ }\bibfield  {title} {\enquote {\bibinfo {title} {A phaseless
  auxiliary-field quantum monte carlo perspective on the uniform electron gas
  at finite temperatures: Issues, observations, and benchmark study},}\ }\href
  {\doibase 10.1063/5.0041378} {\bibfield  {journal} {\bibinfo  {journal} {The
  Journal of Chemical Physics}\ }\textbf {\bibinfo {volume} {154}},\ \bibinfo
  {pages} {064109} (\bibinfo {year} {2021})}\BibitemShut {NoStop}%
\bibitem [{\citenamefont {Yilmaz}\ \emph {et~al.}(2020)\citenamefont {Yilmaz},
  \citenamefont {Hunger}, \citenamefont {Dornheim}, \citenamefont {Groth},\
  and\ \citenamefont {Bonitz}}]{Yilmaz_JCP_2020}%
  \BibitemOpen
  \bibfield  {author} {\bibinfo {author} {\bibfnamefont {A.}~\bibnamefont
  {Yilmaz}}, \bibinfo {author} {\bibfnamefont {K.}~\bibnamefont {Hunger}},
  \bibinfo {author} {\bibfnamefont {T.}~\bibnamefont {Dornheim}}, \bibinfo
  {author} {\bibfnamefont {S.}~\bibnamefont {Groth}}, \ and\ \bibinfo {author}
  {\bibfnamefont {M.}~\bibnamefont {Bonitz}},\ }\bibfield  {title} {\enquote
  {\bibinfo {title} {Restricted configuration path integral monte carlo},}\
  }\href {\doibase 10.1063/5.0022800} {\bibfield  {journal} {\bibinfo
  {journal} {The Journal of Chemical Physics}\ }\textbf {\bibinfo {volume}
  {153}},\ \bibinfo {pages} {124114} (\bibinfo {year} {2020})}\BibitemShut
  {NoStop}%
\bibitem [{\citenamefont {Huotari}\ \emph {et~al.}(2010)\citenamefont
  {Huotari}, \citenamefont {Soininen}, \citenamefont {Pylkk\"anen},
  \citenamefont {H\"am\"al\"ainen}, \citenamefont {Issolah}, \citenamefont
  {Titov}, \citenamefont {McMinis}, \citenamefont {Kim}, \citenamefont {Esler},
  \citenamefont {Ceperley}, \citenamefont {Holzmann},\ and\ \citenamefont
  {Olevano}}]{Huotari_PRL_2010}%
  \BibitemOpen
  \bibfield  {author} {\bibinfo {author} {\bibfnamefont {Simo}\ \bibnamefont
  {Huotari}}, \bibinfo {author} {\bibfnamefont {J.~Aleksi}\ \bibnamefont
  {Soininen}}, \bibinfo {author} {\bibfnamefont {Tuomas}\ \bibnamefont
  {Pylkk\"anen}}, \bibinfo {author} {\bibfnamefont {Keijo}\ \bibnamefont
  {H\"am\"al\"ainen}}, \bibinfo {author} {\bibfnamefont {Arezki}\ \bibnamefont
  {Issolah}}, \bibinfo {author} {\bibfnamefont {Andrey}\ \bibnamefont {Titov}},
  \bibinfo {author} {\bibfnamefont {Jeremy}\ \bibnamefont {McMinis}}, \bibinfo
  {author} {\bibfnamefont {Jeongnim}\ \bibnamefont {Kim}}, \bibinfo {author}
  {\bibfnamefont {Ken}\ \bibnamefont {Esler}}, \bibinfo {author} {\bibfnamefont
  {David~M.}\ \bibnamefont {Ceperley}}, \bibinfo {author} {\bibfnamefont
  {Markus}\ \bibnamefont {Holzmann}}, \ and\ \bibinfo {author} {\bibfnamefont
  {Valerio}\ \bibnamefont {Olevano}},\ }\bibfield  {title} {\enquote {\bibinfo
  {title} {Momentum distribution and renormalization factor in sodium and the
  electron gas},}\ }\href {\doibase 10.1103/PhysRevLett.105.086403} {\bibfield
  {journal} {\bibinfo  {journal} {Phys. Rev. Lett.}\ }\textbf {\bibinfo
  {volume} {105}},\ \bibinfo {pages} {086403} (\bibinfo {year}
  {2010})}\BibitemShut {NoStop}%
\bibitem [{\citenamefont {Zastrau}\ \emph {et~al.}(2014)\citenamefont
  {Zastrau}, \citenamefont {Sperling}, \citenamefont {Harmand}, \citenamefont
  {Becker}, \citenamefont {Bornath}, \citenamefont {Bredow}, \citenamefont
  {Dziarzhytski}, \citenamefont {Fennel}, \citenamefont {Fletcher},
  \citenamefont {F{"o}rster}, \citenamefont {G{"o}de}, \citenamefont {Gregori},
  \citenamefont {Hilbert}, \citenamefont {Hochhaus}, \citenamefont {Holst},
  \citenamefont {Laarmann}, \citenamefont {Lee}, \citenamefont {Ma},
  \citenamefont {Mithen}, \citenamefont {Mitzner}, \citenamefont {Murphy},
  \citenamefont {Nakatsutsumi}, \citenamefont {Neumayer}, \citenamefont
  {Przystawik}, \citenamefont {Roling}, \citenamefont {Schulz}, \citenamefont
  {Siemer}, \citenamefont {Skruszewicz}, \citenamefont {Tiggesb{"a}umker},
  \citenamefont {Toleikis}, \citenamefont {Tschentscher}, \citenamefont
  {White}, \citenamefont {W{"o}stmann}, \citenamefont {Zacharias},
  \citenamefont {D{"o}ppner}, \citenamefont {Glenzer},\ and\ \citenamefont
  {Redmer}}]{Zastrau}%
  \BibitemOpen
  \bibfield  {author} {\bibinfo {author} {\bibfnamefont {U.}~\bibnamefont
  {Zastrau}}, \bibinfo {author} {\bibfnamefont {P.}~\bibnamefont {Sperling}},
  \bibinfo {author} {\bibfnamefont {M.}~\bibnamefont {Harmand}}, \bibinfo
  {author} {\bibfnamefont {A.}~\bibnamefont {Becker}}, \bibinfo {author}
  {\bibfnamefont {T.}~\bibnamefont {Bornath}}, \bibinfo {author} {\bibfnamefont
  {R.}~\bibnamefont {Bredow}}, \bibinfo {author} {\bibfnamefont
  {S.}~\bibnamefont {Dziarzhytski}}, \bibinfo {author} {\bibfnamefont
  {T.}~\bibnamefont {Fennel}}, \bibinfo {author} {\bibfnamefont {L.~B.}\
  \bibnamefont {Fletcher}}, \bibinfo {author} {\bibfnamefont {E.}~\bibnamefont
  {F{"o}rster}}, \bibinfo {author} {\bibfnamefont {S.}~\bibnamefont {G{"o}de}},
  \bibinfo {author} {\bibfnamefont {G.}~\bibnamefont {Gregori}}, \bibinfo
  {author} {\bibfnamefont {V.}~\bibnamefont {Hilbert}}, \bibinfo {author}
  {\bibfnamefont {D.}~\bibnamefont {Hochhaus}}, \bibinfo {author}
  {\bibfnamefont {B.}~\bibnamefont {Holst}}, \bibinfo {author} {\bibfnamefont
  {T.}~\bibnamefont {Laarmann}}, \bibinfo {author} {\bibfnamefont {H.~J.}\
  \bibnamefont {Lee}}, \bibinfo {author} {\bibfnamefont {T.}~\bibnamefont
  {Ma}}, \bibinfo {author} {\bibfnamefont {J.~P.}\ \bibnamefont {Mithen}},
  \bibinfo {author} {\bibfnamefont {R.}~\bibnamefont {Mitzner}}, \bibinfo
  {author} {\bibfnamefont {C.~D.}\ \bibnamefont {Murphy}}, \bibinfo {author}
  {\bibfnamefont {M.}~\bibnamefont {Nakatsutsumi}}, \bibinfo {author}
  {\bibfnamefont {P.}~\bibnamefont {Neumayer}}, \bibinfo {author}
  {\bibfnamefont {A.}~\bibnamefont {Przystawik}}, \bibinfo {author}
  {\bibfnamefont {S.}~\bibnamefont {Roling}}, \bibinfo {author} {\bibfnamefont
  {M.}~\bibnamefont {Schulz}}, \bibinfo {author} {\bibfnamefont
  {B.}~\bibnamefont {Siemer}}, \bibinfo {author} {\bibfnamefont
  {S.}~\bibnamefont {Skruszewicz}}, \bibinfo {author} {\bibfnamefont
  {J.}~\bibnamefont {Tiggesb{"a}umker}}, \bibinfo {author} {\bibfnamefont
  {S.}~\bibnamefont {Toleikis}}, \bibinfo {author} {\bibfnamefont
  {T.}~\bibnamefont {Tschentscher}}, \bibinfo {author} {\bibfnamefont
  {T.}~\bibnamefont {White}}, \bibinfo {author} {\bibfnamefont
  {M.}~\bibnamefont {W{"o}stmann}}, \bibinfo {author} {\bibfnamefont
  {H.}~\bibnamefont {Zacharias}}, \bibinfo {author} {\bibfnamefont
  {T.}~\bibnamefont {D{"o}ppner}}, \bibinfo {author} {\bibfnamefont {S.~H.}\
  \bibnamefont {Glenzer}}, \ and\ \bibinfo {author} {\bibfnamefont
  {R.}~\bibnamefont {Redmer}},\ }\bibfield  {title} {\enquote {\bibinfo {title}
  {Resolving ultrafast heating of dense cryogenic hydrogen},}\ }\href
  {https://journals.aps.org/prl/abstract/10.1103/PhysRevLett.112.105002}
  {\bibfield  {journal} {\bibinfo  {journal} {Phys. Rev. Lett}\ }\textbf
  {\bibinfo {volume} {112}},\ \bibinfo {pages} {105002} (\bibinfo {year}
  {2014})}\BibitemShut {NoStop}%
\bibitem [{\citenamefont {Hamann}\ \emph {et~al.}(2020)\citenamefont {Hamann},
  \citenamefont {Dornheim}, \citenamefont {Vorberger}, \citenamefont
  {Moldabekov},\ and\ \citenamefont {Bonitz}}]{Hamann_PRB_2020}%
  \BibitemOpen
  \bibfield  {author} {\bibinfo {author} {\bibfnamefont {Paul}\ \bibnamefont
  {Hamann}}, \bibinfo {author} {\bibfnamefont {Tobias}\ \bibnamefont
  {Dornheim}}, \bibinfo {author} {\bibfnamefont {Jan}\ \bibnamefont
  {Vorberger}}, \bibinfo {author} {\bibfnamefont {Zhandos~A.}\ \bibnamefont
  {Moldabekov}}, \ and\ \bibinfo {author} {\bibfnamefont {Michael}\
  \bibnamefont {Bonitz}},\ }\bibfield  {title} {\enquote {\bibinfo {title}
  {Dynamic properties of the warm dense electron gas based on $ab initio$ path
  integral monte carlo simulations},}\ }\href {\doibase
  10.1103/PhysRevB.102.125150} {\bibfield  {journal} {\bibinfo  {journal}
  {Phys. Rev. B}\ }\textbf {\bibinfo {volume} {102}},\ \bibinfo {pages}
  {125150} (\bibinfo {year} {2020})}\BibitemShut {NoStop}%
\bibitem [{\citenamefont {Glenzer}\ and\ \citenamefont
  {Redmer}(2009)}]{siegfried_review}%
  \BibitemOpen
  \bibfield  {author} {\bibinfo {author} {\bibfnamefont {S.~H.}\ \bibnamefont
  {Glenzer}}\ and\ \bibinfo {author} {\bibfnamefont {R.}~\bibnamefont
  {Redmer}},\ }\bibfield  {title} {\enquote {\bibinfo {title} {X-ray thomson
  scattering in high energy density plasmas},}\ }\href
  {https://journals.aps.org/rmp/abstract/10.1103/RevModPhys.81.1625} {\bibfield
   {journal} {\bibinfo  {journal} {Rev. Mod. Phys}\ }\textbf {\bibinfo {volume}
  {81}},\ \bibinfo {pages} {1625} (\bibinfo {year} {2009})}\BibitemShut
  {NoStop}%
\bibitem [{\citenamefont {Tschentscher}\ \emph {et~al.}(2017)\citenamefont
  {Tschentscher}, \citenamefont {Bressler}, \citenamefont {Grünert},
  \citenamefont {Madsen}, \citenamefont {Mancuso}, \citenamefont {Meyer},
  \citenamefont {Scherz}, \citenamefont {Sinn},\ and\ \citenamefont
  {Zastrau}}]{Tschentscher_2017}%
  \BibitemOpen
  \bibfield  {author} {\bibinfo {author} {\bibfnamefont {Thomas}\ \bibnamefont
  {Tschentscher}}, \bibinfo {author} {\bibfnamefont {Christian}\ \bibnamefont
  {Bressler}}, \bibinfo {author} {\bibfnamefont {Jan}\ \bibnamefont
  {Grünert}}, \bibinfo {author} {\bibfnamefont {Anders}\ \bibnamefont
  {Madsen}}, \bibinfo {author} {\bibfnamefont {Adrian~P.}\ \bibnamefont
  {Mancuso}}, \bibinfo {author} {\bibfnamefont {Michael}\ \bibnamefont
  {Meyer}}, \bibinfo {author} {\bibfnamefont {Andreas}\ \bibnamefont {Scherz}},
  \bibinfo {author} {\bibfnamefont {Harald}\ \bibnamefont {Sinn}}, \ and\
  \bibinfo {author} {\bibfnamefont {Ulf}\ \bibnamefont {Zastrau}},\ }\bibfield
  {title} {\enquote {\bibinfo {title} {Photon beam transport and scientific
  instruments at the european xfel},}\ }\href {\doibase 10.3390/app7060592}
  {\bibfield  {journal} {\bibinfo  {journal} {Applied Sciences}\ }\textbf
  {\bibinfo {volume} {7}} (\bibinfo {year} {2017}),\
  10.3390/app7060592}\BibitemShut {NoStop}%
\bibitem [{\citenamefont {Sch\"uler}\ and\ \citenamefont
  {Pavlyukh}(2018)}]{Schueler_PRB_2018}%
  \BibitemOpen
  \bibfield  {author} {\bibinfo {author} {\bibfnamefont {M.}~\bibnamefont
  {Sch\"uler}}\ and\ \bibinfo {author} {\bibfnamefont {Y.}~\bibnamefont
  {Pavlyukh}},\ }\bibfield  {title} {\enquote {\bibinfo {title} {Spectral
  properties from matsubara green's function approach: Application to
  molecules},}\ }\href {\doibase 10.1103/PhysRevB.97.115164} {\bibfield
  {journal} {\bibinfo  {journal} {Phys. Rev. B}\ }\textbf {\bibinfo {volume}
  {97}},\ \bibinfo {pages} {115164} (\bibinfo {year} {2018})}\BibitemShut
  {NoStop}%
\bibitem [{\citenamefont {Meijering}(2002)}]{LIP}%
  \BibitemOpen
  \bibfield  {author} {\bibinfo {author} {\bibfnamefont {E.}~\bibnamefont
  {Meijering}},\ }\bibfield  {title} {\enquote {\bibinfo {title} {A chronology
  of interpolation: {F}rom ancient astronomy to modern signal and image
  processing},}\ }\href@noop {} {\bibfield  {journal} {\bibinfo  {journal}
  {Proceedings of the {IEEE}}\ }\textbf {\bibinfo {volume} {90}},\ \bibinfo
  {pages} {319--342} (\bibinfo {year} {2002})}\BibitemShut {NoStop}%
\bibitem [{\citenamefont {Gergonne}(1974)}]{gergonne1974application}%
  \BibitemOpen
  \bibfield  {author} {\bibinfo {author} {\bibfnamefont {Joseph~Diaz}\
  \bibnamefont {Gergonne}},\ }\bibfield  {title} {\enquote {\bibinfo {title}
  {The application of the method of least squares to the interpolation of
  sequences},}\ }\href@noop {} {\bibfield  {journal} {\bibinfo  {journal}
  {Historia Mathematica}\ }\textbf {\bibinfo {volume} {1}},\ \bibinfo {pages}
  {439--447} (\bibinfo {year} {1974})}\BibitemShut {NoStop}%
\bibitem [{\citenamefont {Stigler}(1974)}]{stigler1974gergonne}%
  \BibitemOpen
  \bibfield  {author} {\bibinfo {author} {\bibfnamefont {Stephen~M}\
  \bibnamefont {Stigler}},\ }\bibfield  {title} {\enquote {\bibinfo {title}
  {{G}ergonne's 1815 paper on the design and analysis of polynomial regression
  experiments},}\ }\href@noop {} {\bibfield  {journal} {\bibinfo  {journal}
  {Historia Mathematica}\ }\textbf {\bibinfo {volume} {1}},\ \bibinfo {pages}
  {431--439} (\bibinfo {year} {1974})}\BibitemShut {NoStop}%
\bibitem [{\citenamefont {Platte}\ \emph {et~al.}(2011)\citenamefont {Platte},
  \citenamefont {Trefethen},\ and\ \citenamefont {Kuijlaars}}]{platte:2011}%
  \BibitemOpen
  \bibfield  {author} {\bibinfo {author} {\bibfnamefont {Rodrigo~B}\
  \bibnamefont {Platte}}, \bibinfo {author} {\bibfnamefont {Lloyd~N}\
  \bibnamefont {Trefethen}}, \ and\ \bibinfo {author} {\bibfnamefont {Arno~BJ}\
  \bibnamefont {Kuijlaars}},\ }\bibfield  {title} {\enquote {\bibinfo {title}
  {Impossibility of fast stable approximation of analytic functions from
  equispaced samples},}\ }\href@noop {} {\bibfield  {journal} {\bibinfo
  {journal} {SIAM review}\ }\textbf {\bibinfo {volume} {53}},\ \bibinfo {pages}
  {308--318} (\bibinfo {year} {2011})}\BibitemShut {NoStop}%
\bibitem [{\citenamefont {Runge}(1901)}]{runge}%
  \BibitemOpen
  \bibfield  {author} {\bibinfo {author} {\bibfnamefont {Carl}\ \bibnamefont
  {Runge}},\ }\bibfield  {title} {\enquote {\bibinfo {title} {{\"U}ber
  empirische {F}unktionen und die {I}nterpolation zwischen {\"a}quidistanten
  {O}rdinaten},}\ }\href@noop {} {\bibfield  {journal} {\bibinfo  {journal}
  {Zeitschrift f{\"u}r Mathematik und Physik}\ }\textbf {\bibinfo {volume}
  {46}},\ \bibinfo {pages} {20} (\bibinfo {year} {1901})}\BibitemShut {NoStop}%
\bibitem [{\citenamefont {Hewitt}\ and\ \citenamefont
  {Hewitt}(1979)}]{hewitt1979gibbs}%
  \BibitemOpen
  \bibfield  {author} {\bibinfo {author} {\bibfnamefont {Edwin}\ \bibnamefont
  {Hewitt}}\ and\ \bibinfo {author} {\bibfnamefont {Robert~E}\ \bibnamefont
  {Hewitt}},\ }\bibfield  {title} {\enquote {\bibinfo {title} {The
  {G}ibbs-{W}ilbraham phenomenon: an episode in {F}ourier analysis},}\
  }\href@noop {} {\bibfield  {journal} {\bibinfo  {journal} {Archive for
  history of Exact Sciences}\ ,\ \bibinfo {pages} {129--160}} (\bibinfo {year}
  {1979})}\BibitemShut {NoStop}%
\bibitem [{\citenamefont {Dimarogonas}(1996)}]{dimarogonas1996vibration}%
  \BibitemOpen
  \bibfield  {author} {\bibinfo {author} {\bibfnamefont {Andrew~D}\
  \bibnamefont {Dimarogonas}},\ }\href@noop {} {\emph {\bibinfo {title}
  {Vibration for engineers}}}\ (\bibinfo  {publisher} {Prentice Hall},\
  \bibinfo {year} {1996})\BibitemShut {NoStop}%
\bibitem [{\citenamefont {Veettil}\ \emph {et~al.}(2022)\citenamefont
  {Veettil}, \citenamefont {Zheng}, \citenamefont {Acosta}, \citenamefont
  {Wicaksono},\ and\ \citenamefont {Hecht}}]{REG_arxiv}%
  \BibitemOpen
  \bibfield  {author} {\bibinfo {author} {\bibfnamefont {Sachin K~Thekke}\
  \bibnamefont {Veettil}}, \bibinfo {author} {\bibfnamefont {Yuxi}\
  \bibnamefont {Zheng}}, \bibinfo {author} {\bibfnamefont {Uwe~Hernandez}\
  \bibnamefont {Acosta}}, \bibinfo {author} {\bibfnamefont {Damar}\
  \bibnamefont {Wicaksono}}, \ and\ \bibinfo {author} {\bibfnamefont {Michael}\
  \bibnamefont {Hecht}},\ }\bibfield  {title} {\enquote {\bibinfo {title}
  {Multivariate polynomial regression of {E}uclidean degree extends the
  stability for fast approximations of {T}refethen functions},}\ }\href@noop {}
  {\bibfield  {journal} {\bibinfo  {journal} {arXiv preprint arXiv:2212.11706}\
  } (\bibinfo {year} {2022})}\BibitemShut {NoStop}%
\bibitem [{\citenamefont {Hecht}\ \emph {et~al.}(2017)\citenamefont {Hecht},
  \citenamefont {Cheeseman}, \citenamefont {Hoffmann},\ and\ \citenamefont
  {Sbalzarini}}]{PIP1}%
  \BibitemOpen
  \bibfield  {author} {\bibinfo {author} {\bibfnamefont {M}~\bibnamefont
  {Hecht}}, \bibinfo {author} {\bibfnamefont {Bevan~L.}\ \bibnamefont
  {Cheeseman}}, \bibinfo {author} {\bibfnamefont {Karl~B.}\ \bibnamefont
  {Hoffmann}}, \ and\ \bibinfo {author} {\bibfnamefont {Ivo~F.}\ \bibnamefont
  {Sbalzarini}},\ }\bibfield  {title} {\enquote {\bibinfo {title} {A
  quadratic-time algorithm for general multivariate polynomial
  interpolation},}\ }\href@noop {} {\bibfield  {journal} {\bibinfo  {journal}
  {arXiv preprint arXiv:1710.10846}\ } (\bibinfo {year} {2017})}\BibitemShut
  {NoStop}%
\bibitem [{\citenamefont {Hecht}\ \emph {et~al.}(2018)\citenamefont {Hecht},
  \citenamefont {Hoffmann}, \citenamefont {Cheeseman},\ and\ \citenamefont
  {Sbalzarini}}]{PIP2}%
  \BibitemOpen
  \bibfield  {author} {\bibinfo {author} {\bibfnamefont {Michael}\ \bibnamefont
  {Hecht}}, \bibinfo {author} {\bibfnamefont {Karl~B.}\ \bibnamefont
  {Hoffmann}}, \bibinfo {author} {\bibfnamefont {Bevan~L}\ \bibnamefont
  {Cheeseman}}, \ and\ \bibinfo {author} {\bibfnamefont {Ivo~F}\ \bibnamefont
  {Sbalzarini}},\ }\bibfield  {title} {\enquote {\bibinfo {title} {Multivariate
  {N}ewton interpolation},}\ }\href@noop {} {\bibfield  {journal} {\bibinfo
  {journal} {arXiv preprint arXiv:1812.04256}\ } (\bibinfo {year}
  {2018})}\BibitemShut {NoStop}%
\bibitem [{\citenamefont {Hecht}\ \emph {et~al.}(2020)\citenamefont {Hecht},
  \citenamefont {Gonciarz}, \citenamefont {Michelfeit}, \citenamefont
  {Sivkin},\ and\ \citenamefont {Sbalzarini}}]{MIP}%
  \BibitemOpen
  \bibfield  {author} {\bibinfo {author} {\bibfnamefont {Michael}\ \bibnamefont
  {Hecht}}, \bibinfo {author} {\bibfnamefont {Krzysztof}\ \bibnamefont
  {Gonciarz}}, \bibinfo {author} {\bibfnamefont {Jannik}\ \bibnamefont
  {Michelfeit}}, \bibinfo {author} {\bibfnamefont {Vladimir}\ \bibnamefont
  {Sivkin}}, \ and\ \bibinfo {author} {\bibfnamefont {Ivo~F}\ \bibnamefont
  {Sbalzarini}},\ }\bibfield  {title} {\enquote {\bibinfo {title} {Multivariate
  interpolation in unisolvent nodes--lifting the curse of dimensionality},}\
  }\href@noop {} {\bibfield  {journal} {\bibinfo  {journal} {arXiv preprint
  arXiv:2010.10824}\ } (\bibinfo {year} {2020})}\BibitemShut {NoStop}%
\bibitem [{\citenamefont {Hecht}\ and\ \citenamefont
  {Sbalzarini}(2018)}]{IEEE}%
  \BibitemOpen
  \bibfield  {author} {\bibinfo {author} {\bibfnamefont {Michael}\ \bibnamefont
  {Hecht}}\ and\ \bibinfo {author} {\bibfnamefont {Ivo~F.}\ \bibnamefont
  {Sbalzarini}},\ }\bibfield  {title} {\enquote {\bibinfo {title} {Fast
  interpolation and {F}ourier transform in high-dimensional spaces},}\ }in\
  \href@noop {} {\emph {\bibinfo {booktitle} {Intelligent Computing. Proc. 2018
  IEEE Computing Conf., Vol. 2,}}},\ \bibinfo {series} {Advances in Intelligent
  Systems and Computing}, Vol.\ \bibinfo {volume} {857},\ \bibinfo {editor}
  {edited by\ \bibinfo {editor} {\bibfnamefont {K.}~\bibnamefont {Arai}},
  \bibinfo {editor} {\bibfnamefont {S.}~\bibnamefont {Kapoor}}, \ and\ \bibinfo
  {editor} {\bibfnamefont {R.}~\bibnamefont {Bhatia}}}\ (\bibinfo  {publisher}
  {Springer Nature},\ \bibinfo {address} {London, UK},\ \bibinfo {year}
  {2018})\ pp.\ \bibinfo {pages} {53--75}\BibitemShut {NoStop}%
\bibitem [{\citenamefont {Hernandez~Acosta}\ \emph {et~al.}(2021)\citenamefont
  {Hernandez~Acosta}, \citenamefont {Krishnan Thekke~Veettil}, \citenamefont
  {Wicaksono},\ and\ \citenamefont {Hecht}}]{minterpy}%
  \BibitemOpen
  \bibfield  {author} {\bibinfo {author} {\bibfnamefont {Uwe}\ \bibnamefont
  {Hernandez~Acosta}}, \bibinfo {author} {\bibfnamefont {Sachin}\ \bibnamefont
  {Krishnan Thekke~Veettil}}, \bibinfo {author} {\bibfnamefont {Damar}\
  \bibnamefont {Wicaksono}}, \ and\ \bibinfo {author} {\bibfnamefont {Michael}\
  \bibnamefont {Hecht}},\ }\bibfield  {title} {\enquote {\bibinfo {title} {{\sc
  minterpy} - multivariate interpolation in python},}\ }\href@noop {}
  {\bibfield  {journal} {\bibinfo  {journal}
  {https://github.com/casus/minterpy/}\ } (\bibinfo {year} {2021})}\BibitemShut
  {NoStop}%
\bibitem [{\citenamefont {Trefethen}(2019)}]{trefethen2019}%
  \BibitemOpen
  \bibfield  {author} {\bibinfo {author} {\bibfnamefont {Lloyd~N.}\
  \bibnamefont {Trefethen}},\ }\href@noop {} {\emph {\bibinfo {title}
  {Approximation theory and approximation practice}}},\ Vol.\ \bibinfo {volume}
  {164}\ (\bibinfo  {publisher} {SIAM},\ \bibinfo {year} {2019})\BibitemShut
  {NoStop}%
\end{thebibliography}%
\end{document}